# Dynamic Critical Behavior of a Swendsen-Wang-Type Algorithm for the Ashkin-Teller Model

Jesús Salas
Alan D. Sokal
*Department of Physics*
*New York University*
*4 Washington Place*
*New York, NY 10003 USA*
SALAS@MAFALDA.PHYSICS.NYU.EDU, SOKAL@NYU.EDU

November 20, 1995

**Abstract**

We study the dynamic critical behavior of a Swendsen–Wang–type algorithm for the Ashkin–Teller model. We find that the Li–Sokal bound on the autocorrelation time ($\tau_{\mathrm{int},\mathcal{E}} \geq \mathrm{const} \times C_H$) holds along the self-dual curve of the symmetric Ashkin–Teller model, and is almost but not quite sharp. The ratio $\tau_{\mathrm{int},\mathcal{E}}/C_H$ appears to tend to infinity either as a logarithm or as a small power ($0.05 \lesssim p \lesssim 0.12$). In an appendix we discuss the problem of extracting estimates of the exponential autocorrelation time.

# 1 Introduction

Critical slowing-down has become one of the main limiting factors of the state of art of Monte Carlo simulations [1,2,3,4]. The autocorrelation time $\tau$, which roughly measures the Monte Carlo time between two statistically independent configurations, diverges near a critical point. More precisely, for a finite system of linear size $L$ at criticality, we expect a behavior $\tau \sim L^z$ for large $L$: here the power $z$ is a *dynamic critical exponent*, which characterizes the dynamic universality class of the Monte Carlo algorithm. The traditional local algorithms (such as single-site Metropolis) have a dynamic critical exponent $z \gtrsim 2$. This is a severe critical slowing-down, in which the amount of computer work needed to study a lattice of size $L$ grows approximately $L^2$ times faster than the naive geometrical factor $L^d$ ($d$ being the dimensionality of the lattice). To study the static critical behavior we need high-precision data (run lengths $\gtrsim 10^3\tau$). In practice, it is very difficult to obtain high-precision data for large lattices with this kind of algorithm. To study the dynamic critical behavior, the situation is even worse, as we need much higher statistics (run lengths $\gtrsim 10^4\tau$). The geometrical factor $L^d$ is unavoidable for the usual Monte Carlo simulations, so the elimination (or reduction) of the critical slowing-down is the only way to make Monte Carlo simulations feasible close to a critical point.

Dramatic progress in this direction was stimulated by the introduction of the so-called cluster algorithms [5]. Instead of sequentially updating the whole lattice by single-spin moves, these algorithms employ non-local moves, such as cluster flips. For the ferromagnetic $q$-state Potts model, the Swendsen–Wang (SW) cluster algorithm [5] achieves a a significant reduction in $z$ compared to the local algorithms: one has $z$ between 0 and $\approx 1$, where the exact value depends on the number of Potts states and on the dimensionality of the lattice [4]. The two-dimensional (2D) Ising model is the most favorable case: the critical slowing-down becomes extremely weak, with estimates from different workers ranging from $z = 0.35 \pm 0.01$ [5] to $z = 0.25 \pm 0.01$ [6,7] to $z = 0 \times \log$ (i.e., $\tau \sim \log L$) [8]. Unfortunately, it is very hard to distinguish between the power-law and logarithmic scenarios using only lattices with $L \leq 512$ [6,7]. In other cases, the performance of the SW algorithm is less impressive (though still quite good): e.g., $z = 0.55 \pm 0.03$ for the 2D 3-state Potts model [9], and $z \approx 1$ for the 2D 4-state Potts model [9] and for the 4D Ising model [10,11]. Clearly, we would like to understand why this algorithm works so well in some cases and not in others; we hope in this way to obtain new insights into the dynamics of non-local Monte Carlo algorithms, with the ultimate aim of devising new and more efficient algorithms.

A single-cluster variant of the SW algorithm was introduced by Wolff [12]. Instead of updating all the clusters (with a given probability), only one cluster is selected and updated. It is not known why the dynamic exponents $z_{1C}$ associated to the single-cluster dynamics are very close to those of the SW dynamics in some cases (e.g., 2D $q = 2, 3$ Potts models) but not in other cases (e.g., Ising model in dimension $d \geq 3$) [12,13]. *A priori* one would expect the two algorithms to belong to different dynamic universality classes.

There is at present no adequate theory for predicting the dynamic critical behavior of an SW-type algorithm. However, there is one rigorous lower bound on $z$. In 1989 Li and Sokal [9] showed that the autocorrelation times of the standard (multi-cluster) SW algorithm for the ferromagnetic $q$-state Potts model are bounded below by a multiple of the specific heat:

$$\tau_{\text{int},\mathcal{N}},\ \tau_{\text{int},\mathcal{E}},\ \tau_{\text{exp}} \;\geq\; \text{const} \times C_H \; . \tag{1.1}$$

Here $\mathcal{N}$ is the bond density in the SW algorithm, $\mathcal{E}$ is the energy, and $C_H$ is the specific heat; $\tau_{\text{int}}$ and $\tau_{\text{exp}}$ denote the integrated and exponential autocorrelation times, respectively [1,4]. As a result one has

$$z_{\text{int},\mathcal{N}},\ z_{\text{int},\mathcal{E}},\ z_{\text{exp}} \;\geq\; \frac{\alpha}{\nu} \; , \tag{1.2}$$

where $\alpha$ and $\nu$ are the standard *static* critical exponents. Thus, the SW algorithm for the $q$-state Potts model cannot *completely* eliminate the critical slowing-down in any model in which the specific heat is divergent at criticality (although one might hope to obtain $z = 0 \times \log$ in the the 2D Ising model, where the specific heat is only logarithmically divergent). Now, one would like to know whether the bound (1.2) on critical exponents is *sharp*: that is, does it hold as *equality*, or only as a *strict inequality*? In more detail, one would like to know whether the bound (1.1) on the autocorrelation times is *sharp* ($\tau/C_H$ bounded), *sharp modulo a logarithm* ($\tau/C_H \sim \log^p L$), or *not sharp* ($\tau/C_H \sim L^p$ with $p > 0$).

Unfortunately, the empirical situation for the 2D Potts models is not very clear. For the Ising case, the bound (1.2) would be sharp if (and only if) the autocorrelation time grows like a logarithm; this is consistent with the data but not demanded by it [5,6,7,8]. For the 3-state Potts model, the bound is apparently not sharp: we have $z = 0.55 \pm 0.03$ [9] versus $\alpha/\nu = \frac{2}{5} = 0.4$ [14]. The 4-state Potts model is rather peculiar: the naive fit to the data, $z = 0.89 \pm 0.05$ [9], is *smaller* than the (exactly known) value of $\alpha/\nu = 1$ [15]. The explanation of this paradox is that the true leading term in the specific heat has a multiplicative logarithmic correction, $C_H \sim L \log^{-3/2} L$ [16,17,18]; and indeed the *observed* exponent $\alpha/\nu$ (from a naive power-law fit) is $0.75 \pm 0.01$ [9], consistent with the bound (1.2). It is reasonable to conjecture that the true behavior of the autocorrelation time is likewise of the form $\tau \sim L \log^p L$ (with $p > -\frac{3}{2}$), in which case the bound (1.1) would be sharp modulo a multiplicative logarithm.

So we are in a strange situation: the Li–Sokal bound *might* be sharp (possibly modulo a logarithm) for the 2D Potts models with $q = 2$ and $q = 4$, but it is apparently *not* sharp for $q = 3$.[1]

There is yet another way of "interpolating" between the 2-state (Ising) and 4-state Potts models: both are special cases of the Ashkin–Teller (AT) model [23], which has *two* interacting Ising spins on each lattice site. (For a review of the AT model, see Section 2 below.) The symmetric AT model (which is equivalent to the general $Z_4$ clock model) presents a very rich phase diagram. In particular, one of the critical curves (namely, the self-dual curve) is quite unusual: the critical exponents vary continuously along this curve (as a result of a marginal operator), thus violating the usual notion of universality. One point on this critical curve is precisely the 4-state Potts model at criticality, while another point on this curve corresponds to a pair of decoupled Ising models. Thus, new insights on the sharpness of the Li–Sokal bound in the 2-state and 4-state Potts models might be obtainable by studying the same question on the self-dual curve of the symmetric AT model.

[1]For Ising models in lattice dimensions $d \geq 3$, the bound (1.2) is clearly *not* sharp. For the 3D Ising model, estimates of $z$ range from $0.339 \pm 0.004$ to $0.75 \pm 0.01$ [5,19,20], while $\alpha/\nu \approx 0.17$ [21,22]. For Ising models in dimension $d \geq 4$, we expect $z = 1$ (possibly modulo a logarithm in $d = 4$) [10,11], while of course $\alpha/\nu = 0$ (or $0 \times \log^{2/3}$ in $d = 4$).

Wiseman and Domany [24] devised the first SW–type algorithm for the AT model. Though their method of derivation is rather complicated, the algorithm is simple and reduces to the well-known SW algorithms in the special cases of the Ising and 4-state Potts models. The same algorithm had been independently introduced by Laanait *et al.* in another context [25]: they studied a model closely related to the AT model, and they used the same SW–type algorithm as a tool for their rigorous proofs.

In this paper we would like to address the issue of the sharpness of the Li–Sokal bound (1.1)/(1.2) along the self-dual curve of the symmetric AT model, and in particular for the 4-state Potts model. We propose two variants of the algorithm. The first, which we call the "direct" algorithm, is essentially the same as that of Wiseman and Domany [24]; however, we think that our derivation is simpler. (The reader can judge!) The second variant, which we call the "embedding" algorithm, is somewhat simpler to implement in practice; it is *not* equivalent to the direct algorithm, although we expect it to lie in the same dynamic universality class.

We have studied numerically the multi-cluster ("standard SW") version of the embedding algorithm[2] at three points on the AT self-dual curve: the 4-state Potts model and two additional models (ZF and X2) interpolating between the 4-state Potts and Ising models. We have used lattices up to $L = 512$ (as well as $L = 1024$ for the 4-state Potts model), and have systematically employed finite-size-scaling techniques to analyze the numerical data. We have also reanalyzed the very precise data reported by Baillie and Coddington [7] for the 2D Ising model on lattices up to $L = 512$. Our results are the following:

1. The Li–Sokal bound is satisfied on the AT self-dual curve. (Indeed, for the *direct* algorithm we are able to prove this rigorously.)

2. The bound is rather close to being sharp for a generic point on the AT self-dual curve. Using power-law fits for the quantity $\tau_{\mathrm{int},\mathcal{E}}/C_H$, we obtain estimates of

[2] By contrast, Wiseman and Domany [24] studied the *single-cluster* ("Wolff") version of the direct algorithm (in the single-cluster context the direct and embedding algorithms turn out to be equivalent). It is important that both the multi-cluster and single-cluster versions be studied, as they may well lie in different dynamic universality classes. We concentrate here on the multi-cluster version, because it is only for this version of cluster algorithms that the Li–Sokal bound (1.1)/(1.2) is known.

$z_{\text{int},\mathcal{E}} - \alpha/\nu$ ranging from $\approx 0.05$ to $\approx 0.12$. The lower value corresponds to the Ising and X2 cases, and the higher value to the 4-state Potts model.

3. In all cases the data are *consistent* with a logarithmic growth of $\tau_{\text{int},\mathcal{E}}/C_H$ as $A + B \log L$. For the Ising and the ZF models, a logarithmic behavior $A \log^p L$ with $p \approx 0.31$ also gives a reasonable fit.

4. In all cases, the data are consistent with the boundedness of $\tau_{\text{int},\mathcal{E}}/C_H$ as $L \to \infty$ *only if* one assumes rather strong corrections to scaling, i.e., $A + BL^{-\Delta}$ with $\frac{1}{8} \lesssim \Delta \lesssim \frac{1}{4}$. Moreover, for the 4-state Potts model this scenario implies an implausibly large value for the coefficient $B$ ($|B| \gtrsim 10$). In all cases the $A + BL^{-\Delta}$ fit is inferior to the $AL^p$ and $A + B \log L$ fits.

5. If we believe (on theoretical grounds) that there should be some *continuity* in the behavior of the ratio $\tau_{\text{int},\mathcal{E}}/C_H$ along the self-dual curve of the AT model, then the possible scenarios reduce to the pure power-law and simple logarithmic $A + B \log L$ behaviors.

Thus, the bound (1.1) on the autocorrelation time is not sharp, but it might be sharp modulo a logarithm at some or all points of the AT self-dual curve. Further studies on significantly larger lattices will be required in order to distinguish convincingly between a logarithmic and a small-power-law growth.

This paper is organized as follows: Section 2 reviews the definition and properties of the AT model: phase diagram, critical exponents, etc. In Section 3 we construct our SW–type algorithms for the AT model, and we relate them to other SW-type algorithms. In Section 4 we present and analyze our numerical results for three selected points on the AT self-dual curve. The sharpness of the Li–Sokal bound is also discussed. Finally, in Section 5 we summarize our conclusions. In Appendix A we provide a rigorous proof of the Li–Sokal bound for the direct AT algorithm. In Appendix B we explain in detail how we performed the fits of the autocorrelation functions to extract estimates of the exponential autocorrelation times.

# 2 The Ashkin–Teller model

The Ashkin–Teller (AT) model [23] is a generalization of the Ising model to a four-state model. To each lattice site $x$ we assign two Ising spins $\sigma_x = \pm 1$ and $\tau_x = \pm 1$, and they interact through the Hamiltonian

$$H_{\text{AT}} \;=\; -J \sum_{\langle xy \rangle} \sigma_x \sigma_y - J' \sum_{\langle xy \rangle} \tau_x \tau_y - K \sum_{\langle xy \rangle} \sigma_x \tau_x \sigma_y \tau_y \;, \qquad (2.1)$$

where the sums run over nearest-neighbor pairs $\langle xy \rangle$. It can be interpreted as two Ising models with nearest-neighbor couplings $J$ and $J'$ and interacting via a four-spin coupling $K$. Note that the fields $\sigma$, $\tau$ and $\sigma\tau$ play symmetric roles in this model; we can consider any two of these three as the "fundamental fields".

This model contains as particular cases some other well-known systems. The plane $K = 0$ in the coupling-constant space $(J, J', K)$ corresponds to a pair of decoupled Ising models, one with coupling $J$ and the other with coupling $J'$. At the other extreme, the limit $K \to +\infty$ corresponds to a single Ising model ($\sigma = \tau$) with coupling $J + J'$. Finally, the line $J = J' = K$ is the 4-state Potts model with $J_{\rm Potts} = 4J$.

The family (2.1) of AT Hamiltonians exhibits several symmetries. First of all, we can permute freely the spin variables $(\sigma, \tau, \sigma\tau)$. This implies that the AT model is mapped onto an essentially equivalent model under any permutation of the couplings $(J, J', K)$. Moreover, if the lattice is *bipartite* we can flip $\sigma$ or $\tau$ or both on one of the two sublattices (in other words, we choose *exactly two* of the three variables $\sigma$, $\tau$, $\sigma\tau$ and flip these on the chosen sublattice). This implies that the AT model is mapped onto an essentially equivalent model under the following transformations:

$$(J, J', K) \to (-J, J', -K) \tag{2.2a}$$

$$(J, J', K) \to (J, -J', -K) \tag{2.2b}$$

$$(J, J', K) \to (-J, -J', K) \tag{2.2c}$$

These transformations will be useful in Section 3.

We are mainly interested in one particular case of the Hamiltonian (2.1): the symmetric[3] AT model characterized by $J = J'$,

$$H_{\rm SAT} \;=\; -J \sum_{\langle xy \rangle} (\sigma_x \sigma_y + \tau_x \tau_y) - K \sum_{\langle xy \rangle} \sigma_x \tau_x \sigma_y \tau_y \; . \tag{2.3}$$

This is exactly the general $Z_4$ clock model[4]

$$H_{\rm clock} \;=\; -2J \sum_{\langle xy \rangle} \cos(\theta_x - \theta_y) - K \sum_{\langle xy \rangle} \cos(2\theta_x - 2\theta_y) \; , \tag{2.4}$$

where the dynamical variables take the values $\theta_x \in \{0, \pi/2, \pi, 3\pi/2\}$. The relation can be easily seen by setting $\sigma_x = \sqrt{2}\cos(\theta_x - \pi/4)$ and $\tau_x = \sqrt{2}\cos(\theta_x + \pi/4)$.

The AT model exhibits a rich phase diagram, in both two [26,27,28] and three [26] dimensions. Here we concentrate on the two-dimensional square-lattice symmetric AT model. Although we do not know how to solve this model analytically, we have a fairly good understanding of its phase diagram (see Figure 1). From (2.2c) we see that in the symmetric AT model a sublattice flip of $\sigma$ and $\tau$ corresponds to the change $J \to -J$; it follows that the phase diagram is symmetric under reflection in the $J = 0$ axis, under which ferromagnetic $\sigma$ and $\tau$ ordering becomes antiferromagnetic (AF) and vice versa. For this reason, we show in Figure 1 only the half-plane $J \geq 0$.

The line $K = 0$ corresponds to a pair of decoupled Ising models, so there are Ising critical points at $(J, K) = (\pm\frac{1}{2}\log(1+\sqrt{2}), 0)$. Point DIs in Figure 1 represents the one with the plus sign (i.e., the ferromagnetic one). The model with $J = 0$ is again an Ising model, but in the variable $\sigma\tau$. There are thus additional critical Ising points $(J, K) = (0, \pm\frac{1}{2}\log(1+\sqrt{2}))$. The one with the plus sign (point Is in Figure 1) is ferromagnetic, while the one with the minus sign (AFIs) is antiferromagnetic. Finally, in the limit $K \to +\infty$ we find Ising transition points at $J = \pm\frac{1}{4}\log(1+\sqrt{2})$. The one with the plus sign is ferromagnetic and corresponds to the point Is$'$ of Figure 1. The other one is antiferromagnetic and is not depicted in Figure 1.

---

[3]Baxter [27] calls this model the "isotropic" AT model. We prefer not to use this terminology, in order to avoid confusion with *spatial* isotropy or anisotropy.

[4]More precisely, this is the general $Z_4$ clock model on an *undirected* graph. On a *directed* graph (i.e., one with *oriented* bonds $\langle xy \rangle$), the interaction term $\sin(\theta_x - \theta_y) = \frac{1}{2}(\sigma_x \tau_y - \sigma_y \tau_x)$ is also allowed.

The line $J = K$ corresponds to the 4-state Potts-model subspace (right dash-dotted line in Figure 1). Therefore, there is a ferromagnetic critical point at $J = K = \frac{1}{4}\log 3$ (point P in Figure 1). The AF regime ($J = K < 0$) is more subtle. There is no rigorous result concerning the existence or non-existence of a critical point in the AF 4-state Potts model. However, there is a strong numerical indication [29] that this model is non-critical, even at zero temperature: indeed, the second-moment correlation length $\xi$ is $\lesssim 2$ lattice spacings at all temperatures, uniformly down to $T = 0$.[5] The absence of a critical point along the line $J = -K$ (left dash-dotted line in Figure 1) follows immediately using the $J \to -J$ invariance.

The AT model on any planar graph can be mapped into another AT model on the dual graph [27,30,31]. The duality transformation is best viewed in terms of the Boltzmann weights[6]

$$\omega_0 = e^{J+J'+K+J_0} \quad (2.5a)$$
$$\omega_1 = e^{J-J'-K+J_0} \quad (2.5b)$$
$$\omega_2 = e^{-J+J'-K+J_0} \quad (2.5c)$$
$$\omega_3 = e^{-J-J'+K+J_0} \quad (2.5d)$$

where $J_0$ is an arbitrary constant fixing the zero of energy. The AT model with Boltzmann weights $(\omega_0, \omega_1, \omega_2, \omega_3)$ is mapped by duality to a new AT model with weights $(\widetilde{\omega}_0, \widetilde{\omega}_1, \widetilde{\omega}_2, \widetilde{\omega}_3)$ given by[7,8]

$$\widetilde{\omega}_0 = \tfrac{1}{2}(\omega_0 + \omega_1 + \omega_2 + \omega_3) \quad (2.6a)$$
$$\widetilde{\omega}_1 = \tfrac{1}{2}(\omega_0 + \omega_1 - \omega_2 - \omega_3) \quad (2.6b)$$
$$\widetilde{\omega}_2 = \tfrac{1}{2}(\omega_0 - \omega_1 + \omega_2 - \omega_3) \quad (2.6c)$$
$$\widetilde{\omega}_3 = \tfrac{1}{2}(\omega_0 - \omega_1 - \omega_2 + \omega_3) \quad (2.6d)$$

[5]The critical properties of the antiferromagnetic $q$-state Potts model depend strongly on the lattice structure. For instance, the AF 3-state Potts model has a transition at non-zero temperature on the triangular lattice [32], has a critical point at zero temperature on the square lattice [29,33,34,35], and is expected to be non-critical at all temperatures on the hexagonal lattice.

[6]The four energy states on a bond $\langle xy \rangle$ are labeled 0,1,2,3 as follows: the high-order bit is $(1 - \sigma_x\sigma_y)/2$ and the low-order bit is $(1 - \tau_x\tau_y)/2$.

[7]Note that if the original weights $\omega_i$ are normalized so that $J_0 = 0$, the dual weights $\widetilde{\omega}_i$ do *not* necessarily have this normalization.

[8]This duality transformation corresponds to the Fourier transform on $Z_2 \times Z_2$ *followed by interchange of* $\sigma$ *and* $\tau$ (i.e., interchange of $\omega_1$ and $\omega_2$).

The symmetric AT model (the one with $\omega_1 = \omega_2$) is clearly mapped under duality into another symmetric AT model (i.e., $\tilde{\omega}_1 = \tilde{\omega}_2$). Specializing to the square lattice, the dual graph is again a square lattice, and the self-dual manifold of (2.6) is

$$\omega_0 \;=\; \omega_1 + \omega_2 + \omega_3 \; . \tag{2.7}$$

For the symmetric AT model on the square lattice, the self-duality condition (2.7) can be easily written in terms of the coupling constants:

$$e^{-2K} \;=\; \sinh 2J \; . \tag{2.8}$$

This is represented in Figure 1 by the curve B–DIs–P–C.

Furthermore, the AT model on any planar graph can be mapped onto an 8-vertex model on the medial graph [31]. In particular, the AT model on the square lattice can be mapped [27] onto a *staggered* 8-vertex model on the square lattice (which has not been exactly solved in general). As a special case, the AT model *on the self-dual manifold* (2.7) maps onto a *homogeneous* 8-vertex model, which is exactly soluble. Furthermore, the *symmetric* self-dual AT model (2.8) maps (after a simple further transformation) onto a homogeneous 6-vertex model. In this way, Baxter showed that the self-dual curve (2.8) is critical only for $K \leq \frac{1}{4}\log 3$ (solid curve in Figure 1), and is non-critical for $K > \frac{1}{4}\log 3$ (dotted curve in Figure 1). The critical part belongs to the universality classes of the conformal field theories with central charge $c = 1$ (i.e., it can be related to the Gaussian model [36]). Along this line the critical exponents vary continuously and they are known exactly (see below).

From series expansions [26], mean-field theory and approximate real-space renormalization-group calculations [28], we know that two critical curves emerge from the Potts point P, one going to the Ising critical point Is and the other one going to the Ising critical point Is$'$ at $K = +\infty$.[9] The critical curves P–Is and P–Is$'$ map into one another under the duality relation (2.6) [27]. Finally, there is another critical curve emerging from the point AFIs and pointing towards $K \to -\infty$. The exact location of these three curves is unknown, as is their universality class. However, most people believe that they are Ising-like. In [38] it is argued that these critical curves should be given by non-algebraic functions.

The four critical curves mentioned above are the borderlines of the four phases appearing in this model for $J > 0$. These phases are[10]:

I. This is the so-called Baxter phase [26]. The spins $\sigma$ and $\tau$ are independently ferromagnetically ordered. There are thus four extremal infinite-volume Gibbs measures according to the signs of $\langle\sigma\rangle$ and $\langle\tau\rangle$ (which may be chosen independently); the sign of $\langle\sigma\tau\rangle$ is then equal to that of $\langle\sigma\rangle\langle\tau\rangle$.

[9] Pfister [37] has proven the existence of two phase transitions (i.e., of the three phases II, III and I in succession) along any ray in the quadrant $J, K > 0$ with slope $0 < J/K < \frac{1}{2}$. More generally, this applies in the full AT model along any ray in the octant $J, J', K > 0$ with slope $0 < (J + J')/K < 1$.

[10] We follow the terminology used in Baxter's book [27].

II. This is the paramagnetic phase, in which all three spins $\sigma$, $\tau$ and $\sigma\tau$ are disordered. There is a unique infinite-volume Gibbs measure.

III. In this phase both the spins $\sigma$ and $\tau$ are disordered (i.e., $\lim_{|x-y|\to\infty}\langle\sigma_x\sigma_y\rangle = \lim_{|x-y|\to\infty}\langle\tau_x\tau_y\rangle = 0$), but their product $\sigma\tau$ is ferromagnetically ordered (i.e., $\lim_{|x-y|\to\infty}\langle\sigma_x\tau_x\sigma_y\tau_y\rangle > 0$). There are two extremal infinite-volume Gibbs measures according to the sign of $\langle\sigma\tau\rangle$.

IV. This is the antiferromagnetic analogue of Phase III: the spins $\sigma$ and $\tau$ are disordered, while the product $\sigma\tau$ is antiferromagnetically ordered (i.e., $\lim_{|x-y|\to\infty} (-1)^{|x-y|}\langle\sigma_x\tau_x\sigma_y\tau_y\rangle > 0$). There are again two extremal infinite-volume Gibbs measures, according to the sign of the sublattice magnetization $(-1)^{|x|}\langle\sigma_x\tau_x\rangle$.

This picture has been proven rigorously for low temperature and arbitrary spatial dimension, using Pigorov–Sinai theory [39]. In particular, it was shown that deep in region I there exist four periodic extremal Gibbs measures, and deep in regions III and IV there exist two periodic extremal Gibbs measures.[11]

The critical exponents along the self-dual curve can be obtained by relating the AT model to the 8-vertex model or to the Gaussian model [36,41,42]. We parametrize the critical part of the self-dual curve by

$$e^{4J} = \frac{\sqrt{2+2\cos\mu}+1}{\sqrt{2+2\cos\mu}-1} \tag{2.9a}$$

$$e^{4K} = 1+2\cos\mu \tag{2.9b}$$

where $0 \le \mu \le 2\pi/3$. This parameter $\mu$ is related to the coupling constant $g$ of the Gaussian model [43][12] by $\mu = \pi(1-g/4)$, so that $\frac{4}{3} \le g \le 4$. Thus, $g = \frac{4}{3}$ corresponds to the point at $K = -\infty$ (B in Figure 1), $g = 2$ is the decoupled Ising model, $g = 3$ is the model considered by Zamolodchikov and Fateev [45], and $g = 4$ is the 4-state Potts model. The critical exponents along the self-dual curve are given by

$$\nu = \frac{2-y}{3-2y} \tag{2.10a}$$

$$\frac{\alpha}{\nu} = \frac{2-2y}{2-y} \tag{2.10b}$$

$$\frac{\gamma}{\nu} = \frac{7}{4} \tag{2.10c}$$

$$\frac{\gamma'}{\nu} = \frac{7-4y}{4-2y} \tag{2.10d}$$

[11]Note the order of adjectives: Pigorov–Sinai theory studies extremal Gibbs measures *that happen to be periodic*; it says nothing about nonperiodic Gibbs measures (e.g., those with interfaces) or about periodic Gibbs measures that are extremal only within the restricted class of *periodic* Gibbs measures. For further discussion, see [40, Section B.3.1].

[12]Our $g$ corresponds to that of Saleur [43], and equals $2\pi$ times the $K$ of Kadanoff and Brown [36] and Yang [44].

where the parameter $y$ is related to $\mu$ by

$$y = \frac{2\mu}{\pi} = 2 - \frac{g}{2} \tag{2.11}$$

with $0 \leq y \leq \frac{4}{3}$. Here $\alpha$ is the specific-heat exponent, while $\gamma$ (resp. $\gamma'$) is the susceptibility exponent for $\sigma$ and $\tau$ (resp. for $\sigma\tau$). We have chosen to specify the ratios (2.10b-d) because these are directly measurable by our Monte Carlo methods. For $y > 1$ (corresponding to $K < 0$), the specific heat has a cusp singularity ($\alpha < 0$) rather than a divergence, but both susceptibilities remain divergent.

The region between the decoupled Ising model (DIs) and the critical 4-state Potts model (P) is the most interesting. It is worth mentioning that both the $q$-state Potts model at criticality and the symmetric AT on the self-dual curve can be represented as certain 6-vertex models [27,41,46]. By relating these two 6-vertex models, we can map the former model onto the latter one, and use $q$ as a parametrization of this subset of the AT self-dual curve. For the square lattice, the $q$-state Potts model at criticality is mapped to the point given by (2.9) with $2\cos\mu = \sqrt{q}$. Thus, the case $q = 0$ is mapped to $K = 0$ (i.e., the decoupled Ising model), $q = 2$ to the model considered by Zamolodchikov and Fateev [45], and $q = 4$ to the point $J = K = \frac{1}{4}\log 3$ (i.e., the 4-state Potts model).

# 3 The Algorithm

## 3.1 Direct algorithm

The idea behind this new algorithm is the same as that of all Swendsen–Wang–type algorithms [47][13]: we decompose the Boltzmann weight by introducing new dynamical variables (living on the bonds of the lattice), and we then simulate the joint model of old and new variables by alternately updating one set of variables conditional on the other set. As we have two distinct sets of Ising spins, we expect to introduce two distinct sets of auxiliary variables.

We begin by enumerating the possible energy values which can occur on a given bond $\langle xy \rangle$. Out of the 16 spin configurations on each bond, there are only four different energy values (see Table 1). We can order these energies in increasing order, but this ordering of course depends on the relative values of the coupling constants $J$, $J'$ and $K$. Instead of developing a different algorithm for each possible ordering, we can use the symmetries (2.2a–c) of the general AT Hamiltonian (2.1) and choose[14] an equivalent general AT model satisfying

$$J, J' \geq |K| \,. \tag{3.1}$$

For such a model, the energies satisfy

$$0 = E_0 \;\leq\; E_1, E_2 \;\leq\; E_3 \,. \tag{3.2}$$

[13] For a pedagogical presentation, see [1, Section 6] or [4, Section 4].

[14] At least if the lattice is bipartite.

In particular, for a given bond, the lowest-energy state is the one with the two $\sigma$ spins parallel and the two $\tau$ spins parallel. We shall hereafter assume that (3.1) holds.

**Remark.** The following algorithm is also valid for a non-homogeneous AT model on an arbitrary finite graph, with Hamiltonian

$$H_{AT} \;=\; -\sum_{\langle xy\rangle} J_{xy}\sigma_x\sigma_y - \sum_{\langle xy\rangle} J'_{xy}\tau_x\tau_y - \sum_{\langle xy\rangle} K_{xy}\sigma_x\tau_x\sigma_y\tau_y \tag{3.3}$$

satisfying

$$J_{xy}, J'_{xy} \ge |K_{xy}| \tag{3.4}$$

for every bond $\langle xy\rangle$. It suffices to make the obvious notational alterations.

The Boltzmann weight associated with the bond $\langle xy\rangle$ is equal to

$$\begin{aligned} W_{\rm bond}(\sigma_x,\sigma_y,\tau_x,\tau_y) \;=\;& e^{-2(J+J')} \;+ \\ & e^{-2J'}\left[e^{-2K} - e^{-2J}\right]\delta_{\sigma_x,\sigma_y} \;+ \\ & e^{-2J}\left[e^{-2K} - e^{-2J'}\right]\delta_{\tau_x,\tau_y} \;+ \\ & \left[1 - e^{-2(J'+K)} - e^{-2(J+K)} + e^{-2(J+J')}\right]\delta_{\sigma_x,\sigma_y}\delta_{\tau_x,\tau_y}\;. \end{aligned} \tag{3.5}$$

Let us now introduce the auxiliary variables $m_{xy}, n_{xy} = 0,1$ associated with the spins $\sigma$, $\tau$ respectively, and define the joint-model Boltzmann weight for the bond $\langle xy\rangle$ to be

$$\begin{aligned} W_{\rm bond}^{\rm joint}(\sigma_x,\sigma_y,\tau_x,\tau_y;m_{xy},n_{xy}) \;=\;& e^{-2(J+J')}\delta_{m_{xy},0}\delta_{n_{xy},0} \;+ \\ & e^{-2J'}\left[e^{-2K} - e^{-2J}\right]\delta_{\sigma_x,\sigma_y}\delta_{m_{xy},1}\delta_{n_{xy},0} \;+ \\ & e^{-2J}\left[e^{-2K} - e^{-2J'}\right]\delta_{\tau_x,\tau_y}\delta_{m_{xy},0}\delta_{n_{xy},1} \;+ \\ & \left[1 - e^{-2(J'+K)} - e^{-2(J+K)} + e^{-2(J+J')}\right]\delta_{\sigma_x,\sigma_y}\delta_{\tau_x,\tau_y}\delta_{m_{xy},1}\delta_{n_{xy},1}\;. \end{aligned} \tag{3.6}$$

If we sum (3.6) over $m_{xy}$ and $n_{xy}$, we obtain (3.5), as desired. The complete joint probability is then

$$W^{\rm joint}(\{\sigma,\tau\};\{m,n\}) \;=\; \frac{1}{Z}\prod_{xy} W_{\rm bond}^{\rm joint}(\sigma_x,\sigma_y,\tau_x,\tau_y;m_{xy},n_{xy})\;, \tag{3.7}$$

where $Z$ is the partition function of both the joint model and the original model:

$$Z \;=\; \sum_{\sigma,\tau=\pm1}\;\sum_{m,n=0,1}\;\prod_{\langle xy\rangle} W_{\rm bond}^{\rm joint}(\sigma_x,\sigma_y,\tau_x,\tau_y;m_{xy},n_{xy}) \tag{3.8a}$$

$$\;=\; \sum_{\sigma,\tau=\pm1}\;\prod_{\langle xy\rangle} W_{\rm bond}(\sigma_x,\sigma_y,\tau_x,\tau_y) \tag{3.8b}$$

Our SW–type algorithm consists in simulating the joint probability distribution (3.7) by alternately applying the conditional distributions on $\{\sigma,\tau\}$ and on $\{m,n\}$.

These two steps can be read off immediately from (3.6)/(3.7); in detail, they are the following:

**Step 1: Update of $\{m,n\}$ given $\{\sigma,\tau\}$.** Conditional on the $\{\sigma,\tau\}$ configuration, the bond variables $\{m,n\}$ are given independently for each bond. For a bond $\langle xy\rangle$ with spins $\sigma_x$, $\sigma_y$, $\tau_x$, $\tau_y$, we obtain the new bond variables $m_{xy}$ and $n_{xy}$ (independently of the old values) by the following rules:

1a) If $\sigma_x = \sigma_y$ and $\tau_x = \tau_y$, then we choose $(m_{xy}, n_{xy})$ with the following probabilities

$$(m_{xy}, n_{xy}) = (1,1) \text{ with } p_1 = 1 - e^{-2(J'+K)} - e^{-2(J+K)} + e^{-2(J+J')}$$
$$(m_{xy}, n_{xy}) = (1,0) \text{ with } p_2 = e^{-2J'}\left[e^{-2K} - e^{-2J}\right].$$
$$(m_{xy}, n_{xy}) = (0,1) \text{ with } p_3 = e^{-2J}\left[e^{-2K} - e^{-2J'}\right].$$
$$(m_{xy}, n_{xy}) = (0,0) \text{ with } p_4 = e^{-2(J+J')} = 1 - p_1 - p_2 - p_3.$$

1b) If $\sigma_x = \sigma_y$ and $\tau_x = -\tau_y$, then the probabilities are

$$(m_{xy}, n_{xy}) = (1,0) \text{ with } q_1 = 1 - e^{-2(J+K)}.$$
$$(m_{xy}, n_{xy}) = (0,0) \text{ with } q_2 = e^{-2(J+K)} = 1 - q_1.$$

1c) If $\sigma_x = -\sigma_y$ and $\tau_x = \tau_y$, the probabilities are

$$(m_{xy}, n_{xy}) = (0,1) \text{ with } r_1 = 1 - e^{-2(J'+K)}.$$
$$(m_{xy}, n_{xy}) = (0,0) \text{ with } r_2 = e^{-2(J'+K)} = 1 - r_1.$$

1d) If $\sigma_x = -\sigma_y$ and $\tau_x = -\tau_y$, we choose $(m_{xy}, n_{xy}) = (0,0)$ with probability 1.

All these choices are made independently for each bond $\langle xy\rangle$.

**Step 2: Update of $\{\sigma,\tau\}$ given $\{m,n\}$.** Given the bond-configuration $\{m,n\}$, we build all the connected clusters of $\sigma$ spins (resp. $\tau$ spins) joined by bonds with $m_{xy} = 1$ (resp. $n_{xy} = 1$). Within each cluster, the spin values are required to be equal, but this common value may be either $+1$ or $-1$. The spin value for each cluster is chosen randomly, independently of the old value and of the choices made for the other clusters.

One iteration of the direct algorithm consists of an application of Step 1 followed by an application of Step 2.

**Remarks.** 1. Wiseman and Domany [24] have introduced essentially this same decomposition of the Boltzmann weight, although their derivation is in our opinion more complicated. They then studied numerically the single-cluster ("Wolff") version of this algorithm. Here we study the many-cluster ("Swendsen–Wang") version.

2. This direct SW-type algorithm satisfies the Li–Sokal bound (1.1)/(1.2). The proof is a straightforward generalization of the one given in [9] for the Potts case; we present it in Appendix A.

3. We can generalize our SW–type algorithm to a "generalized Ashkin–Teller model" consisting of a $q$-state Potts variable $\sigma$ and an $r$-state Potts variable $\tau$ interacting through the Hamiltonian

$$H_{\rm GAT} \;=\; -2(J-K)\sum_{\langle xy\rangle}\delta_{\sigma_x\sigma_y} - 2(J'-K)\sum_{\langle xy\rangle}\delta_{\tau_x\tau_y} - 4K\sum_{\langle xy\rangle}\delta_{\sigma_x\sigma_y}\delta_{\tau_x\tau_y}\;. \qquad (3.9)$$

It is clear that from this Hamiltonian we obtain again the joint probability distribution (3.6). This model has been considered in [25] when $J' = K$.

## 3.2 The embedding algorithm

The algorithm presented in the preceding subsection is perfectly legal, but it is somewhat complicated to write the computer code for its Step 1 in an efficient way. In this section we introduce a variant algorithm in which we deal with only one kind of spin ($\sigma$ or $\tau$) at a time.

Consider the Boltzmann weight of a given bond $\langle xy\rangle$, *conditional on the* $\{\tau\}$ *configuration* (i.e., the $\tau$ spins are kept fixed): it is

$$W_{\rm bond}(\sigma_x,\sigma_y,\tau_x,\tau_y) \;=\; e^{-2(J+K\tau_x\tau_y)} + \left[1-e^{-2(J+K\tau_x\tau_y)}\right]\delta_{\sigma_x,\sigma_y}\;. \qquad (3.10)$$

We can simulate this system of $\sigma$ spins using a standard SW algorithm. The effective nearest-neighbor coupling

$$J^{\rm eff}_{xy} \;=\; J + K\tau_x\tau_y \qquad (3.11)$$

is no longer translation-invariant, but this does not matter. The key point is that the effective coupling is always *ferromagnetic*, due to the condition (3.1). An exactly analogous argument applies to the $\{\tau\}$ spins when the $\{\sigma\}$ spins are held fixed.

The embedding algorithm for the AT model has therefore two parts:

**Step 1: Update of $\{\sigma\}$ spins.** Given the $\{\tau\}$ configuration (which we hold fixed), we perform a standard SW iteration on the $\sigma$ spins. The probability $p_{xy}$ arising in the SW algorithm takes the value $p_{xy} = 1-\exp[-2(J+K\tau_x\tau_y)]$.

**Step 2: Update of $\{\tau\}$ spins.** Given the $\{\sigma\}$ configuration (which we hold fixed), we perform a standard SW iteration on the $\tau$ spins. The probability $p_{xy}$ arising in the SW algorithm takes the value $p_{xy} = 1-\exp[-2(J'+K\sigma_x\sigma_y)]$.

One iteration of the embedding algorithm consists, by definition, of a single application of Step 1 followed by a single application of Step 2.

Wiseman and Domany [24] also constructed a embedding version of their single-cluster algorithm. Furthermore, they showed that, in the single-cluster context, the direct and embedding algorithms define the *same* dynamics[15]; only the computer implementation is different. However, this equivalence does *not* hold for our many-cluster algorithm. In the direct algorithm we have independent clusters of $\sigma$ spins and $\tau$ spins that could be flipped simultaneously. In the embedding algorithm we have at each step only one of the two types of clusters.

[15]More precisely, this equivalence holds when the embedding algorithm is defined by making a *random* choice of Step 1 or Step 2 at each iteration.

The embedding algorithm, due to its simplicity, is the one used in our MC study of the AT model (see Section 4).

**Remark:** This algorithm is closely related to an embedding algorithm for general $Z_n$ clock models on an undirected graph. We can consider $Z_n$ as a subgroup of $U(1)$ and then apply Wolff's embedding algorithm for the $XY$ model [12,48,49]. Let us specify the reflection plane by a vector $\vec{s} = (\cos\phi, \sin\phi)$ in this plane; clearly $\phi$ is specified only modulo $\pi$. If $n$ is odd, there is a unique type of reflection: the reflection plane passes through one spin value and one point bisecting two spin values, and it corresponds to $\phi = 2\pi k/n$ with $k$ either integer or half-integer. However, if $n$ is even, there are two types of reflections: the reflection plane can either pass through two spin values or else through two bisector values; these correspond to $\phi = 2\pi k/n$ with $k$ integer or half-integer, respectively. Thus, for the 4-state clock model we have two reflections of the first type and two of the second type: with the identifications $\sigma_x = \sqrt{2}\cos(\theta_x - \pi/4)$ and $\tau_x = \sqrt{2}\cos(\theta_x + \pi/4)$ [taking $\theta_x \in \{0, \pi/2, \pi, 3\pi/2\}$], these reflections are

$$\begin{array}{ll} \phi = 0: & (\sigma,\tau) \to (\tau,\sigma) \\ \phi = \pi/2: & (\sigma,\tau) \to (-\tau,-\sigma) \end{array} \tag{3.12}$$

and

$$\begin{array}{ll} \phi = -\pi/4: & (\sigma,\tau) \to (-\sigma,\tau) \\ \phi = \pi/4: & (\sigma,\tau) \to (\sigma,-\tau) \end{array} \tag{3.13}$$

respectively. The last two moves (i.e., those of bisector type) are precisely the moves allowed in our algorithm. The first two moves correspond to the interchange of $\sigma$ and $\tau$, either without or with a simultaneous flip of both spins.

So let us fix $\phi$ to be one of the four values listed above, and let us embed Ising spins $\varepsilon_x = \pm 1$ into the 4-state clock model via the Wolff update

$$\theta_x \;\to\; \phi + \varepsilon_x(\theta_x - \phi)\,. \tag{3.14}$$

Plugging this into the clock-model Hamiltonian (2.4), we obtain an induced Ising system in the $\{\varepsilon\}$ variables with interaction

$$J_{xy}^{\rm eff} \;=\; 2\,|\sin(\theta_x - \phi)\sin(\theta_y - \phi)|\;[J + 2K\cos(\theta_x - \phi)\cos(\theta_y - \phi)] \tag{3.15}$$

This Ising system can then be simulated by means of an SW algorithm. The cases $\phi = \mp\pi/4$ correspond to Steps 1 and 2 of our embedding algorithm. On the other hand, the case $\phi = 0$ is characterized by an effective coupling

$$J_{xy}^{\rm eff} \;=\; 2J\delta_{\sigma_x,-\tau_x}\delta_{\sigma_y,-\tau_y}\,. \tag{3.16}$$

This means that only bonds joining sites with $\sigma_x \neq \tau_x$ and $\sigma_y \neq \tau_y$ can be "activated" [with probability $p = 1 - \exp(-4J)$]; the move $(\sigma,\tau) \to (\tau,\sigma)$ is then equivalent to flipping both $\sigma$ and $\tau$ within each such cluster. An analogous conclusion applies to the case $\phi = \pi/2$: here only bonds joining sites with $\sigma_x = \tau_x$ and $\sigma_y = \tau_y$ can be "activated". Thus, the moves with $\phi = 0, \pi/2$ are in a sense merely combinations of the moves $\phi = \pm\pi/4$ already contained in our AT embedding algorithm. For this reason, we think that the introduction of the moves $\phi = 0, \pi/2$ into our embedding algorithm will not further reduce the dynamic critical exponent. Indeed, the algorithm with *only* the $\phi = 0, \pi/2$ moves is not even ergodic: at each site the product $\sigma_x \tau_x$ is conserved.

### 3.3 Particular cases

As pointed out in Section 2, the AT model reduces to two decoupled Ising models at $K = 0$ and to the 4-state Potts model al $J = J' = K$. It is worth mentioning that the above-discussed algorithms reduce to the well-known SW algorithms for those particular cases.

When $K = 0$, it is easy to verify that (3.6) reduces to

$$W^{\rm joint}_{\rm bond}(\sigma_x, \sigma_y, \tau_x, \tau_y; m_{xy}, n_{xy}) \;=\; \left[e^{-2J}\delta_{m_{xy},0} + \left(1 - e^{-2J}\right)\delta_{\sigma_x,\sigma_y}\delta_{m_{xy},1}\right] \times \left[e^{-2J'}\delta_{n_{xy},0} + \left(1 - e^{-2J'}\right)\delta_{\tau_x,\tau_y}\delta_{n_{xy},1}\right] \;. \qquad (3.17)$$

This means that the Boltzmann weight for any bond is just the product of the weights of the two independent Ising models. As a result, our direct AT algorithm reduces to two independent SW algorithms on the systems $\{\sigma, m\}$ and $\{\tau, n\}$. Of course, the same holds for the embedding algorithm, as the $\sigma$ spins are decoupled from the $\tau$ spins.

When $J = J' = K \geq 0$, (3.6) can be written as

$$W^{\rm joint}_{\rm bond}(\sigma_x, \sigma_y, \tau_x, \tau_y; m_{xy}, n_{xy}) \;=\; e^{-4J}\delta_{m_{xy},0}\delta_{n_{xy},0} + \left[1 - e^{-4J}\right]\delta_{\sigma_x,\sigma_y}\delta_{\tau_x,\tau_y}\delta_{m_{xy},1}\delta_{n_{xy},1} \;, \qquad (3.18)$$

which is exactly the standard SW decomposition for the Boltzmann weight of the 4-state ferromagnetic Potts model. As a result, our direct AT algorithm on the line $J = J' = K \geq 0$ reduces to the standard SW algorithm for the 4-state ferromagnetic Potts model. However, the embedding algorithm does *not* reduce to the standard SW algorithm in this case.

## 4 Numerical Results

### 4.1 Autocorrelation functions and autocorrelation times

We are interested in the dynamic behavior of the embedding SW algorithm described in Section 3.2. Thus, we need to study the autocorrelation functions and autocorrelation times for each measured observable. Given an observable $\mathcal{O}$, we define the corresponding unnormalized autocorrelation function as

$$C_{\mathcal{O}\mathcal{O}}(t) \;=\; \langle \mathcal{O}_s \mathcal{O}_{s+t} \rangle - \langle \mathcal{O} \rangle^2 \;, \qquad (4.1)$$

where all the expectation values $\langle\cdot\rangle$ are taken in equilibrium and $t$ is the "time" in units of MC steps.[16] The associated normalized autocorrelation function is

$$\rho_{\mathcal{O}\mathcal{O}}(t) \;=\; C_{\mathcal{O}\mathcal{O}}(t)/C_{\mathcal{O}\mathcal{O}}(0)\,. \tag{4.2}$$

The integrated autocorrelation time for the observable $\mathcal{O}$ is defined as

$$\begin{aligned}\tau_{\text{int},\mathcal{O}} &= \frac{1}{2}\sum_{t=-\infty}^{\infty}\rho_{\mathcal{O}\mathcal{O}}(t)\\ &= \frac{1}{2}+\sum_{t=1}^{\infty}\rho_{\mathcal{O}\mathcal{O}}(t)\,.\end{aligned} \tag{4.3}$$

Here the factor $\frac{1}{2}$ is purely a matter of convention (if the normalized autocorrelation function is a pure exponential, $\rho_{\mathcal{O}\mathcal{O}}(t)\approx e^{-|t|/\tau}$ with $\tau\gg 1$, then this definition implies that $\tau_{\text{int},\mathcal{O}}\approx\tau$). Finally, the *exponential autocorrelation time* for the observable $\mathcal{O}$ is defined as

$$\tau_{\text{exp},\mathcal{O}} \;=\; \limsup_{t\to\infty}\frac{|t|}{-\log|\rho_{\mathcal{O}\mathcal{O}}(t)|}\,, \tag{4.4}$$

and the exponential autocorrelation time ("slowest mode") for the system as a whole is defined as

$$\tau_{\text{exp}} \;=\; \sup_{\mathcal{O}}\,\tau_{\text{exp},\mathcal{O}}\,. \tag{4.5}$$

Note that $\tau_{\text{exp}}=\tau_{\text{exp},\mathcal{O}}$ whenever the observable $\mathcal{O}$ is not orthogonal to the slowest mode of the system.

The integrated autocorrelation time controls the statistical error in Monte Carlo estimates of the mean $\langle\mathcal{O}\rangle$. In particular, given a sequence of $n$ Monte Carlo measurements of the observable $\mathcal{O}$ — call them $\{\mathcal{O}_1,\ldots,\mathcal{O}_n\}$ — the sample mean

$$\overline{\mathcal{O}} \;\equiv\; \frac{1}{n}\sum_{t=1}^{n}\mathcal{O}_t \tag{4.6}$$

has a variance

$$\operatorname{var}(\overline{\mathcal{O}}) \;=\; \frac{1}{n^2}\sum_{r,s=1}^{n}C_{\mathcal{O}\mathcal{O}}(r-s) \tag{4.7a}$$

$$=\; \frac{1}{n}\sum_{t=-(n-1)}^{n-1}\left(1-\frac{|t|}{n}\right)C_{\mathcal{O}\mathcal{O}}(t) \tag{4.7b}$$

$$\approx\; \frac{1}{n}\,2\tau_{\text{int},\mathcal{O}}\,C_{\mathcal{O}\mathcal{O}}(0)\quad\text{for}\quad n\gg\tau_{\text{int},\mathcal{O}} \tag{4.7c}$$

This means that the variance is a factor $2\tau_{\text{int},\mathcal{O}}$ larger than it would be if the measurements were uncorrelated. It is therefore, very important to estimate the autocorrelation times for all the interesting observables in order to ensure a correct determination of the statistical errors. The integrated autocorrelation time $\tau_{\mathrm{int},\mathcal{O}}$ can be estimated using standard procedures of statistical time-series analysis [50,51]. In this way, we obtain reliable error bars for both $\tau_{\mathrm{int},\mathcal{O}}$ and $\langle \mathcal{O} \rangle$. We have used a self-consistent truncation window of width $6\tau_{\mathrm{int},\mathcal{O}}$ [52, Appendix C]. This window width is sufficient whenever the autocorrelation function $\rho_{\mathcal{O}\mathcal{O}}(t)$ decays roughly exponentially, a behavior that we will confirm explicitly here (see Section 4.5).

[16] One "Monte Carlo step" consists of one application of "step 1" of Section 3.2 followed by one application of "step 2".

The exponential autocorrelation times $\tau_{\mathrm{exp},\mathcal{O}}$ are extracted by fitting the autocorrelation function $\rho_{\mathcal{O}\mathcal{O}}(t)$, for $t$ large enough, to a pure exponential $A\exp(-t/\tau_{\mathrm{exp},\mathcal{O}})$. The statistical details of this fit are described in Appendix B. However, it should be emphasized that it is in principle *impossible* to obtain a statistically valid estimate of $\tau_{\mathrm{exp},\mathcal{O}}$, as there could always be a very-slowly-decaying component of $\rho_{\mathcal{O}\mathcal{O}}(t)$ with arbitrarily small amplitude (which would thus be invisible under the statistical noise). Thus, our estimates of $\tau_{\mathrm{exp},\mathcal{O}}$ are really *lower bounds*. They can be taken as estimates of $\tau_{\mathrm{exp},\mathcal{O}}$ only if one *assumes* that $\rho_{\mathcal{O}\mathcal{O}}(t)$ is roughly an exponential, with no very-slowly-decaying components.

## 4.2 Observables to be measured

Let us begin by defining some basic observables. The observables of interest involving only the $\sigma$ spins are

$$\mathcal{M}_\sigma \equiv \sum_x \sigma_x \tag{4.8}$$

$$\mathcal{E}_\sigma \equiv \sum_{\langle xy \rangle} \sigma_x \sigma_y \tag{4.9}$$

$$\mathcal{F}_\sigma \equiv \frac{1}{2}\left[ \left| \sum_x \sigma_x e^{2\pi x_1/L} \right|^2 + \left| \sum_x \sigma_x e^{2\pi x_2/L} \right|^2 \right] \tag{4.10}$$

where $L$ is the linear size of the system (we always use periodic boundary conditions) and $(x_1, x_2)$ are the Cartesian coordinates of the point $x$. The last observable can be also seen as the square of the Fourier transform of $\sigma$ at the smallest allowed non-zero momenta (i.e., $(\pm 2\pi/L, 0)$ and $(0, \pm 2\pi/L)$ for the square lattice); it is normalized to be comparable to its zero-momentum analogue $\mathcal{M}_\sigma^2$. We define analogous observables for the $\tau$ spins and for the composite operator $\sigma\tau$.

We have run the MC algorithm on three different points of the self-dual curve of the symmetric AT model (see Table 2). One is the 4-state Potts model at criticality, where the three variables ($\sigma$, $\tau$, $\sigma\tau$) are related by symmetry. But for the rest of the points of the self-dual curve, only $\sigma$ and $\tau$ are related by symmetry. Since we wish to exploit the symmetries of the model in our data analysis, our choice of observables to measure will depend on which model we are studying.

### 4.2.1 Observables for the critical 4-state Potts model

For AT models on the 4-state Potts line $J = J' = K$, the natural choice of the observables are those having the symmetries of the original Potts model, namely those invariant under permutations of $(\sigma, \tau, \sigma\tau)$. We have measured the expectations and autocorrelation times for the following observables:

$$\begin{aligned}
\mathcal{M}^2 &\equiv \tfrac{1}{3}\left(\mathcal{M}_\sigma^2 + \mathcal{M}_\tau^2 + \mathcal{M}_{\sigma\tau}^2\right) &(4.11)\\
\mathcal{E} &\equiv \tfrac{1}{3}\left(\mathcal{E}_\sigma + \mathcal{E}_\tau + \mathcal{E}_{\sigma\tau}\right) &(4.12)\\
\mathcal{F} &\equiv \tfrac{1}{3}\left(\mathcal{F}_\sigma + \mathcal{F}_\tau + \mathcal{F}_{\sigma\tau}\right) &(4.13)
\end{aligned}$$

These observables coincide with the usual ones for the 4-state Potts model up to some multiplicative constants. We can then define the magnetic susceptibility

$$\chi = \frac{1}{V}\langle \mathcal{M}^2 \rangle \, , \tag{4.14}$$

the total energy[17]

$$E = \frac{1}{2V}\langle \mathcal{E} \rangle \, , \tag{4.15}$$

the specific heat

$$C_H = \frac{1}{2V}\left(\langle \mathcal{E}^2 \rangle - \langle \mathcal{E} \rangle^2\right) \, , \tag{4.16}$$

and the second-moment correlation length

$$\xi = \frac{\left(\frac{\chi}{F} - 1\right)^{1/2}}{2 \sin \frac{\pi}{L}} \tag{4.17}$$

where $F$ is defined as

$$F = \frac{1}{V}\langle \mathcal{F} \rangle \, . \tag{4.18}$$

In all these formulae, $V$ is the number of lattice sites (i.e., $V = L^2$) and $2V$ is the number of bonds. This definition of the correlation length is *not* equal to the exponential correlation length (= 1/mass gap), but it is expected that both correlation lengths scale in the same way as we approach the critical point.

**Remark.** To compute the error bar of the specific heat $C_H$ we have first computed the mean energy $\langle \mathcal{E} \rangle$, and then considered the observable $\mathcal{O} \equiv (\mathcal{E} - \langle \mathcal{E} \rangle)^2$ using the procedures described in this section.

### 4.2.2 Observables for a general point on the self-dual curve

A generic point on the self-dual curve does not enjoy the 4-state Potts symmetry, but it does of course enjoy the $\sigma \leftrightarrow \tau$ symmetry characteristic of all symmetric AT models. The natural choice is thus to define two sets of observables: one for the $\sigma$

[17]We have normalized the energy such that $-1 \leq E \leq 1$ (i.e., $E$ is the energy density *per bond*), and the same normalization has been taken for the specific heat. However, in the literature it is more common to find the energy and specific heat normalized *per site*, i.e., with a factor $1/V$ rather than our $1/(2V)$.

and $\tau$ variables, and another for the composite operator $\sigma\tau$. The first set is given by

$$\mathcal{M}_\omega^2 \equiv \tfrac{1}{2}\left(\mathcal{M}_\sigma^2 + \mathcal{M}_\tau^2\right) \tag{4.19}$$

$$\mathcal{E}_\omega \equiv \tfrac{1}{2}\left(\mathcal{E}_\sigma + \mathcal{E}_\tau\right) \tag{4.20}$$

$$\mathcal{F}_\omega \equiv \tfrac{1}{2}\left(\mathcal{F}_\sigma + \mathcal{F}_\tau\right) \tag{4.21}$$

and the second one by $\mathcal{M}_{\sigma\tau}^2$, $\mathcal{E}_{\sigma\tau}$ and $\mathcal{F}_{\sigma\tau}$. We then define

$$\chi_\omega = \frac{1}{V}\langle \mathcal{M}_\omega^2 \rangle \, , \quad E_\omega = \frac{1}{2V}\langle \mathcal{E}_\omega \rangle \, , \quad F_\omega = \frac{1}{V}\langle \mathcal{F}_\omega \rangle \, , \quad \xi_\omega = \frac{\left(\frac{\chi_\omega}{F_\omega} - 1\right)^{1/2}}{2\sin\frac{\pi}{L}} \tag{4.22}$$

and

$$\chi_{\sigma\tau} = \frac{1}{V}\langle \mathcal{M}_{\sigma\tau}^2 \rangle \, , \quad E_{\sigma\tau} = \frac{1}{2V}\langle \mathcal{E}_{\sigma\tau} \rangle \, , \quad F_{\sigma\tau} = \frac{1}{V}\langle \mathcal{F}_{\sigma\tau} \rangle \, , \quad \xi_{\sigma\tau} = \frac{\left(\frac{\chi_{\sigma\tau}}{F_{\sigma\tau}} - 1\right)^{1/2}}{2\sin\frac{\pi}{L}} \, . \tag{4.23}$$

Finally, we define a specific-heat matrix $\widehat{C}_H$

$$\widehat{C}_H = \frac{1}{2V}\begin{pmatrix} \mathrm{var}(\mathcal{E}_\omega) & \mathrm{cov}(\mathcal{E}_\omega, \mathcal{E}_{\sigma\tau}) \\ \mathrm{cov}(\mathcal{E}_\omega, \mathcal{E}_{\sigma\tau}) & \mathrm{var}(\mathcal{E}_{\sigma\tau}) \end{pmatrix} \, , \tag{4.24}$$

with eigenvalues $C_{H,\mathrm{min}}$ and $C_{H,\mathrm{max}}$ and corresponding eigenvectors

$$\vec{w}_{\mathrm{min}} = \begin{pmatrix} 1 \\ a \end{pmatrix} \tag{4.25a}$$

$$\vec{w}_{\mathrm{max}} = \begin{pmatrix} a \\ -1 \end{pmatrix} \tag{4.25b}$$

where $a$ is some real number. These two eigenvalues have distinct critical exponents: $C_{H,\mathrm{max}}$ corresponds to a relevant operator (with exponent $\alpha > 0$), while $C_{H,\mathrm{min}}$ is expected to correspond to a marginal operator (with exponent 0). The marginal operator arises from the existence of the self-dual curve (2.8), along which the critical exponents vary continuously. In particular, *on* the self-dual curve (2.8), we expect that

$$a = -\tfrac{1}{2}\coth 2J \tag{4.26}$$

(in the infinite-volume limit), as can easily be computed from the tangent vector to (2.8), taking into account the normalization $\mathcal{E} = 2J\mathcal{E}_\omega + K\mathcal{E}_{\sigma\tau}$.

## 4.3 Summary of our simulations

We have run our MC program for the embedding algorithm (Section 3.2) at three different points of the self-dual curve (2.8): see Table 2 for details. One of the points is the critical point of the 4-state Potts model (i.e., $J = J' = K = \frac{1}{4}\log 3$). The second point is the image of the $q = 2$ Potts model via the transformation discussed at the end of Section 2 [i.e., (2.9a–b) with $2\cos\mu = \sqrt{2}$]: this point will be denoted as ZF and corresponds to the model studied by Zamolodchikov and Fateev [45]. The third point is one of the ones studied in [24] and will be denoted as in that paper (X2); it corresponds to a Potts model with $q \approx 0.651287$. We also notice that the point X3 of [24] is rather close to our choice ZF. Finally, we have used the extensive MC data of Baillie and Coddington [7][18] for the critical Ising model (which corresponds to the point DIs of the AT model). In Table 3 we include the static and dynamic data corresponding to this point.

For each of these points we have run the MC program at different lattice sizes ranging from $L = 16$ to $L = 1024$ for the 4-state Potts model and from $L = 16$ to $L = 512$ for the other two points. In all cases we have started the simulations with a random configuration and we have discarded the first $10^5$ iterations to allow the system to reach thermodynamic equilibrium. This discard interval is sufficient for equilibration: in the worst case (i.e., the 4-state Potts model with $L = 1024$), it is roughly equal to $190\tau_{\text{int},\mathcal{E}}$ (or about $160\tau_{\text{exp},\mathcal{E}}$). The number of measurements ranges between $8 \times 10^5$ and $4.4 \times 10^6$. In all cases except the $L = 1024$ Potts, the number of measurements is greater than $10^4\tau_{\text{int},\mathcal{E}}$. This is sufficient to obtain good estimates (errors $\sim$ 1–4%) for the autocorrelation times, and excellent estimates (errors $\sim$ 0.1–1%) for the static observables. On the other hand, for the 4-state Potts model with $L = 1024$ we were able to achieve only $\sim 1500\tau_{\text{int},\mathcal{E}}$. The error bars on this point are therefore rather large.

To test the program we have compared the MC results to the exact solution for small lattices ($3 \times 3$ and $4 \times 4$). We performed this test over a wide range of couplings $(J, J', K)$, including both high- and low-temperature regions as well as the critical region. We have also compared the results for larger lattices to previous MC computations [9] for the critical 4-state Potts model. In all cases the agreement was good.

The CPU time required by our program is approximately $10L^2$ $\mu$s/iteration on an IBM RS-6000/370. The total CPU time used in this project was approximately 1.2 years on this same machine.

The estimates for the observables discussed in Section 4.2 are shown in Tables 4–8. In all of them, the quoted errors correspond to one standard deviation (i.e., confidence level of 68%). In reporting the number of measurements (MCS), we have already subtracted the MC steps discarded for equilibration.

A summary of our estimates of critical exponents, together with the theoretically predicted exponents, can be found in Table 9. The data analysis leading to these estimates is the subject of the next two subsections.

---

[18] We thank Paul Coddington for communicating to us the numerical values of these data, which formed the basis for the graphs in [7]. For the lattices $L \leq 128$ these data coincide with the data reported by Baillie and Coddington in an earlier paper [6], while for $L = 256, 512$ they improve the statistics somewhat.

## 4.4 Static quantities

For each quantity $\mathcal{O}$, we carry out a fit to the power-law Ansatz $\mathcal{O} = AL^p$ using the standard weighted least-squares method. As a precaution against corrections to scaling, we impose a lower cutoff $L \geq L_{min}$ on the data points admitted in the fit, and we study systematically the effects of varying $L_{min}$. In general, our preferred fit corresponds to the smallest $L_{min}$ for which the goodness of fit is reasonable (e.g., the confidence level[19] is $\gtrsim$ 10–20%), and for which subsequent increases in $L_{min}$ do not cause the $\chi^2$ to drop vastly more than one unit per degree of freedom.

### 4.4.1 Susceptibilities

We have fitted the values of the susceptibilities to the power-law functions $\chi_\omega = AL^{\gamma/\nu}$ and $\chi_{\sigma\tau} = AL^{\gamma'/\nu}$ as described above. The estimates for $\gamma/\nu$ and $\gamma'/\nu$ are, in all cases, very stable as we vary $L_{min}$. This means that the corrections to scaling for these observables are not statistically significant (to the degree of accuracy we have attained here).

For the 4-state Potts point (Table 4), our preferred estimate is obtained for $L_{min} = 16$:[20]

$$\frac{\gamma}{\nu}(\mathrm{P}) \;=\; \frac{\gamma'}{\nu}(\mathrm{P}) \;=\; 1.744 \pm 0.001 \tag{4.27}$$

with $\chi^2 = 2.21$ for 5 degrees of freedom (DF), confidence level = 82%. The difference from the exact result ($\gamma/\nu = \frac{7}{4}$) is small, but it is roughly equal to 6 standard deviations. Possibly this is due to a small correction-to-scaling effect (which has become statistically significant due to the very high precision we have obtained on the smaller lattices). For $L_{min} = 128$ we have

$$\frac{\gamma}{\nu}(\mathrm{P}) \;=\; \frac{\gamma'}{\nu}(\mathrm{P}) \;=\; 1.744 \pm 0.004 \tag{4.28}$$

with $\chi^2 = 1.54$ (2 DF, level = 46%), which is now fully compatible with the exact result.

For the ZF point (Table 5), our preferred estimates are obtained for $L_{min} = 128$:

$$\frac{\gamma}{\nu}(\mathrm{ZF}) \;=\; 1.750 \pm 0.004 \tag{4.29}$$

with $\chi^2 = 1.54$ (1 DF, level = 22%), and

$$\frac{\gamma'}{\nu}(\mathrm{ZF}) \;=\; 1.668 \pm 0.005 \tag{4.30}$$

[19]"Confidence level" is the probability that $\chi^2$ would exceed the observed value, assuming that the underlying statistical model is correct. An unusually low confidence level (e.g., less than 5%) thus suggests that the underlying statistical model is *incorrect* — the most likely cause of which would be corrections to scaling.

[20]We will hereafter use the indices P, ZF and X2 to designate the points where the results apply. The index DIs will be used to denote the Ising model.

with $\chi^2 = 1.57$ (1 DF, level = 21%). The agreement with the exact results ($\gamma/\nu = \frac{7}{4}$ and $\gamma'/\nu = \frac{5}{3}$) is extremely good, and the $\chi^2$ is acceptable.

For the point X2 (Table 7), we get our preferred estimates for $L_{min} = 32$:

$$\frac{\gamma}{\nu}(\mathrm{X2}) \;=\; 1.750 \pm 0.001 \tag{4.31}$$

with $\chi^2 = 0.98$ (3 DF, level = 81%), and

$$\frac{\gamma'}{\nu}(\mathrm{X2}) \;=\; 1.605 \pm 0.001 \tag{4.32}$$

with $\chi^2 = 1.24$ (3 DF, level = 74%). The agreement with the exact values $\gamma/\nu = \frac{7}{4}$ and $\gamma'/\nu = 1.6045$ is again extremely good, as is the $\chi^2$.

Finally, we have re-analyzed the MC data of Baillie and Coddington [7] for the Ising model (Table 3). Our preferred fit is for $L_{min} = 64$:

$$\frac{\gamma}{\nu}(\mathrm{DIs}) \;=\; 1.7501 \pm 0.0002 \tag{4.33}$$

with $\chi^2 = 0.75$ (3 DF, level = 86%). Both the high accuracy of the result and the agreement with the exact answer ($\gamma/\nu = \frac{7}{4}$) are remarkable.

### 4.4.2 Specific Heat

Here we have to distinguish between the 4-state Potts model and the other two points (ZF and X2). For the latter points the analysis is a little bit more complicated, as we have to deal with a specific-heat matrix $\widehat{C}_H$ [cf. (4.24)] instead of a single number.

For the 4-state Potts model (Table 4) we first tried to fit the data to a pure power-law function $C_H = AL^{\alpha/\nu}$. The results of this fit (as a function of $L_{min}$) are contained in Table 10. We observe a systematic trend toward higher values of $\alpha/\nu$ as we increase $L_{min}$. For $L_{min} \geq 64$ the $\chi^2$ values are acceptable. Nevertheless, being conservative, we take as our preferred fit $L_{min} = 128$ (boldfaced in Table 10):

$$\frac{\alpha}{\nu}(\mathrm{P}) \;=\; 0.768 \pm 0.009 \tag{4.34}$$

with $\chi^2 = 1.40$ (2 DF, level = 50%). Clearly, the agreement with the exact known result ($\alpha/\nu = 1$) leaves something to be desired! A similar result was reported by Wiseman and Domany [24]: $\alpha/\nu = 0.747 \pm 0.003$, using lattices $16 \leq L \leq 128$. As a matter of fact, if we fit our own data restricted to the interval $16 \leq L \leq 128$, we obtain $\alpha/\nu = 0.741 \pm 0.004$ ($\chi^2 = 3.07$, 2 DF, level = 22%), which is consistent with the value of Wiseman and Domany. Thus, as we go to larger lattices we obtain estimates of $\alpha/\nu$ that are closer to the exact value, but the improvement from $L_{max} = 128$ to $L_{max} = 1024$ is extremely slow.

However, we already know on theoretical grounds [16,17,18] that the true leading behavior of the specific heat involves a multiplicative logarithmic correction

$$C_H \;\sim\; L \log^{-3/2} L\,. \tag{4.35}$$

In the range of $L$ considered here, the logarithmic factor could be mimicked by a power-law function, thus yielding an "effective" critical exponent $(\alpha/\nu)_{\rm eff}$ lower than $\alpha/\nu = 1$ (in agreement with our numerical results). To check this, we tried to fit our data to the function $C_H = AL^{\alpha/\nu}\log^{-3/2} L$ (see Table 11). We observe that the estimates for $\alpha/\nu$ are much closer to the exact value. These estimates are slightly *higher* than 1, but there is a clear systematic trend towards smaller values of $\alpha/\nu$ as $L_{min}$ is increased. For $L_{min} = 128$ we obtain a reasonable fit with

$$\frac{\alpha}{\nu}({\rm P}) \;=\; 1.039 \pm 0.009 \tag{4.36}$$

with $\chi^2 = 0.45$ (2 DF, level = 80%). However, this value still differs from the true one by four standard deviations.

Alternatively, we can fit the data to the trial function $C_H = AL\log^{-p} L$ (see again Table 11). We also observe a systematic trend towards the exact value $p = \frac{3}{2} = 1.5$, but the estimates of $p$ are not compatible within errors with the exact value. Our preferred fit occurs for $L_{min} = 128$,

$$p({\rm P}) \;=\; 1.286 \pm 0.052 \tag{4.37}$$

with $\chi^2 = 0.34$ (2 DF, level = 84%). This estimate is again four standard deviations away from the expected result.

In summary, it is extremely difficult is to obtain a reliable estimate of $\alpha/\nu$ when the leading term of the specific heat behaves like (4.35). For lattices up to $L = 1024$ it is impossible to see anything even resembling the correct exponent $\alpha/\nu = 1$. On the other hand, if we introduce in the fits some theoretical information (e.g., the power of the logarithmic term) the estimate improves a lot, but it is still four standard deviations away from the expected result. We need to go beyond $L = 1024$ to disentangle the true asymptotic behavior of the specific heat.

For the ZF and X2 models, we have to deal with a specific-heat matrix $\widehat{C}_H$ as in (4.24). First we computed all its matrix elements; then we diagonalized it to obtain the eigenvalues $C_{H,\rm min}$ and $C_{H,\rm max}$ and the corresponding eigenvectors $\vec{w}_{\rm min} = (1, a)$ and $\vec{w}_{\rm max} = (a, -1)$ [cf. (4.25)]. The error bars on the eigenvalues and on the eigenvector parameter $a$ are computed by using the standard error-propagation formulae. The minimum eigenvalue $C_{H,\rm min}$ is expected to tend to a finite constant as $L \to \infty$ (i.e., to have critical exponent zero), as there should be a marginal operator responsible for movements along the critical self-dual curve (2.8). The maximum eigenvalue $C_{H,\rm max}$ is expected to grow as $L^{\alpha/\nu}$ with the power given by (2.10b). In general the value of the parameter $a$ varies with the lattice size $L$, and in the limit $L \to \infty$ we expect $a$ to tend to the value

$$a_\infty \;\equiv\; -\tfrac{1}{2}\coth 2J \tag{4.38}$$

[cf. (4.26)]. In Table 12 we show the evolution with the lattice size of the parameter $a$ corresponding to the models ZF and X2, while in Tables 5 and 7 we report the eigenvalues $C_{H,\rm max}$ and $C_{H,\rm min}$.

*ZF model.* If we try to fit $C_{H,\rm max}$ for the point ZF (Table 5) to a pure power-law function, we obtain estimates of $\alpha/\nu$ which are quite consistent among themselves for $L_{min} \geq 64$. The fit for $L_{min} = 64$ gives

$$\frac{\alpha}{\nu}({\rm ZF}) \;=\; 0.663 \pm 0.006 \tag{4.39}$$

with $\chi^2 = 1.12$ (2 DF, level = 57%). This result agrees well with the the theoretical prediction $\alpha/\nu = \frac{2}{3}$.

The minimum eigenvalue of $\widehat{C}_H$ is almost constant within statistical errors for $L \geq 128$. As a matter of fact, the fit to a constant $C^\star_{H,\rm min}$ is very good for $L_{min} = 128$: the result is $C^\star_{H,\rm min} = 0.562 \pm 0.004$ with $\chi^2 = 0.55$ (2 DF, level = 76%).

Finally, from Table 12 we see that the eigenvector parameter $a$ appears to be tending as $L \to \infty$ to the predicted exact value (4.38). We can test this convergence quantitatively, and also extract an estimate of the correction-to-scaling exponent $\Delta$, by fitting the data to $a - a_\infty = AL^{-\Delta}$. We obtain a reasonable fit already for $L_{min} = 16$:

$$\Delta({\rm ZF}) = 0.715 \pm 0.008 \tag{4.40}$$

with $\chi^2 = 1.08$ (4 DF, level = 90%).

*X2 model.* The estimate of $\alpha/\nu$ for the point X2 (Table 7) is not well stabilized: it decreases systematically as $L_{min}$ increases (see Table 13), and for $L_{min} \leq 64$ the $\chi^2$ values are horrible. Our preferred fit is obtained for $L_{min} = 128$:

$$\frac{\alpha}{\nu}({\rm X2}) = 0.438 \pm 0.008 \tag{4.41}$$

with $\chi^2 = 0.32$ (1 DF, level = 57%). Notice that the $\chi^2$ value is now very reasonable. This estimate is still three standard deviations away from the exact result $\alpha/\nu = 0.4183$, but the trend is in the right direction. Moreover, for $L_{min} = 256$ we obtain a slightly lower estimate, $\alpha/\nu = 0.430 \pm 0.017$, which is now consistent with the the exact result. The smallness of the difference between the observed and the true values (less than 0.02) suggests that multiplicative logarithmic corrections (as occur in the Potts case) are absent at this point; we are most likely seeing the effects of additive corrections to scaling of the form $L^{-\Delta}$ with $\Delta$ on the order of 0.5 (or conceivably $1/\log L$). Wiseman and Domany [24] reported a value $\alpha/\nu = 0.542 \pm 0.008$, which is much larger than the exact one and than ours. This is surely due to the fact that they considered only rather smaller lattices ($16 \leq L \leq 128$). To check this, we fit the subset of our data corresponding to $16 \leq L \leq 128$, and obtained $\alpha/\nu = 0.502 \pm 0.003$ with $\chi^2 = 8.81$ (2 DF, level = 1.2%).

The smallest eigenvalue grows is again consistent with a constant $C^\star_{H,\rm min}$ within statistical errors for $L \geq 128$. For $L_{min} = 128$ the result is $C^\star_{H,\rm min} = 0.303 \pm 0.002$ with $\chi^2 = 1.17$ (2 DF, level = 56%).

Finally, the behavior of the eigenvectors is rather similar as the ZF case: they approach the predicted exact value (4.38) as $L$ grows. If we fit the data to $a - a_\infty = AL^{-\Delta}$, we obtain for $L_{min} = 128$ the value

$$\Delta({\rm X2}) = 0.518 \pm 0.024 \tag{4.42}$$

with $\chi^2 = 0.02$ (1 DF, level = 88%). This value for the exponent $\Delta$ agrees well with the rough estimate obtained from the corrections to scaling in the specific heat.

*Ising model.* We have also computed the *effective* exponents $(\alpha/\nu)_{\rm eff}$ associated with a pure-power-law fit to the specific heat of the Ising model, for various intervals of $L$. Such an effective exponent will be useful as a standard of comparison for the numerically extracted estimates of the dynamic critical exponent $z_{{\rm int},\mathcal{E}}$ (see Section 4.5.1 below). First we computed the *exact* values of the specific heat for a finite lattice, using the formulae of Ferdinand and Fisher [53]: see Table 3. We then performed a power-law fit, with the (fake) error bars chosen so as to give all points the same statistical weight; we took $L_{max} = 512$ and considered various values of $L_{min}$. As expected, $(\alpha/\nu)_{\rm eff}$ is not stable: it decreases as $L_{min}$ grows, ranging from $(\alpha/\nu)_{\rm eff} = 0.244$ for $L_{min} = 8$ to $(\alpha/\nu)_{\rm eff} = 0.162$ when $L_{min} = 256$.

### 4.4.3 Second-moment correlation length

The quantity $x(L) \equiv \xi(L)/L$ is expected to approach a constant $x^\star$ as $L \to \infty$. This constant is characteristic of the massless Ashkin–Teller field theory on a continuum torus with aspect ratio 1, and in principle it should be calculable via conformal field theory (although to our knowledge this calculation has not yet been done).

Our Monte Carlo data are consistent with this behavior. First we tried to fit the values for $L \geq L_{min}$ to a constant, using the weighted least-squares method. In Table 14 (fits marked C) we report our best estimates for $x^\star$.

It is intriguing to note that the values of $x^\star_\omega$ are all quite close to 1, though the differences from 1 are 7–10 standard deviations for the points ZF and X2. This (near-)agreement might be due to the fact that all these three models have the same central charge ($c = 1$). However, a detailed study is needed to understand the observed variations. On the other hand, the value of $x^\star_{\sigma\tau}$ manifestly decreases as we move towards the point $K = 0$, where the $\sigma$ spins are decoupled from the $\tau$ spins. At that point we expect $x^\star_{\sigma\tau} = 0$.

We can also study the rate at which $\xi(L)/L$ approaches $x^\star$ for these three models. For the 4-state Potts model we already know [54] the answer for the case of the exponential correlation length ($\xi_{\rm exp} = 1/\text{mass gap}$) on an $L \times \infty$ cylinder. The behavior for large $L$ is

$$\frac{\xi_{\rm exp}(L)}{L} \;=\; \frac{4}{\pi^2} - \frac{24}{\pi^2}\frac{1}{\log 2L} + o\left(\frac{1}{\log L}\right) \;. \tag{4.43}$$

Of course, there is no guarantee that the asymptotic behavior of $\xi_{\rm second-moment}$ on a torus is the same as that of $\xi_{\rm exp}$ on a cylinder, but it is a plausible guess. Therefore, we tried to fit our data to the function

$$\xi(L)/L \;=\; x^\star \,+\, \frac{A}{\log 2L} \,. \tag{4.44}$$

The result can be found in Table 14 (fit marked L), and in more detail in Table 15. The $\chi^2$ of this fit is excellent, but roughly the same values of $\chi^2$ could have been obtained using corrections of the type $1/\sqrt{L}$ or $1/L$ (see Table 15). The values of $x^\star$ vary a little bit from one fit to the next (in some cases by several standard deviations). For a full understanding of these results, it would be very helpful to have a theoretical prediction analogous to (4.43) for the second-moment correlation length $\xi$ on a torus.

For the rest of the self-dual curve we have no theoretical hint analogous to (4.43), so we must fit our data empirically. For the ZF model the fits of $x_\omega$ and $x_{\sigma\tau}$ to a constant are so perfect ($L_{min} = 16$) that no further conclusions could be obtained (see Table 14). For the X2 model, by contrast, the fits to a constant need a much higher value of $L_{min}$, indicating that the corrections to scaling are statistically significant. First we tried a correction of the type $x = x^\star + AL^{-\Delta}$ with $\Delta = \frac{1}{2}$; this value is suggested by the result (4.42) obtained in the preceding subsection. The fits are very good, even for $L_{min} = 16$ (see Table 14). Similarly good fits are obtained also with $\Delta = \frac{1}{4}$, 1, 2. Finally, we also tried a logarithmic fit as in the 4-state Potts model. The results (marked with L in Table 14) are also excellent. It is worth mentioning that the estimates of $x^\star_\omega$ and $x^\star_{\sigma\tau}$ vary slightly from one type of fit to another, in some cases by several standard deviations.

## 4.5 Dynamic quantities

In this section we analyze both the integrated autocorrelation times $\tau_{\mathrm{int},\mathcal{O}}$ and the exponential autocorrelation times $\tau_{\mathrm{exp},\mathcal{O}}$. Using standard power-law fits to the Ansätze $\tau_{\mathrm{int},\mathcal{O}} = AL^{z_{\mathrm{int},\mathcal{O}}}$ and $\tau_{\mathrm{exp},\mathcal{O}} = AL^{z_{\mathrm{exp},\mathcal{O}}}$, we can extract the dynamic critical exponents.[21] However, the existence of multiplicative logarithmic corrections to the specific heat for the Ising and 4-state Potts models suggests, in view of the Li–Sokal bound, that similar multiplicative logarithmic corrections *might* occur also in the autocorrelation times. We shall look for such corrections in two ways:

(a) by fitting to an explicit logarithmic Ansatz $\tau = AL^z(\log L)^p$; and

(b) by studying the ratio $\tau/C_H$ (for the X2 and ZF models we consider the ratio $\tau/C_{H,\mathrm{max}}$).

### 4.5.1 Integrated autocorrelation time: power-law fits

From Tables 4, 6 and 8, we see that the integrated autocorrelation time for $\mathcal{M}^2$ (resp. $\mathcal{M}^2_\omega$ and $\mathcal{M}^2_{\sigma\tau}$) is always slightly smaller than the integrated autocorrelation time for $\mathcal{E}$ (resp. $\mathcal{E}_\omega$ and $\mathcal{E}_{\sigma\tau}$). Furthermore, we see that the ratio of those autocorrelation times is a constant (i.e., independent of $L$) within statistical errors:

$$\frac{\tau_{\mathrm{int},\mathcal{M}^2}}{\tau_{\mathrm{int},\mathcal{E}}}(\mathrm{P}) \quad = \quad 0.978 \pm 0.011 \tag{4.45a}$$

[21] We emphasize that in general $z_{\mathrm{int},\mathcal{O}}$ need not be equal to $z_{\mathrm{exp},\mathcal{O}}$. However, in SW-type algorithms it does appear that the autocorrelation function of the energy is very close to a pure exponential, so that $\tau_{\mathrm{int},\mathcal{E}}/\tau_{\mathrm{exp},\mathcal{E}}$ approaches a constant (in fact, a constant very close to 1) as $L \to \infty$. So in *these* algorithms we do empirically seem to have $z_{\mathrm{int},\mathcal{E}} = z_{\mathrm{exp},\mathcal{E}}$.

$$\frac{\tau_{\mathrm{int},\mathcal{M}^2_\omega}}{\tau_{\mathrm{int},\mathcal{E}_\omega}}(\mathrm{ZF}) \;=\; 0.944 \pm 0.018\;; \quad \frac{\tau_{\mathrm{int},\mathcal{M}^2_{\sigma\tau}}}{\tau_{\mathrm{int},\mathcal{E}_{\sigma\tau}}}(\mathrm{ZF}) = 0.887 \pm 0.020 \qquad (4.45\mathrm{b})$$

$$\frac{\tau_{\mathrm{int},\mathcal{M}^2_\omega}}{\tau_{\mathrm{int},\mathcal{E}_\omega}}(\mathrm{X2}) \;=\; 0.857 \pm 0.018\;; \quad \frac{\tau_{\mathrm{int},\mathcal{M}^2_{\sigma\tau}}}{\tau_{\mathrm{int},\mathcal{E}_{\sigma\tau}}}(\mathrm{X2}) \;=\; 0.803 \pm 0.017 \qquad (4.45\mathrm{c})$$

Likewise, if we look at the ratio $\tau_{\mathrm{int},\mathcal{E}_{\sigma\tau}}/\tau_{\mathrm{int},\mathcal{E}_\omega}$ for the points ZF and X2, we see that in both cases that ratio is also consistent with a constant:

$$\frac{\tau_{\mathrm{int},\mathcal{E}_{\sigma\tau}}}{\tau_{\mathrm{int},\mathcal{E}_\omega}}(\mathrm{ZF}) \;=\; 0.960 \pm 0.017 \qquad (4.46\mathrm{a})$$

$$\frac{\tau_{\mathrm{int},\mathcal{E}_{\sigma\tau}}}{\tau_{\mathrm{int},\mathcal{E}_\omega}}(\mathrm{X2}) \;=\; 0.962 \pm 0.013 \qquad (4.46\mathrm{b})$$

It therefore suffices to consider the critical behavior of $\tau_{\mathrm{int},\mathcal{E}}$ (for Potts) and $\tau_{\mathrm{int},\mathcal{E}_\omega}$ (for ZF and X2); all other quantities will have the same dynamic critical exponent.

*Potts model.* If we fit the data from the 4-state Potts model point to a pure power-law function $\tau_{\mathrm{int},\mathcal{E}} = AL^{z_{\mathrm{int},\mathcal{E}}}$, we obtain a good fit already for $L_{min} = 16$ (see Table 10). However, there seems to be a weak upward trend with $L_{min}$, so to be conservative we choose $L_{min} = 32$ as our preferred fit:

$$z_{\mathrm{int},\mathcal{E}}(\mathrm{P}) \;=\; 0.876 \pm 0.012 \qquad (4.47)$$

with $\chi^2 = 3.16$ (4 DF, level = 53%). Notice that this value of $z_{\mathrm{int},\mathcal{E}}$ is strictly greater than the effective exponent $(\alpha/\nu)_{\mathrm{eff}} = 0.768$ obtained by a pure power-law fit [see (4.34)]. This implies that the Li–Sokal bound (1.1) is satisfied, but it is apparently not sharp. We can also compare the power-law estimates $z_{\mathrm{int},\mathcal{E}}$ and $\alpha/\nu$ for each $L_{min}$ separately (see Table 10), and again conclude that the Li–Sokal bound is always satisfied but that it is not sharp. Note, however, that this effective exponent $(z_{\mathrm{int},\mathcal{E}} - \alpha/\nu)_{\mathrm{eff}} \approx 0.11$ is consistent with the true behavior of $\tau_{\mathrm{int},\mathcal{E}}/C_H$ being *either* a small positive power *or* a logarithm.

If we compare our results for the 4-state Potts model with the embedding algorithm (see Table 4) to those quoted in [9] (which correspond to the direct algorithm of Section 3.1), we see that the ratio of the two autocorrelation times is more or less constant within statistical errors, and that there is no systematic trend as $L$ grows. For these reasons, we conclude that they are proportional in the limit $L \to \infty$:[22]

$$\frac{\tau^{\mathrm{direct}}_{\mathrm{int},\mathcal{E}}}{\tau^{\mathrm{embedd}}_{\mathrm{int},\mathcal{E}}}(\mathrm{P}) \;=\; 1.516 \pm 0.035\;. \qquad (4.48)$$

In particular, the direct and embedding algorithms belong to the same dynamic universality class (as was of course to be expected). It follows from (4.48) that our algorithm is 150% as effective as the standard SW algorithm for this model in terms of autocorrelation times. However, our algorithm needs roughly twice the CPU time for a complete sweep over the lattice (there are twice the number of points as in the Potts formulation)[23]; so, in the end, the embedding algorithm is about 25% less efficient than the standard SW algorithm for the 4-state Potts model at criticality.

[22] It is therefore hardly surprising that the dynamic critical exponent reported in [9] for a pure power-law fit, $z_{\mathrm{int},\mathcal{E}} = 0.87 \pm 0.02$, is virtually identical to our value (4.47). If we fit our data for $L \leq 256$ as in [9], we obtain (for $L_{min} = 64$) $z_{\mathrm{int},\mathcal{E}} = 0.900 \pm 0.025$ with $\chi^2 = 0.63$ (1 DF, level = 43%). This result is also compatible within errors with the one reported in [9].

Our estimate of $z_{\mathrm{int},\mathcal{E}}$ for a pure power-law fit, $0.876 \pm 0.012$, is very close to the result $z_{\mathrm{int},\mathcal{E},\mathrm{1C}} = 0.92 \pm 0.01$ found by Wiseman and Domany [24] for the single-cluster version of the direct algorithm.[24,25] Thus, the 2D Potts model with $q = 4$ conforms to the behavior found previously [6] for the 2D Potts models with $q = 2, 3$, in which $z_{\mathrm{int},\mathcal{E},\mathrm{1C}} \approx z_{\mathrm{int},\mathcal{E},\mathrm{SW}}$. On the other hand, this almost-equality seems *not* to occur for Ising models in dimension $d \geq 3$ [12,13].

*ZF model.* For the ZF point (Table 6), the estimates for $z_{\mathrm{int},\mathcal{E}_\omega}$ are quite stable, giving for $L_{min} = 32$

$$z_{\mathrm{int},\mathcal{E}_\omega}(\mathrm{ZF}) \;=\; 0.733 \pm 0.014 \tag{4.49}$$

with $\chi^2 = 1.48$ (3 DF, level = 68%). Once again, the Li–Sokal bound holds but it is not quite sharp, as here $\alpha/\nu = \frac{2}{3}$.

*X2 model.* At the point X2 (Table 8), the corrections to scaling for $\tau_{\mathrm{int},\mathcal{E}_\omega}$ seem to be significant, just as they are for $C_{H,\mathrm{max}}$ (see Table 13). The estimates for $z_{\mathrm{int},\mathcal{E}_\omega}$ tend to be systematically smaller as $L_{min}$ grows (although this effect is only on the borderline of statistical significance, due to the rather large error bars). Our preferred fit uses $L_{min} = 128$:

$$z_{\mathrm{int},\mathcal{E}_\omega}(\mathrm{X2}) \;=\; 0.477 \pm 0.028 \tag{4.50}$$

with $\chi^2 = 0.39$ (1 DF, level = 53%). Once again, the Li–Sokal bound holds but is not quite sharp, as $\alpha/\nu = 0.438$ (our numerical estimate) or $0.4183\ldots$ (exact value).

Also here we can compare our result with the one found by Wiseman and Domany [24] for the single-cluster algorithm. They obtain $z_{\mathrm{int},\mathcal{E},\mathrm{1C}} = 0.61 \pm 0.01$. This value is very far from ours, even if we fit our data for $16 \leq L \leq 128$ to facilitate the comparison with their data: in that case our preferred fit is $z_{\mathrm{int},\mathcal{E}} = 0.553 \pm 0.012$ for $L_{min} = 16$ and $\chi^2 = 0.32$ (2 DF, level = 85%). Here the difference between $z_{\mathrm{int},\mathcal{E},\mathrm{1C}}$ and $z_{\mathrm{int},\mathcal{E},\mathrm{SW}}$ is nearly four standard deviations, so it seems that the two algorithms belong to different dynamic universality classes.

---

[23]We have directly measured the CPU ratio between these two algorithms and it is $\approx 1.9$, i.e., very close to the guessed value of 2.

[24]We emphasize that $z_{\mathrm{int},\mathcal{E},\mathrm{1C}}$ is the dynamic critical exponent for the autocorrelation time of the single-cluster algorithm *measured in units of "equivalent sweeps"*. This is $d - \gamma/\nu = 1/4$ less than the dynamic critical exponent for the autocorrelation time measured in units of "cluster hits".

[25]Actually, the Wiseman–Domany result was obtained with $16 \leq L \leq 128$. If we fit our own data with this constraint, we get for $L_{min} = 32$ the value $z_{\mathrm{int},\mathcal{E}} = 0.876 \pm 0.021$ with $\chi^2 = 1.96$ (1 DF, level = 16%). So our estimate of $z_{\mathrm{int},\mathcal{E}}$ is *not* sensitive to $L_{max}$; and it differs from the Wiseman–Domany estimate of $z_{\mathrm{int},\mathcal{E},\mathrm{1C}}$ by two standard deviations.

*Ising model.* Finally, we have reanalyzed the data of Baillie and Coddington [7] for the 2D Ising model (see Table 3). Our preferred fit is for $L_{min} = 100$, giving

$$z_{\mathrm{int},\mathcal{E}}(\mathrm{DIs}) \;=\; 0.240 \pm 0.004 \tag{4.51}$$

with $\chi^2 = 1.67$ (2 DF, level = 43%). In this case the Li–Sokal bound is clearly satisfied, as $(\alpha/\nu)_{\mathrm{eff}} \approx 0.173$ when $L_{min} = 128$ (see Section 4.4.2); and once again the bound appears to be not quite sharp.

In summary, the Li–Sokal bound is fulfilled in all cases, but apparently not as an equality: the effective exponents $(z_{\mathrm{int},\mathcal{E}} - \alpha/\nu)_{\mathrm{eff}}$ range from $\approx 0.04$ to $\approx 0.11$. However, we know that in two of the four cases — namely, the Ising model and the 4-state Potts model — the leading behavior of the specific heat is not merely a power law, but rather contains multiplicative logarithmic corrections. So the question of the sharpness of the bound cannot be answered until we take into account the exact leading term of the specific heat. This will be done in the next subsection.

### 4.5.2 Might the Li–Sokal bound be sharp modulo a logarithm?

It is well known that the true leading behavior of the specific heat is not given by $\alpha/\nu = 0.768$ (or any other power law) for the 4-state Potts model, nor by $\alpha/\nu = 0.173$ (or any other power law) for the Ising model. Rather, the behavior is $L \log^{-3/2} L$ and $\log L$, respectively. A similar problem arises for the autocorrelation times: to answer the question of the sharpness of the Li–Sokal bound, it is necessary to guess the "exact" leading behavior of $\tau_{\mathrm{int},\mathcal{E}}$. In analogy with $C_H$, we may entertain the possibility that $\tau_{\mathrm{int},\mathcal{E}}$ contains a multiplicative logarithmic term (at least for the Ising and 4-state Potts models).

Our first approach is to consider the ratio $\tau_{\mathrm{int},\mathcal{E}}/C_H$.[26] For all four models, we find that this ratio is an increasing function of the lattice size $L$. We then distinguish three possible asymptotic behaviors: If $\tau_{\mathrm{int},\mathcal{E}}/C_H$ tends to a constant as $L \to \infty$, then the Li–Sokal bound (1.1) is sharp; if it grows like some logarithmic function, then the bound is sharp modulo a logarithm; and if it grows like some power law, then the bound is not sharp. It is reasonable to hope that the ratio $\tau_{\mathrm{int},\mathcal{E}}/C_H$ will be less affected by corrections to scaling than either $C_H$ or $\tau_{\mathrm{int},\mathcal{E}}$ separately. (Of course, this hope may also be false!) For the Ising model we took $\tau_{\mathrm{int},\mathcal{E}}$ from Ref. [7] and the specific heat from the *exact* finite-volume solution given in Ref. [53]. For the other three models, we used our numerical data; in computing the error bar on the ratio, we used the triangle inequality, which of course yields only an upper bound on the true error bar. This overestimate of the error bars in the three non-Ising cases should be taken into account when interpreting the results.

We tried to fit the ratio $\tau_{\mathrm{int},\mathcal{E}}/C_H$ to various different Ansätze (see Table 16). The first is a pure power-law function. In all cases the fit is very good and the estimates of the power are very small (between 0.05 and 0.12); the power seems to increase slightly as we go from the Ising model to the 4-state Potts model. Note also that the fit for the Ising model requires $L_{min} = 100$ in order to get a reasonable $\chi^2$, while for the other three models an excellent $\chi^2$ is obtained already for $L_{min} = 16$. This arises from the very accurate data in the Ising case, which permit the observation of very small corrections to scaling, in contrast to the rather larger (and overestimated!) error bars in the three non-Ising cases.

[26] More precisely, we consider $\tau_{\mathrm{int},\mathcal{E}}/C_H$ for the Ising and 4-state Potts models, and $\tau_{\mathrm{int},\mathcal{E}_\omega}/C_{H,\mathrm{max}}$ for the X2 and ZF models.

Often such small powers indicate that the true behavior is logarithmic. Indeed, a logarithm can be very well mimicked by a power law over any not-too-wide range of $L$. We therefore tried various combinations of logarithmic functions. The first one was a function $A \log^p L$. The quality of fit for the Ising model is rather inferior to that obtained with a power-law function, although the confidence level is still reasonable (27%). However, for the three other models, the fits are very stable and give excellent values of the $\chi^2$. Next we tried a fit to $A + B \log L$; this gives very good results for all four models.

Finally, we have tried to check whether the ratio $\tau_{\mathrm{int},\mathcal{E}}/C_H$ approaches a constant as $L \to \infty$. First we tried a fit to a pure constant $A$. Even by eye one can see that the values of $\tau_{\mathrm{int},\mathcal{E}}/C_H$ for different $L$ exhibit statistically significant deviations (e.g., consecutive values differing by at least two standard deviations) for $L \leq 128$, even for the non–Ising models where we already know that the error bars are overestimated. So, not surprisingly, it is impossible to get a decent fit to a constant with $L_{min} \leq 128$. However, for the non–Ising models with $L_{min} \geq 256$, the ratios $\tau_{\mathrm{int},\mathcal{E}}/C_H$ are consistent with a constant $A$, at least if one takes at face value the overestimated error bars: see Table 16. It follows that no reliable information can be obtained on the *manner* of convergence to a constant for these models, if one takes $L_{min} \geq 256$. Next we tried fitting the ratio $\tau_{\mathrm{int},\mathcal{E}}/C_H$ to $A + BL^{-\Delta}$ with $\Delta = 2, 1, \frac{1}{2}, \frac{1}{4}$, and $\frac{1}{8}$, and also to $A + B/\log L$ ("$\Delta = 0 \times \log$"). For all the models, the fits with $\Delta = 1, 2$ are both implausible and unreliable, as the parameter $B$ becomes very large ($|B| > 10$) and does not stabilize as $L_{min}$ grows. For the Ising model, only the fits with $\Delta = \frac{1}{4}, \frac{1}{8}$ have reasonable $\chi^2$ values (confidence levels of 14% and 28%, respectively). Note, in particular, that the logarithmic-correction Ansatz gives a poor fit (level = 5%). For the 4-state Potts model we always obtain unreasonably large values of $|B|$ ($\gtrsim 10$), so we cannot trust these fits, even if they have reasonable values of $\chi^2$. We therefore rule out this scenario for the 4-state Potts model. The good values of the $\chi^2$ are surely due to the overestimated error bars. This is a warning against trusting the $\chi^2$ values for the other two non–Ising models. Finally, we get reasonable fits for the X2 and ZF models for $0 \leq \Delta \leq \frac{1}{2}$; but perhaps the good $\chi^2$ values are not to be trusted.

Let us discuss the above fits. For the Ising model we have been able to associate a reliable error bar to the ratio $\tau_{\mathrm{int},\mathcal{E}}/C_H$, so we can trust the $\chi^2$ values. We conclude that an asymptotically bounded ratio with additive corrections like either $\Delta = 0 \times \log$ or $\Delta \geq \frac{1}{2}$ is very unlikely to occur. The most plausible scenarios (i.e., highest confidence levels) are the pure power law with $p = 0.060 \pm 0.004$ or a simple logarithmic growth $A + B \log L$. However, a logarithmic power-law behavior with $p = 0.315 \pm 0.020$, or an asymptotically bounded behavior with additive corrections given by $\frac{1}{8} \lesssim \Delta \lesssim \frac{1}{4}$, cannot be completely ruled out.

For the other three models we cannot trust the $\chi^2$ values, as the error bars are overestimated. We have followed three criteria to interpret our results and decide which are the "best" fits:

- *Absolute $\chi^2$ value.* If the $\chi^2$ value of a given fit is *not* good, then the fit is surely poor, as all the error bars are *over*estimated.
- *Relative $\chi^2$ values.* If two fits to the same data exhibit vastly different values of $\chi^2/\mathrm{DF}$, then the fit with the larger $\chi^2/\mathrm{DF}$ can be considered less plausible (all other things being equal).
- *Reasonable results.* When fitting to a constant plus additive corrections ($A + BL^{-\Delta}$), we expect that the parameter $B$ should remain not too large (e.g., $|B| \lesssim 10$). If this does not occur, we tend to distrust the fit.
- *Value of $L_{min}$.* If $L_{min}$ is taken large enough (i.e., $\geq 256$), then we can fit the data for the three non-Ising models with almost any additive-correction Ansatz, as all the values of $\tau_{\mathrm{int},\mathcal{E}}/C_H$ are already consistent with a constant within the (overestimated) errors. Thus, given two fits with similar $\chi^2/\mathrm{DF}$ values, we tend to trust more the one with smaller $L_{min}$.

The 4-state Potts model is the clearest case. The asymptotic constancy of $\tau_{\mathrm{int},\mathcal{E}}/C_H$ is clearly ruled out due to unreasonably large parameter $B$ in all the fits. Moreover, all the asymptotic-constant fits need $L_{min} = 64$ to obtain a decent $\chi^2$ (even with the overestimated error bars) whereas $L_{min} = 16$ is sufficient for the pure power-law and $A + B\log L$ scenarios. Even the logarithmic power-law Ansatz needs $L_{min} = 32$ to achieve a comparably good $\chi^2$. If we consider the $\chi^2$ when $L_{min} = 16$, we see that the best fits have $\chi^2 = 1.39$ (power-law) and 1.98 ($A + B\log L$), while the others have 3.47 ($\Delta = \frac{1}{8}$), 3.64 ($A\log^p L$) and 5.61 ($\Delta = \frac{1}{4}$). We conclude that the two most likely scenarios are the pure power-law with $p = 0.018 \pm 0.012$ and the simple logarithmic behavior $A + B\log L$.

For the X2 model we arrive at the same conclusion. The best fits need $L_{min} = 16$, and the rest need at least $L_{min} = 32$. If we consider the $\chi^2$ when $L_{min} = 16$, we see that the best fits have $\chi^2 = 1.28$ (power-law) and 1.43 ($A + B\log L$), while the others have $\chi^2 = 1.96$ ($\Delta = \frac{1}{8}$), 2.39 (logarithmic power-law) and 2.68 ($\Delta = \frac{1}{4}$).

Finally, the ZF model is the least clear case. Here most of the fits are reasonable with $L_{min} = 16$. Thus, we cannot decide among the pure-power scenario, the two logarithmic scenarios, and the scenario of asymptotic constancy with corrections to scaling given by $\frac{1}{8} \lesssim \Delta \lesssim \frac{1}{4}$.

On theoretical grounds one might expect some kind of "continuity" in the behavior of the ratio $\tau_{\mathrm{int},\mathcal{E}}/C_H$ along the self-dual curve: that is, one might expect the *same* scenario to hold everywhere along the curve (except perhaps at the Ising and 4-state-Potts points, where there might be additional logarithmic effects). There are only two scenarios that are consistent with the data in all four cases:

$$\tau_{\mathrm{int},\mathcal{E}}/C_H \;\approx\; \begin{cases} AL^p & \text{with small } p \\ A + B\log L \end{cases} \tag{4.52}$$

Using the theoretically known exact behavior of $C_H$, we can investigate the validity of the Ansätze (4.52) directly on $\tau_{\mathrm{int},\mathcal{E}}$. For the Ising model these Ansätze become

$$\tau_{\mathrm{int},\mathcal{E}}(\mathrm{DIs}) \;\approx\; \begin{cases} A' L^p \log L \\ A' \log L + B' \log^2 L \end{cases} \tag{4.53}$$

A fit to the first Ansatz with $L_{min} = 100$ has $\chi^2 = 1.35$ (2 DF, level = 51%) and gives $p = 0.051 \pm 0.004$, while a fit to the second Ansatz with $L_{min} = 100$ has $\chi^2 = 1.69$ (2 DF, level = 43%). For the X2 model the Ansätze are

$$\tau_{\mathrm{int},\mathcal{E}_\omega}(\mathrm{X2}) \;\approx\; \begin{cases} A' L^{p+0.4183} \\ (A + B\log L) L^{0.4183} \end{cases} \tag{4.54}$$

The first (pure power-law) Ansatz has already been studied in the preceding subsection, yielding $\chi^2 = 0.39$ (1 DF, level = 53%) with $L_{min} = 128$ [see (4.50)]; the second Ansatz yields $\chi^2 = 0.42$ (1 DF, level = 52%), also with $L_{min} = 128$. So both fits are good, and there is nothing to distinguish them. For the ZF model, the Ansätze are

$$\tau_{\mathrm{int},\mathcal{E}_\omega}(\mathrm{ZF}) \;\approx\; \begin{cases} A' L^{p+2/3} \\ (A + B\log L) L^{2/3} \end{cases} \tag{4.55}$$

The first (pure power-law) Ansatz has already been studied in the preceding subsection, yielding $\chi^2 = 1.48$ (3 DF, level = 68%) with $L_{min} = 32$ [see (4.49)]; the second Ansatz yields $\chi^2 = 1.53$ (4 DF, level = 82%) with $L_{min} = 16$. So in this case there is a slight preference for the logarithmic Ansatz. Finally, for the 4-state Potts model the Ansätze are

$$\tau_{\mathrm{int},\mathcal{E}_\omega}(\mathrm{P}) \;\approx\; \begin{cases} A' L^{p+1} \log^{-3/2} L \\ (A + B\log L) L \log^{-3/2} L \end{cases} \tag{4.56}$$

The first one gives the power $p = 0.13 \pm 0.03$ for $L_{min} = 128$ with $\chi^2 = 0.96$ (2 DF, level = 62%). The second one yields $\chi^2 = 0.98$ for the same $L_{min}$ (2 DF, level = 61%). Finally, one can try the fit

$$\tau_{\mathrm{int},\mathcal{E}}(\mathrm{P}) \;\approx\; A L \log^{-p'} L \,, \tag{4.57}$$

in which one imposes $z_{\mathrm{int},\mathcal{E}} = 1 \times \log^{-p'}$ and attempts to find the multiplicative logarithmic exponent $p'$. Here the stability of the results is not very good (see Table 17). One could argue that the fit with $L_{min} = 16$ already has a reasonable $\chi^2$ value, but we are inclined to be conservative and prefer the fit with $L_{min} = 128$ (which of course has much larger error bars):

$$p'(\mathrm{P}) \;=\; 0.776 \pm 0.180 \tag{4.58}$$

with $\chi^2 = 1.09$ (2 DF, level = 58%). This result is compatible (within 2 standard deviations) with the expected value of $p' = \frac{1}{2}$. If we take into account the value reported in Table 16 for the fit $\tau_{\mathrm{int},\mathcal{E}}/C_H = A \log^p L$ [$p = 0.543 \pm 0.073$], we see that our estimates for $p$ and $p'$ are compatible within errors (i.e., $p' = \frac{3}{2} - p =$ $0.957 \pm 0.073$). However, that value of $p$ is not compatible with the value of $p' = \frac{1}{2}$ corresponding to Ansatz (4.56b). Actually, our estimate of $p'$ lies midway between the Ansatz and the value coming from our estimate of $p$.

From the above results it is difficult to tell which is the true asymptotic behavior of $\tau_{\mathrm{int},\mathcal{E}}/C_H$ (and thus of $\tau_{\mathrm{int},\mathcal{E}}$). To disentangle this we would need much larger lattices, as well as much better statistics on the lattices $L \gtrsim 128$.

**Remark.** Let us also test the conjecture proposed by Baillie and Coddington [7] for the Ising model:

$$\tau_{\mathrm{int},\mathcal{E}} \sim (A + B \log L) L^{\beta/\nu} \,, \tag{4.59}$$

where $\beta$ is the static critical exponent for the spontaneous magnetization; note that $\beta/\nu = \frac{1}{8}$ *everywhere* on the AT self-dual curve. For the Ising model, the fit is fairly good for $L_{min} = 100$, giving $\chi^2 = 0.50$ (2 DF, level = 78%). The same conclusion applies to the X2 model ($L_{min} = 128$, $\chi^2 = 0.42$, 1 DF, level = 52%) and to the ZF model ($L_{min} = 128$, $\chi^2 = 0.44$, 1 DF, level = 51%). However, the fit is rather poor for the 4-state Potts model: for $L_{min} = 256$ we only get $\chi^2 = 4.43$ (1 DF, level = 4%).

### 4.5.3 Exponential autocorrelation time

The exponential autocorrelation time is for an observable $\mathcal{O}$ is defined as[27]

$$\tau_{\mathrm{exp},\mathcal{O}} \;=\; \lim_{t\to\infty} \frac{|t|}{-\log|\rho_{\mathcal{O}\mathcal{O}}(t)|} \,. \tag{4.60}$$

This autocorrelation time measures the decay rate of the "slowest mode" of the system, provided that this mode is not orthogonal to $\mathcal{O}$.

The critical behavior of $\tau_{\mathrm{exp},\mathcal{O}}$ is, in general, different from the behavior of $\tau_{\mathrm{int},\mathcal{O}}$. This fact can be seen from the standard dynamic finite-size-scaling Ansatz for the autocorrelation function $\rho_{\mathcal{O}\mathcal{O}}(t)$:

$$\rho_{\mathcal{O}\mathcal{O}}(t;L) \approx |t|^{-p_{\mathcal{O}}} h_{\mathcal{O}}\!\left(\frac{t}{\tau_{\mathrm{exp},\mathcal{O}}}; \frac{\xi(L)}{L}\right) \,. \tag{4.61}$$

(Here the dependence on the coupling constants has been suppressed for notational simplicity.) Summing (4.61) over $t$, it follows that

$$\tau_{\mathrm{int},\mathcal{O}} \;\sim\; \tau_{\mathrm{exp},\mathcal{O}}^{1-p_{\mathcal{O}}} \,, \tag{4.62}$$

or equivalently,

$$z_{\mathrm{int},\mathcal{O}} \;=\; (1 - p_{\mathcal{O}}) z_{\mathrm{exp},\mathcal{O}} \,. \tag{4.63}$$

Thus, only when $p_{\mathcal{O}} = 0$ do we have $z_{\mathrm{int},\mathcal{O}} = z_{\mathrm{exp},\mathcal{O}}$ [1,4].

Here we consider the exponential autocorrelation time of the energy $\mathcal{E}$ for the 4-state Potts model and of the observable $\mathcal{E}_\omega$ for the X2 and ZF models. We expect that a similar behavior would be found for $\mathcal{E}_{\sigma\tau}$ for the X2 and ZF models, and for the squared magnetizations in all three models; but it did not seem worthwhile to us to carry out this analysis in detail.

[27] Strictly speaking, the "lim" should be replaced by "lim sup", as in (4.4). But in virtually all practical applications, the limit really exists.

If we plot the estimated autocorrelation function $\rho_{\mathcal{E}\mathcal{E}}(t)$, we see that it fits beautifully to a pure exponential for $t \gtrsim \tau_{\text{int},\mathcal{E}}/2$: as examples, see Figures 2 and 3, which represent the data for the 4-state Potts model with $L = 16, 32$ and $L = 256$, respectively. So it makes sense to extract estimates of $\tau_{\text{exp},\mathcal{E}}$ from the tail of the autocorrelation function. More precisely, for each run we estimated $\tau_{\text{exp},\mathcal{E}}$ by fitting $\log \rho_{\mathcal{E}\mathcal{E}}(t) = -A - Bt$ for the interval $t_{min} \leq t \leq t_{max}$; obviously $\tau_{\text{exp},\mathcal{E}} = 1/B$. By studying the goodness of fit (i.e., the $\chi^2$ value and the corresponding confidence level) as a function of $t_{min}$ and $t_{max}$, we can choose the "best" fit.

This fit is, however, extremely subtle, because the Monte Carlo estimates of $\rho_{\mathcal{E}\mathcal{E}}(t)$ for different $t$ are in general (highly) correlated. The full covariance matrix $\widehat{C}_\rho$ for these random variables can in principle be computed using the autocorrelation function itself, at least in the approximation that neglects the fourth cumulant of the stochastic process: see equation (B.1) in Appendix B. If one assumes for simplicity that the autocorrelation function is a pure exponential, then the formula simplifies further: see equation (B.2). From this formula we already see that the off-diagonal terms in $\widehat{C}_\rho$ are comparable in magnitude to the diagonal ones; therefore, it is unlikely to be sensible to neglect them.

Nevertheless, one may try the crude approximation in which all off-diagonal terms in $\widehat{C}_\rho$ are dropped, and see what happens. Using the self-consistent procedure explained in detail in Appendix B, we obtained both an estimate of $\tau_{\text{exp},\mathcal{E}}$ (along with its error bar) and the error bars for the autocorrelation function $\rho_{\mathcal{E}\mathcal{E}}(t)$. For almost all values of $t_{min}$ and $t_{max}$, we obtained a perfect fit (confidence level $\approx 100\%$), with values of the $\chi^2$ much smaller than the number of degrees of freedom. However, the variation of the estimates of $\tau_{\text{exp},\mathcal{E}}$ as a function of $t_{min}$ and $t_{max}$ was much larger than the error bars given by the fit. This indicates that the error bar for the estimated $\tau_{\text{exp},\mathcal{E}}$ was not correctly computed. Clearly, the neglect of the off-diagonal covariances is unjustified, as we already expected on theoretical grounds.

We therefore redid the fit using the *full* covariance matrix $\widehat{C}_\rho$, again using the self-consistent procedure described in Appendix B. As before, we systematically varied $t_{min}$ and $t_{max}$. The dependence of the results on $t_{max}$ is usually slight; but the dependence on $t_{min}$ is moderately strong. We report our results in Tables 18–19 for the 4-state Potts model[28], in Tables 20–21 for the X2 model, and in Tables 22–23 for the ZF model. For each model we have presented two different tables, one with $t_{max} = 4\tau_{\text{int},\mathcal{E}}$ and the other with $t_{max} = 3\tau_{\text{int},\mathcal{E}}$.

In all cases we have a bad fit (level $\ll 1\%$) whenever $t_{min} \ll 0.5\tau_{\text{int},\mathcal{E}}$ (this happens irrespective of the value of $t_{max}$). As an example of such a bad fit, we show in each table the fit with $t_{min} = 1$. Clearly, the energy autocorrelation function is significantly different from a pure exponential for very small $t$. However, we obtain reasonable $\chi^2$ values as soon as $t_{min} \gtrsim 0.5\tau_{\text{int},\mathcal{E}}$, indicating that the autocorrelation function becomes very close to a pure exponential for $t \gtrsim 0.5\tau_{\text{int},\mathcal{E}}$.

[28]The results for $L = 1024$ have not been quoted, as the statistic is rather poor ($\sim 1500\tau_{\text{int},\mathcal{E}}$).

For the 4-state Potts model (Tables 18–19) we have rather stable results for $L = 16, 32, 64$. In most of these fits the confidence level remains reasonable, so we think we can fully trust these results. However, for $L \geq 128$ we begin to have difficulties: the self-consistent procedure does not always converge, and the results with $t_{max} = 4\tau_{\mathrm{int},\mathcal{E}}$ sometimes differ from those with $t_{max} = 3\tau_{\mathrm{int},\mathcal{E}}$ by about two standard deviations. Furthermore, for these large lattices we consistently obtain unusually high confidence levels (mostly $> 99\%$); we do not understand why this happens. For these lattice sizes the troubles seem to be somewhat less severe when $t_{max} = 3\tau_{\mathrm{int},\mathcal{E}}$. So we tend to trust better these latter fits when $L \geq 128$.

For the X2 model (Tables 20–21) both the stability of the results and the goodness of fit are excellent for $L = 16$ and $L = 32$. For $L = 64$ we have bad fits when $t_{max} = 4\tau_{\mathrm{int},\mathcal{E}_\omega}$ (level $< 1\%$), and somewhat better ones when $t_{max} = 3\tau_{\mathrm{int},\mathcal{E}_\omega}$ (level $\approx 8\%$). Also for $L = 128$ we obtain somewhat more stable and consistent results when $t_{max} = 3\tau_{\mathrm{int},\mathcal{E}_\omega}$. For $L = 256$ the fits are rather poorer; for $L = 512$ they are good, but the results for the two different values of $t_{max}$ differ by three standard deviations.

Finally, for the ZF model (Tables 22–23) we have good stability and goodness of fit for $L = 16$ and $L = 32$. For $L = 64$ the fits are usually very poor (level $\lesssim$ 5–8%), but they are generally less bad (and also more stable) for $t_{max} = 3\tau_{\mathrm{int},\mathcal{E}_\omega}$ than for $t_{max} = 4\tau_{\mathrm{int},\mathcal{E}_\omega}$. For $L \geq 128$ we also find the results with $t_{max} = 3\tau_{\mathrm{int},\mathcal{E}_\omega}$ somewhat better than for $t_{max} = 4\tau_{\mathrm{int},\mathcal{E}_\omega}$; again we find inexplicably high confidence levels (often $> 99\%$).

From the above discussion we can already see that we have good and stable fits only for the smaller lattices ($L \leq 32$ and in some cases $L = 64$), where the statistics are better. Apparently, in order to have a decent estimate of $\tau_{\mathrm{exp},\mathcal{E}}$ we need a run length of at least $6 \times 10^4\ \tau_{\mathrm{int},\mathcal{E}}$ and possibly more.

To decide which are the "best" fits we use the following criteria: we choose the largest interval $[t_{min}, t_{max}]$ such that

(a) the $\chi^2$ value is reasonable (e.g., confidence level $\gtrsim 10\%$); and

(b) there is some consistency (within error bars) with the values coming from "nearby" choices of $t_{min}$ and $t_{max}$.

In Table 24 we present what we consider to be the "best" fits for each model and each lattice size $L$. We include the interval $[t_{min}, t_{max}]$ and the ratio $\tau_{\mathrm{int},\mathcal{E}}/\tau_{\mathrm{exp},\mathcal{E}}$. The error bar on this ratio was computed using the triangle inequality, as we do not know what is the covariance between our estimates of $\tau_{\mathrm{int},\mathcal{E}}$ and $\tau_{\mathrm{exp},\mathcal{E}}$. We think these estimates are reasonably reliable for $L = 16$ and 32, somewhat reliable for $L = 64$, and possibly unreliable for $L \geq 128$.

From Table 24 we see that for each model, the values of the ratio $\tau_{\mathrm{int},\mathcal{E}}/\tau_{\mathrm{exp},\mathcal{E}}$ are more or less constant when $L \leq 64$. If we fit the ratios for $16 \leq L \leq 64$ to a constant, we obtain reasonable fits for the three models: $\tau_{\mathrm{int},\mathcal{E}}/\tau_{\mathrm{exp},\mathcal{E}} = 0.936 \pm 0.015$ for the 4-state Potts model ($\chi^2 = 1.50$, 2 DF, level = 47%), $\tau_{\mathrm{int},\mathcal{E}_\omega}/\tau_{\mathrm{exp},\mathcal{E}_\omega} = 0.966 \pm 0.015$ for the ZF model ($\chi^2 = 0.10$, 2 DF, level = 95%) and $\tau_{\mathrm{int},\mathcal{E}_\omega}/\tau_{\mathrm{exp},\mathcal{E}_\omega} = 0.980 \pm 0.013$ for the X2 model ($\chi^2 = 0.50$, 2 DF, level = 78%). However, for the 4-state Potts and ZF models, this ratio seems to decrease when $L \geq 128$. Now, these points are precisely those where we had troubles obtaining the value of $\tau_{\mathrm{exp},\mathcal{E}}$, so this decrease might be due simply to a poor determination of the exponential autocorrelation time; and it is in any case only about two standard deviations.[29] On the other hand, this decrease *might* indicate that $\tau_{\mathrm{int},\mathcal{E}}/\tau_{\mathrm{exp},\mathcal{E}}$ is tending to *zero* as $L \to \infty$; in this case the dynamic critical exponents $z_{\mathrm{int},\mathcal{E}}$ and $z_{\mathrm{exp},\mathcal{E}}$ would be different (the latter would be larger), and the exponent $p_{\mathcal{E}}$ in (4.61) would be strictly positive.

The result for the X2 model, by contrast, is completely consistent with a constant ratio $\tau_{\mathrm{int},\mathcal{E}_\omega}/\tau_{\mathrm{exp},\mathcal{E}_\omega}$, indicating that $p_{\mathcal{E}_\omega} = 0$. This fit to a constant, using all the data ($16 \leq L \leq 512$), gives

$$\frac{\tau_{\mathrm{int},\mathcal{E}_\omega}}{\tau_{\mathrm{exp},\mathcal{E}_\omega}}(\mathrm{X2}) = 0.976 \pm 0.011 \tag{4.64}$$

with $\chi^2 = 2.14$ (5 DF, level = 83%).

Due to the ambiguities in the determination of $\tau_{\mathrm{exp},\mathcal{E}}$ for $L \geq 128$ in all the models, we are unable to come to any definitive conclusion on whether $z_{\mathrm{int},\mathcal{E}} = z_{\mathrm{exp},\mathcal{E}}$. But it does seem likely.

# 5 Conclusions

We have performed a high-precision Monte Carlo study of the symmetric Ashkin–Teller model at several points on the self-dual (critical) curve, using a Swendsen–Wang–type algorithm. We have considered both the static behavior of the models (known exactly) and the dynamic behavior of the algorithm.

We have had great difficulties in obtaining the correct leading behavior whenever this is not simply a power law plus additive power-law corrections. These difficulties occurred both for static quantities (specific heat in the Ising and 4-state Potts models) and for dynamic quantities (autocorrelation times in all models). Unless we have some theoretical input, it is almost impossible to distinguish between power-law and logarithmic behaviors when the range of lattice sizes $L$ is not extremely large (in our case, $16 \leq L \leq 1024$).

This issue makes it very problematic to tell whether the Li–Sokal bound (1.1)/(1.2) is sharp or not. Our results seem to indicate that there are only two likely scenarios: the Li–Sokal bound fails to be sharp either by a small power [i.e., $\tau_{\mathrm{int},\mathcal{E}}/C_H \approx AL^p$ with $0.05 \lesssim p \lesssim 0.12$] or by only a logarithm [e.g., $\tau_{\mathrm{int},\mathcal{E}}/C_H \approx A + B\log L$]. Either one of these scenarios is consistent with our data at all four points on the AT self-dual curve. Larger lattices and much better statistics will be needed to distinguish between them.

We have also presented a new method for estimating the exponential autocorrelation time, which takes into account the full covariance matrix for the sample autocorrelation function. To do so is essential to obtain reliable results, as the values of the sample autocorrelation function are strongly positively correlated. The quality of the estimates of $\tau_{\mathrm{exp},\mathcal{O}}$ depends strongly on the accuracy of the available data: we seem to get reliable estimates of $\tau_{\mathrm{exp},\mathcal{O}}$ only when the run length is at least $\approx 60000\tau_{\mathrm{int},\mathcal{O}}$.

[29] If we fit all the data ($16 \leq L \leq 512$) we get the following ratios: $\tau_{\mathrm{int},\mathcal{E}}/\tau_{\mathrm{exp},\mathcal{E}} = 0.914 \pm 0.013$ for the 4-state Potts model ($\chi^2 = 13.84$, 5 DF, level = 2%) and $\tau_{\mathrm{int},\mathcal{E}_\omega}/\tau_{\mathrm{exp},\mathcal{E}_\omega} = 0.945 \pm 0.014$ for the ZF model ($\chi^2 = 10.84$, 5 DF, level = 5%). These confidence levels on the order of 5% are as expected from the two-standard-deviation discrepancies.

# A Proof of Li–Sokal bound for the direct Ashkin–Teller algorithm

We can easily extend the proof of the Li–Sokal bound for the $q$-state Potts model [9] to the direct algorithm for the AT model (defined in Section 3.1). Also in this latter case, the transition matrix can be written as a product

$$P_{SW} \;=\; P_{\mathrm{bond}} P_{\mathrm{spin}} \;, \tag{A.1}$$

where $P_{\mathrm{bond}}$ (the update of the bond variables) and $P_{\mathrm{spin}}$ (the update of the spin variables) are given by the conditional expectation operators $E(\,\cdot\,|\{\sigma,\tau\})$ and $E(\,\cdot\,|\{m,n\})$, respectively.

As in [9], we are going to compute explicitly the autocorrelation function at time lags 0 and 1 for several bond densities. Then, using some general properties of reversible Markov chains, we will deduce lower bounds for the autocorrelation times $\tau_{\mathrm{int},A}$ (for certain observables $A$) and $\tau_{\mathrm{exp}}$. These will in turn imply lower bounds on the dynamic critical exponents $z_{\mathrm{int},A}$ and $z_{\mathrm{exp}}$.

Let us consider the bond occupations[30]

$$\mathcal{M} \;=\; \sum_{\langle xy\rangle} m_{xy} \tag{A.2a}$$

$$\mathcal{N} \;=\; \sum_{\langle xy\rangle} n_{xy} \tag{A.2b}$$

$$\mathcal{O} \;=\; \sum_{\langle xy\rangle} m_{xy} n_{xy} \tag{A.2c}$$

We will follow the notation of [9] and henceforth write the Kronecker deltas for a bond $b=\langle xy\rangle$ as $\delta_{\sigma_b} \equiv \delta_{\sigma_x,\sigma_y}$ and $\delta_{\tau_b} \equiv \delta_{\tau_x,\tau_y}$.

From (3.6) we can read off the expectation value of the bond variable $m_b$ conditional on the spin configuration $\{\sigma,\tau\}$: it is

$$E(m_b|\{\sigma,\tau\}) \;=\; q_1\,\delta_{\sigma_b} \;=\; (p_1+p_2)\,\delta_{\sigma_b} \;, \tag{A.3}$$

where $q_1$, $p_1$ and $p_2$ are the probabilities appearing in Steps 1a and 1b of the direct algorithm (Section 3.1).[31] The important fact here is that $q_1 = p_1 + p_2$: this means

[30] Do not confuse this $\mathcal{M}$ with a magnetization!

[31] In this appendix we are assuming that the system is homogeneous, i.e., that the couplings $J$, $J'$ and $K$ do not vary from bond to bond. An inhomogeneous system can be treated by an obvious generalization.

that $E(m_b|\{\sigma,\tau\})$ does *not* depend on the $\tau$ configuration. Likewise we have

$$E(n_b|\{\sigma,\tau\}) \;=\; r_1\,\delta_{\tau_b} \;=\; (p_1+p_3)\,\delta_{\tau_b} \qquad \text{(A.4a)}$$

$$E(m_b n_b|\{\sigma,\tau\}) \;=\; p_1\,\delta_{\sigma_b}\delta_{\tau_b} \qquad \text{(A.4b)}$$

The other conditional probabilities we need are

$$E(m_b m_{b'}|\{\sigma,\tau\}) \;=\; \begin{cases} q_1^2\,\delta_{\sigma_b}\delta_{\sigma_{b'}} & \text{if } b\neq b' \\ q_1\,\delta_{\sigma_b} & \text{if } b=b' \end{cases} \qquad \text{(A.5a)}$$

$$E(n_b n_{b'}|\{\sigma,\tau\}) \;=\; \begin{cases} r_1^2\,\delta_{\tau_b}\delta_{\tau_{b'}} & \text{if } b\neq b' \\ r_1\,\delta_{\tau_b} & \text{if } b=b' \end{cases} \qquad \text{(A.5b)}$$

$$E(m_b n_{b'}|\{\sigma,\tau\}) \;=\; \begin{cases} q_1 r_1\,\delta_{\sigma_b}\delta_{\tau_{b'}} & \text{if } b\neq b' \\ p_1\,\delta_{\sigma_b}\delta_{\tau_b} & \text{if } b=b' \end{cases} \qquad \text{(A.5c)}$$

$$E(m_b n_b m_{b'}|\{\sigma,\tau\}) \;=\; \begin{cases} p_1 q_1\,\delta_{\sigma_b}\delta_{\tau_b}\delta_{\sigma_{b'}} & \text{if } b\neq b' \\ p_1\,\delta_{\sigma_b}\delta_{\tau_b} & \text{if } b=b' \end{cases} \qquad \text{(A.5d)}$$

$$E(m_b n_b n_{b'}|\{\sigma,\tau\}) \;=\; \begin{cases} p_1 r_1\,\delta_{\sigma_b}\delta_{\tau_b}\delta_{\tau_{b'}} & \text{if } b\neq b' \\ p_1\,\delta_{\sigma_b}\delta_{\tau_b} & \text{if } b=b' \end{cases} \qquad \text{(A.5e)}$$

$$E(m_b n_b m_{b'} n_{b'}|\{\sigma,\tau\}) \;=\; \begin{cases} p_1^2\,\delta_{\sigma_b}\delta_{\tau_b}\delta_{\sigma_{b'}}\delta_{\tau_{b'}} & \text{if } b\neq b' \\ p_1\,\delta_{\sigma_b}\delta_{\tau_b} & \text{if } b=b' \end{cases} \qquad \text{(A.5f)}$$

These simply reflect the fact that in Step 1 each bond is updated independently of each other bond.

From (A.3) and (A.5a) it is easy to compute the mean values $\langle\mathcal{M}\rangle$ and $\langle\mathcal{M}^2\rangle$, and hence $\text{var}(\mathcal{M}) \equiv \langle\mathcal{M}^2\rangle - \langle\mathcal{M}\rangle^2$:

$$\langle\mathcal{M}\rangle \;=\; 2V\,\frac{q_1}{2}(1+E_\sigma) \qquad \text{(A.6a)}$$

$$\langle\mathcal{M}^2\rangle \;=\; \frac{q_1^2}{4}\langle\mathcal{E}_\sigma^2\rangle + q_1^2V^2(1+2E_\sigma) + q_1(1-q_1)V(1+E_\sigma) \qquad \text{(A.6b)}$$

$$\text{var}(\mathcal{M}) \;=\; 2V\left[\frac{q_1^2}{4}C_{H,\sigma\sigma} + \frac{q_1(1-q_1)}{2}(1+E_\sigma)\right] \qquad \text{(A.6c)}$$

where $E_\sigma = (1/2V)\langle\mathcal{E}_\sigma\rangle$ is one of the energies defined in Section 4.2, and $C_{H,\sigma\sigma} = (1/2V)\text{var}(\mathcal{E}_\sigma)$ is the corresponding specific heat. The unnormalized autocorrelation function of $\mathcal{M}$ at time lag 0 is precisely $C_{\mathcal{M}\mathcal{M}}(0) = \text{var}(\mathcal{M})$.

The corresponding autocorrelation function at time lag 1 is given by

$$C_{\mathcal{M}\mathcal{M}}(1) \;=\; \langle\mathcal{M}(0)\mathcal{M}(1)\rangle - \langle\mathcal{M}\rangle^2 \;=\; \text{var}(E(\mathcal{M}|\{\sigma\tau\})) \equiv \text{var}(P_{\text{bond}}\mathcal{M})\,. \qquad \text{(A.7)}$$

Now, the energy-like operator $P_{\text{bond}}\mathcal{M}$ is equal to

$$P_{\text{bond}}\mathcal{M} \;\equiv\; E(\mathcal{M}|\{\sigma,\tau\}) \;=\; \frac{q_1}{2}(2V+\mathcal{E}_\sigma)\,. \qquad \text{(A.8)}$$

This implies that

$$C_{\mathcal{M}\mathcal{M}}(1) \;=\; \frac{q_1^2}{4}\text{var}(\mathcal{E}_\sigma) \;=\; 2V\,\frac{q_1^2}{4}C_{H,\sigma\sigma}\,. \qquad \text{(A.9)}$$

Thus, the normalized autocorrelation function for the operator $\mathcal{M}$ at time lag 1 is given by

$$\rho_{\mathcal{M}\mathcal{M}}(1) \;=\; \frac{C_{\mathcal{M}\mathcal{M}}(1)}{C_{\mathcal{M}\mathcal{M}}(0)} \;=\; 1 - \frac{2(1-q_1)(1+E_\sigma)}{q_1 C_{H,\sigma\sigma} + 2(1-q_1)(1+E_\sigma)}\,. \tag{A.10}$$

Note now that when we approach any point on the critical curve, the quantity $q_1 \equiv 1 - e^{-2(J+K)}$ remains positive and less than 1 [i.e., $q_1 \to q_{1,crit} \in (0,1)$], while the energy $E_\sigma$ remains greater than $-1$ [i.e., $E_\sigma \to E_{\sigma,crit} > -1$]. It follows that

$$\rho_{\mathcal{M}\mathcal{M}}(1) \;\geq\; 1 - \frac{\text{const}}{C_{H,\sigma\sigma}} \tag{A.11}$$

uniformly in a neighborhood of that critical point.

The rest of the argument given in [9] can now be transcribed verbatim. The correlation functions of $\mathcal{M}$ under $P_{SW}$ are the same as under the positive-semidefinite self-adjoint operator $P'_{SW} \equiv P_{\rm spin} P_{\rm bond} P_{\rm spin}$. This fact implies that we have a spectral representation

$$\rho_{\mathcal{M}\mathcal{M}}(t) = \int_0^1 \lambda^{|t|}\, d\nu(\lambda) \tag{A.12}$$

with a positive measure $d\nu$. From this equation we conclude that

$$\rho_{\mathcal{M}\mathcal{M}}(t) \geq \rho_{\mathcal{M}\mathcal{M}}(1)^{|t|}\,. \tag{A.13}$$

Using the definition (4.3) of the integrated autocorrelation time and the definition (4.4) of the exponential autocorrelation time, we arrive at the following bounds

$$\tau_{{\rm int},\mathcal{M}} \;\geq\; \frac{1}{2}\,\frac{1+\rho_{\mathcal{M}\mathcal{M}}(1)}{1-\rho_{\mathcal{M}\mathcal{M}}(1)} \;\geq\; \text{const} \times C_{H,\sigma\sigma} \tag{A.14a}$$

$$\tau_{{\rm exp},\mathcal{M}} \;\geq\; \frac{-1}{\log \rho_{\mathcal{M}\mathcal{M}}(1)} \;\geq\; \text{const} \times C_{H,\sigma\sigma} \tag{A.14b}$$

Let us now assume that the autocorrelation times diverge (for a finite system of size $L$ at criticality) as $L^{z_{{\rm int},\mathcal{M}}}$ and $L^{z_{{\rm exp},\mathcal{M}}}$, respectively, and that the matrix element $C_{H,\sigma\sigma}$ of the specific-heat matrix diverges as $L^{\alpha/\nu}$. We then conclude that

$$z_{{\rm int},\mathcal{M}},\; z_{{\rm exp},\mathcal{M}} \;\geq \frac{\alpha}{\nu}\,, \tag{A.15}$$

which is the result of Ref. [9].

The only way that the bound (A.15) could fail is in case the matrix element $C_{H,\sigma\sigma}$ fails to diverge as $L^{\alpha/\nu}$ (e.g., by diverging with a smaller power or by being bounded). Now, the matrix $\widehat{C}_H$ does have an exactly marginal eigenvalue everywhere on the self-dual curve (2.8) of the symmetric AT model, so that the component of $\widehat{C}_H$ tangent to this self-dual curve (namely, $C_{H,\min}$) is bounded as $L \to \infty$.[32] However, this marginal direction is never exactly $\sigma$, so the preceding proof does in fact always succeed: $C_{H,\sigma\sigma}$ does always diverge as $L^{\alpha/\nu}$. Nevertheless, some extra generality — as well as slightly sharper *constants* in the lower bound (A.14b) on $\tau_{\rm exp}$— can be obtained by choosing in place of $\mathcal{M}$ a more general bond observable

$$\mathcal{X} = c_1\mathcal{M} + c_2\mathcal{N} + c_3\mathcal{O}\ , \tag{A.16}$$

where $c_1$, $c_2$ and $c_3$ are arbitrary real constants. This case trivially includes the preceding one. Using the techniques described above and after some algebra we arrive at the following results

$$\begin{aligned}
\langle \mathcal{X} \rangle &= 2V\ \left[\frac{c_1 r_1}{2}(1+E_\tau) + \frac{c_2 q_1}{2}(1+E_\sigma) + \frac{c_3 p_1}{4}(1+E_T)\right] && \text{(A.17a)}\\
C_{\mathcal{X}\mathcal{X}}(0) &= 2V\ (C_{H,P_{\rm bond}\mathcal{X}P_{\rm bond}\mathcal{X}} + \widetilde{E}_{\mathcal{X}}) && \text{(A.17b)}\\
C_{\mathcal{X}\mathcal{X}}(1) &= 2V\ C_{H,P_{\rm bond}\mathcal{X}P_{\rm bond}\mathcal{X}} && \text{(A.17c)}\\
\rho_{\mathcal{X}\mathcal{X}}(1) &= 1 - \frac{\widetilde{E}_{\mathcal{X}}}{C_{H,P_{\rm bond}\mathcal{X}P_{\rm bond}\mathcal{X}} + \widetilde{E}_{\mathcal{X}}} && \text{(A.17d)}
\end{aligned}$$

where we have defined the following quantities

$$\begin{aligned}
E_T &= E_\tau + E_\sigma + E_{\sigma\tau} && \text{(A.18a)}\\
P_{\rm bond}\mathcal{X} &= E(\mathcal{X}|\{\sigma,\tau\}) \\
&= \left(\frac{r_1 c_1}{2} + \frac{c_3 p_1}{4}\right)\mathcal{E}_\tau + \left(\frac{q_1 c_2}{2} + \frac{c_3 p_1}{4}\right)\mathcal{E}_\sigma + \frac{c_3 p_1}{4}\mathcal{E}_{\sigma\tau} \\
&\qquad +2V\ \frac{2c_1 r_1 + 2c_2 q_1 + c_3 p_1}{4} && \text{(A.18b)}\\
C_{H,P_{\rm bond}\mathcal{X}P_{\rm bond}\mathcal{X}} &= \frac{1}{2V}\text{var}(P_{\rm bond}\mathcal{X}) && \text{(A.18c)}\\
\widetilde{E}_{\mathcal{X}} &= \frac{c_1^2 r_1(1-r_1)}{2}(1+E_\tau) + \frac{c_2^2 q_1(1-q_1)}{2}(1+E_\sigma) \\
&\qquad + \frac{1+E_T}{4}[c_3^2 p_1(1-p_1) + 2c_1c_2(p_1 - q_1^2) \\
&\qquad\qquad + 2c_1c_3p_1(1-r_1) + 2c_2c_3p_1(1-q_1)]\ , && \text{(A.18d)}
\end{aligned}$$

and the rest of the observable quantities are defined in Section 4.

[32] More generally, in the non-symmetric AT model on the self-dual manifold (2.7), there are *two* marginal eigenvalues, corresponding to the two tangent vectors to (2.7).

Using the parameters $(c_1, c_2, c_3)$ we can choose the energy-like operator $P_{\rm bond}\mathcal{X}$ in such a way that it contains a non-zero projection on the most divergent eigenvector of the matrix $\widehat{C}_H$. This is always possible as we have enough freedom. On the other hand, the term $\widetilde{E}_{\mathcal{X}}$ should be bounded from below by a strictly positive number.

Let us analyze an interesting particular case: the symmetric model ($r_1 = q_1$) on the self-dual curve. The natural choice is $c_1 = c_2$ and the previous formulae can be simplified to

$$P_{\rm bond}\mathcal{X} = \left(\frac{c_3 p_1}{2} + c_1 q_1\right)\mathcal{E}_\omega + \frac{c_3 p_1}{4}\mathcal{E}_{\sigma\tau} + 2V\ \left(c_1 q_1 + \frac{c_3 p_1}{4}\right) \tag{A.19a}$$

$$
\begin{aligned}
\widetilde{E}_{\mathcal{X}} &= c_1^2 q_1(1-q_1)(1+E_\omega) \\
&\quad + \frac{1+E_T}{4}[c_3^2 p_1(1-p_1) + 4c_1c_3p_1(1-q_1) + 2c_1^2(p_1-q_1^2)] \qquad \text{(A.19b)}
\end{aligned}
$$

We can choose any ratio $c_3/c_1$ such that $P_{\rm bond}\mathcal{X}$ is not a multiple of $\mathcal{X}_{\rm min} = \mathcal{E}_\omega + a\mathcal{E}_{\sigma\tau}$, where $a$ is the parameter defined in (4.26). This means that on the self-dual curve the ratio $c_3/c_1$ could be anything different from $-\frac{1}{2}\sinh 2J$. In particular, we can make the choice $(c_1, c_3) = (1, 0)$, leading to $P_{\rm bond}\mathcal{X} = 2q_1\mathcal{E}_\omega + \text{const}$ and to the bound

$$
\rho_{\mathcal{X}\mathcal{X}}(1) = 1 - \frac{\widetilde{E}_X}{q_1^2 C_{H,\omega\omega} + \widetilde{E}_X} \qquad \text{(A.20a)}
$$

$$
\widetilde{E}_{\mathcal{X}} = q_1(1-q_1)(1+E_\omega) + \frac{p_1 - q_1^2}{2}(1+E_T) \qquad \text{(A.20b)}
$$

Let us now restrict attention to the part of the self-dual curve for which the specific heat is divergent, namely the interval between the decoupled Ising point (DIs) and the 4-state Potts point (P). In this interval we have $J, K \geq 0$, so Griffiths' first inequality implies that the energies $E_\omega$ and $E_{\sigma\tau}$ are $\geq 0$ (and hence so is $E_T$). Moreover, both $q_1$ and $p_1 - q_1^2$ are strictly positive on this interval. Since the specific heat $C_{H,\omega\omega}$ is divergent everywhere on this interval, the proof of the bound (1.2) is complete.

# B Fitting (highly correlated) autocorrelation functions

Let $\widehat{\rho}(t)$ be the normalized sample autocorrelation function for some particular observable, measured in a Monte Carlo run of length $N$. These *measured* values $\{\widehat{\rho}(t)\}_{t=-(N-1)}^{N-1}$ are of course random variables; as such they have a covariance matrix $\widehat{C}_\rho$, which is given by a standard formula from time-series analysis [51]:

$$
\begin{aligned}
\text{cov}[\widehat{\rho}(r), \widehat{\rho}(s)] &= \frac{1}{N}\sum_{m=-\infty}^{+\infty} \Big[\rho(m)\rho(m+r-s) + \rho(m+s)\rho(m-r) + 2\rho(r)\rho(s)\rho(m)^2 \\
&\qquad - 2\rho(r)\rho(m)\rho(m-s) - 2\rho(s)\rho(m)\rho(m-r)\Big] + O\left(\frac{1}{N^2}\right) , \qquad \text{(B.1)}
\end{aligned}
$$

where $\{\rho(t)\}$ are the *true* values of the autocorrelation function. This formula is valid in the limit $N \to \infty$, provided that the stochastic process is Gaussian. [If the process is not Gaussian, then we have to include also terms proportional to the fourth cumulant $\kappa_4(m, r, r-s)$.] In our case the stochastic process is of course not Gaussian, but we may hope that it is "not too far from Gaussian". So let us for simplicity take formula (B.1) as correct.

The qualitative import of (B.1) can be understood by examining the special case of a pure exponential decay $\rho(t) = e^{-|t|/\tau}$. In this case (B.1) reduces to

$$
\text{cov}[\widehat{\rho}(r), \widehat{\rho}(s)] = \frac{1}{N}\left[e^{-|r-s|/\tau}\left(|r-s| + \frac{1+e^{-2/\tau}}{1-e^{-2/\tau}}\right) - \right.
$$

$$e^{-(r+s)/\tau}\left(r+s+\frac{1+e^{-2/\tau}}{1-e^{-2/\tau}}\right)\Bigg] + O\left(\frac{1}{N^2}\right) \quad (B.2)$$

We thus see that the off-diagonal terms in $\widehat{C}_\rho$ (i.e., $r \neq s$) are comparable in magnitude to the diagonal ones ($r = s$). In other words, the sample autocorrelations $\widehat{\rho}(t)$ for different time lags $t$ are *strongly positively correlated*, and any valid analysis method must take proper account of this correlation.[33]

Our Ansatz for the autocorrelation function $\rho(t)$ will be the following:

$$\rho(t) \;=\; \begin{cases} \widehat{\rho}(t) & \text{for } 0 \leq t < t_{min} \\ Ae^{-t/\tau_{\rm exp}} & \text{for } t \geq t_{min} \end{cases} \quad (B.3a)$$

$$\rho(t) \;=\; \rho(-t) \quad (B.3b)$$

where $\widehat{\rho}(t)$ is the autocorrelation function at time lag $t$ measured in the Monte Carlo simulation, and $\tau_{\rm exp}$ will be chosen by least-squares fitting (see below); here $t_{min}$ is some chosen cut point, to take account of the fact that the behavior of $\rho(t)$ for *small* $t$ need not be exponential. Now, for each $t_{min}$ we can compute explicitly the sums appearing in (B.1), when $\rho(t)$ is given by (B.3). Indeed, all the terms in (B.1) can be written in terms of $\rho(t)$ and its convolution

$$\mu(s) \equiv \sum_{m=-\infty}^{+\infty} \rho(m)\rho(m-s)\,. \quad (B.4)$$

The sum (B.4) can be split in two pieces: one piece contains only $\rho(t)$'s with $t \geq t_{min}$, and the other piece contains the rest. The first piece can be summed exactly, giving the result

$$\mu(s) = \sum_{m=1}^{s+t_{min}-1} \rho(m)\rho(m-s) + \sum_{m=s}^{s+t_{min}-1} \rho(m)\rho(m-s) + 2A^2\frac{e^{-(2t_{min}+s)/\tau_{\rm exp}}}{1-e^{-2/\tau_{\rm exp}}} \quad (B.5)$$

Thus, given $\{\widehat{\rho}(t)\}$, $t_{min}$, $A$ and $\tau_{\rm exp}$ we can compute the covariance matrix $\widehat{C}_\rho$ given by (B.1). With this matrix, we can perform the standard weighted least-squares fit [56] to $\log\rho(t) = a + bt$ for the interval $t_{min} \leq t \leq t_{max}$ and obtain new estimates for $A \equiv e^a$ and $\tau_{\rm exp} \equiv -1/b$. We iterate this process until we reach a fixed point, for which the input and output values of $A$ and $\tau_{\rm exp}$ are equal. In practice, we initialized this self-consistent process by supposing that $\rho(t) = e^{-|t|/\tau_{\rm exp}^{(0)}}$ with

$$\tau_{\rm exp}^{(0)} \;=\; -\log\frac{2\tau_{\rm int}-1}{2\tau_{\rm int}+1}\;; \quad (B.6)$$

here the value of $\tau_{\rm int}$ is of course our estimate from the Monte Carlo simulation, using the usual self-consistent truncation window of width $6\tau_{\rm int}$ [52, Appendix C]. We have followed this procedure both for the fits with the full covariance matrix and for those with only a diagonal covariance matrix.

[33] A similar problem arises in fitting the *spatial* correlation function to an asymptotic exponential decay in order to extract the mass gap [55].

This procedure was implemented using Mathematica, which allowed us to control accurately the numerical precision of the calculation. This is especially important when inverting the full covariance matrix, as in some cases we obtained nearly-singular matrices. In practice, this method converges well for the smaller lattices (the number of iterations needed is usually $\lesssim 10$). However, when the data are sufficiently poor (run length $\lesssim 60000\tau_{\rm int}$), we noticed some cases of cyclic behavior instead of convergence to a single fixed point. This appears to happen when, due to a statistical fluctuation, the sample autocorrelation function $\widehat{\rho}(t)$ has a sharp bend somewhere in its tail.

# Acknowledgments

We wish to thank Andrea Pelissetto for collaborating in the derivation of the algorithm described in Section 3.1, and to thank Paul Coddington for communicating to us his unpublished data. In addition, we wish to thank Paul Coddington, Hubert Saleur and Doug Toussaint for helpful correspondence.

The computations reported here were carried out principally on the CAPC Cluster at New York University, on the Alpha Cluster at the Pittsburgh Supercomputing Center, and on the SP2 cluster at the Cornell Theory Center.

The authors' research was supported in part by a M.E.C. (Spain)/Fulbright fellowship (J.S.), and by U.S. National Science Foundation grants DMS-9200719 and PHY-9520978 (J.S. and A.D.S.).

# References

[1] A.D. Sokal, *Monte Carlo Methods in Statistical Mechanics: Foundations and New Algorithms*, Cours de Troisième Cycle de la Physique en Suisse Romande (Lausanne, June 1989).

[2] S.L. Adler, Nucl. Phys. B (Proc. Suppl.) **9**, 437 (1989).

[3] U. Wolff, Nucl. Phys. B (Proc. Suppl.) **17**, 93 (1990).

[4] A.D. Sokal, Nucl. Phys. B (Proc. Suppl.) **20**, 55 (1991).

[5] R.H. Swendsen and J.-S. Wang, Phys. Rev. Lett. **58**, 86 (1987).

[6] C.F. Baillie and P.D. Coddington, Phys. Rev. B **43**, 10617 (1991).

[7] C.F. Baillie and P.D. Coddington, Phys. Rev. Lett. **68**, 962 (1992); and private communication.

[8] D.W. Heermann and A.N. Burkitt, Physica A **162**, 210 (1990).

[9] X.–J. Li and A.D. Sokal, Phys. Rev. Lett. **63**, 827 (1989).

[10] W. Klein, T. Ray and P. Tamayo, Phys. Rev. Lett. **63**, 827 (1989).

[11] T. Ray, P. Tamayo and W. Klein, Phys. Rev. A **39**, 5949 (1989).

[12] U. Wolff, Phys. Rev. Lett. **62**, 361 (1989).

[13] P. Tamayo, R.C. Brower and W. Klein, J. Stat. Phys. **58**, 1083 (1990).

[14] S. Alexander, Phys. Lett. A **54**, 353 (1975).

[15] E. Domany and E.K. Riedel, J. Appl. Phys. **49**, 1315 (1978).

[16] M. Nauenberg and D.J. Scalapino, Phys. Rev. Lett. **44**, 837 (1980).

[17] J.L. Cardy, M. Nauenberg and D.J. Scalapino, Phys. Rev. B **22**, 2560 (1980).

[18] J.L. Black and V.J. Emery, Phys. Rev. B **23**, 429 (1981).

[19] U. Wolff, Phys. Lett. **B228**, 379 (1989).

[20] J.–S. Wang, Physica **A164**, 240 (1990).

[21] J.C. Le Guillou and J. Zinn-Justin, J. Phys. France **50**, 1365 (1989).

[22] B.G. Nickel, Physica **A177**, 189 (1991).

[23] J. Ashkin and J. Teller, Phys. Rev. **64**, 178 (1943).

[24] S. Wiseman and E. Domany, Phys. Rev. E **48**, 4080 (1993).

[25] L. Laanait, N. Masaif and J. Ruiz, J. Stat. Phys. **72**, 721 (1993).

[26] R.V. Ditzian, J.R. Banavar, G.S. Grest and L.P. Kadanoff, Phys. Rev. B **22**, 2542 (1980).

[27] R.J. Baxter, *Exactly Solved Models in Statistical Mechanics* (Academic Press, New York, 1982).

[28] H.J.F. Knops, J. Phys. A: Math. Gen. **8**, 1508 (1975).

[29] S.J. Ferreira and A.D. Sokal, Phys. Rev. B **51**, 6727 (1995).

[30] F.Y. Wu and Y.K. Wang, J. Math. Phys. **17**, 439 (1976).

[31] F.Y. Wu, J. Math. Phys. **18**, 611 (1977).

[32] A. van Enter, R. Fernández and A.D. Sokal, unpublished.

[33] R.J. Baxter, Proc. Roy. Soc. Lond. A **383**, 43 (1982).

[34] A. Lenard, cited in E.H. Lieb, Phys. Rev. **162**, 162 (1967) on pp. 169 and 170.

[35] R.J. Baxter, J. Math. Phys. **11**, 3116 (1970).

[36] L.P. Kadanoff and A.C. Brown, Ann. of Physics, **121**, 318 (1979).

[37] C. Pfister, Comm. Math. Phys. **29**, 113 (1982).

[38] J.M. Maillard, P. Rujan and T.T. Truong, J. Phys. A: Math. Gen. **18**, 3399 (1985).

[39] A. Benyoussef, L. Laanait and M. Loulidi, J. Stat. Phys. **74**, 1185 (1994).

[40] A.C.D. van Enter, R. Fernández and A.D. Sokal, J. Stat. Phys. **72**, 879 (1993).

[41] M.P.M. den Nijs, J. Phys. A: Math. Gen. **12**, 1857 (1979).

[42] H.J.F. Knops, Ann. of Physics, **128**, 448 (1980).

[43] H. Saleur, J. Stat. Phys. **50**, 475 (1988).

[44] S.–K. Yang, Nucl. Phys. **B285 [FS19]**, 183 (1987).

[45] A.B. Zamolodchikov and V.A. Fateev, Sov. Phys. JETP **62**, 215 (1985).

[46] H.N.V. Temperley and S. Ashley, Proc. Roy. Soc. London A **265**, 371 (1979).

[47] R.G. Edwards and A.D. Sokal, Phys. Rev. D **38**, 2009 (1988).

[48] R.G. Edwards and A.D. Sokal, Phys. Rev. D **40**, 1374 (1989).

[49] S. Caracciolo, R.G. Edwards, A. Pelissetto and A.D. Sokal, Nucl. Phys. B **403**, 475 (1993).

[50] T.W. Anderson, *The Statistical Analysis of Time Series* (Wiley, New York, 1971).

[51] M.B. Priestley, *Spectral Analysis and Time Series*, 2 Vols (Academic Press, London, 1981).

[52] N. Madras and A.D. Sokal, J. Stat. Phys. **50**, 109 (1988).

[53] A.E. Ferdinand and M.E. Fisher, Phys. Rev. **185**, 832 (1969).

[54] C.J. Hamer, M.T. Batchelor and M.N. Barber, J. Stat. Phys. **52**, 679 (1988).

[55] D. Toussaint, in *From Actions to Answers*, edited by T. DeGrand and D. Toussaint (World Scientific, Singapore, 1990).

[56] S.D. Silvey, *Statistical Inference* (Chapman and Hall, London, 1975), Chapter 3.

| $\sigma\ \sigma'$ | $\tau\ \tau'$ | Energy |
|---|---|---|
| $\uparrow\uparrow$ | $\uparrow\uparrow$ | $E_0 = 0$ |
| $\uparrow\uparrow$ | $\uparrow\downarrow$ | $E_1 = 2(J' + K)$ |
| $\uparrow\downarrow$ | $\uparrow\uparrow$ | $E_2 = 2(J + K)$ |
| $\uparrow\downarrow$ | $\uparrow\downarrow$ | $E_3 = 2(J + J')$ |

Table 1: Energies for a bond joining the spins $(\sigma, \tau)$ and $(\sigma', \tau')$. These energies are invariant under the transformations $\sigma, \sigma' \leftrightarrow -\sigma, -\sigma'$ and $\tau, \tau' \leftrightarrow -\tau, -\tau'$. On each bond we have added a constant energy $J + J' + K$ to (2.1), in order to set $E_0 = 0$.

| Point | $J = J'$ | $K$ | $y$ |
|---|---|---|---|
| 4-state Potts model | $\frac{1}{4}\log 3 \quad \approx 0.274653$ | $\frac{1}{4}\log 3 \quad \approx 0.274653$ | 0 |
| ZF | $\frac{1}{4}\log\frac{\sqrt{2+\sqrt{2}}+1}{\sqrt{2+\sqrt{2}}-1} \approx 0.302923$ | $\frac{1}{4}\log(1+\sqrt{2}) \approx 0.220343$ | $\frac{1}{2}$ |
| X2 | $-\frac{1}{4}\log\left(\frac{5}{3}-\sqrt{2}\right) \approx 0.344132$ | $\frac{1}{4}\log\frac{6(5-3\sqrt{2})}{11-6\sqrt{2}} \approx 0.147920$ | $\approx 0.735579$ |
| Ising model | $\frac{1}{2}\log(1+\sqrt{2}) \quad \approx 0.440687$ | 0 | 1 |

Table 2: Points of the self-dual curve of the symmetric AT model where our MC simulations were performed. The parameter $y$ is defined in (2.9)/(2.11). We also include the values corresponding to the Ising model (DIs); the dynamic data corresponding to this point have been taken from Baillie and Coddington [7].

| $L$ | $\chi$ | $C_H$ (exact) | $\tau_{\text{int},\mathcal{E}}$ |
|---|---|---|---|
| 8 | $41.392 \pm 0.008$ | 1.145559240 | $2.589 \pm 0.005$ |
| 16 | $139.58 \pm 0.04$ | 1.498704959 | $3.258 \pm 0.005$ |
| 32 | $470.12 \pm 0.20$ | 1.846767590 | $4.016 \pm 0.005$ |
| 50 | $1025.9 \pm 0.4$ | 2.069384825 | $4.585 \pm 0.005$ |
| 64 | $1581.4 \pm 0.5$ | 2.192211393 | $4.899 \pm 0.010$ |
| 100 | $3453.7 \pm 1.4$ | 2.413876309 | $5.510 \pm 0.017$ |
| 128 | $5319.2 \pm 2.4$ | 2.536331335 | $5.874 \pm 0.016$ |
| 256 | $17900 \pm 7.0$ | 2.879786255 | $6.928 \pm 0.030$ |
| 512 | $60185 \pm 28.$ | 3.222907954 | $8.144 \pm 0.055$ |

Table 3: Data for the 2D Ising model. The susceptibility $\chi$ and the integrated autocorrelation time $\tau_{\text{int},\mathcal{E}}$ are taken from Ref. [7]. The value of the specific heat $C_H$ is obtained from the exact formula of Ferdinand and Fisher [53].

| $L$ | MCS | $\chi$ | $C_H$ | $\xi$ | $\tau_{\text{int},\mathcal{E}}$ | $\tau_{\text{int},\mathcal{M}^2}$ |
|---|---|---|---|---|---|---|
| 16 | 0.9 | $141.41 \pm 0.29$ | $5.027 \pm 0.027$ | $15.758 \pm 0.056$ | $12.86 \pm 0.24$ | $12.77 \pm 0.24$ |
| 32 | 1.9 | $474.23 \pm 0.94$ | $8.341 \pm 0.040$ | $31.681 \pm 0.100$ | $23.13 \pm 0.40$ | $22.75 \pm 0.39$ |
| 64 | 4.4 | $1589.67 \pm 2.84$ | $13.937 \pm 0.060$ | $63.827 \pm 0.172$ | $41.42 \pm 0.62$ | $40.34 \pm 0.60$ |
| 128 | 2.9 | $5316.35 \pm 16.49$ | $23.576 \pm 0.170$ | $127.974 \pm 0.575$ | $78.79 \pm 2.01$ | $76.19 \pm 1.92$ |
| 256 | 2.9 | $17771.34 \pm 74.91$ | $39.876 \pm 0.388$ | $256.175 \pm 1.523$ | $142.54 \pm 4.90$ | $136.57 \pm 4.59$ |
| 512 | 2.9 | $59876.54 \pm 334.67$ | $67.938 \pm 0.934$ | $519.102 \pm 4.078$ | $252.83 \pm 11.57$ | $241.40 \pm 10.79$ |
| 1024 | 0.8 | $196872.43 \pm 3137.60$ | $120.270 \pm 4.183$ | $1020.767 \pm 21.680$ | $534.44 \pm 67.68$ | $505.27 \pm 62.22$ |

Table 4: Results of the MC simulations at the critical point of the 4-state Potts model. For each lattice size ($L$) we include the number of performed measurements (MCS) in units of $10^6$, the susceptibility ($\chi$), the specific heat ($C_H$), the second-moment correlation length ($\xi$), and the integrated autocorrelation times for the energy ($\tau_{\text{int},\mathcal{E}}$) and the susceptibility ($\tau_{\text{int},\mathcal{M}^2}$). The quoted errors correspond to one standard deviation (i.e., confidence level $\approx 68\%$).

| $L$ | $MCS$ | $\chi_\omega$ | $\chi_{\sigma\tau}$ | $\xi_\omega$ | $\xi_{\sigma\tau}$ | $C_{H,\text{max}}$ | $C_{H,\text{min}}$ |
|---|---|---|---|---|---|---|---|
| 16 | 0.9 | $146.53 \pm 0.24$ | $123.83 \pm 0.25$ | $16.252 \pm 0.049$ | $13.751 \pm 0.039$ | $9.902 \pm 0.043$ | $0.4366 \pm 0.0014$ |
| 32 | 0.9 | $494.14 \pm 1.10$ | $395.77 \pm 1.08$ | $32.374 \pm 0.119$ | $27.289 \pm 0.094$ | $15.922 \pm 0.089$ | $0.4549 \pm 0.0025$ |
| 64 | 0.9 | $1668.42 \pm 4.80$ | $1264.13 \pm 4.47$ | $64.818 \pm 0.296$ | $54.410 \pm 0.229$ | $24.949 \pm 0.181$ | $0.4645 \pm 0.0044$ |
| 128 | 0.9 | $5665.88 \pm 21.00$ | $4067.21 \pm 18.60$ | $131.094 \pm 0.749$ | $109.699 \pm 0.577$ | $39.068 \pm 0.362$ | $0.4700 \pm 0.0050$ |
| 256 | 0.9 | $18925.33 \pm 94.12$ | $12814.29 \pm 78.80$ | $259.066 \pm 1.930$ | $216.909 \pm 1.488$ | $62.652 \pm 0.749$ | $0.4743 \pm 0.0086$ |
| 512 | 1.9 | $64118.12 \pm 274.65$ | $41094.67 \pm 216.67$ | $521.901 \pm 3.349$ | $436.890 \pm 2.548$ | $98.966 \pm 1.034$ | $0.4769 \pm 0.0085$ |

Table 5: Static data from the runs for the point ZF. For each lattice size $L$, we report the number of measurements (MCS) in units of $10^6$, the susceptibilities ($\chi_\omega$ and $\chi_{\sigma\tau}$), the second-moment correlation lengths ($\xi_\omega$ and $\xi_{\sigma\tau}$) and the maximum ($C_{H,\text{max}}$) and the minimum ($C_{H,\text{min}}$) eigenvalues of the specific-heat matrix $\widehat{C}_H$.

| $L$ | MCS | $\tau_{\text{int},\mathcal{E}_\omega}$ | $\tau_{\text{int},\mathcal{E}_{\sigma\tau}}$ | $\tau_{\text{int},\mathcal{M}^2_\omega}$ | $\tau_{\text{int},\mathcal{M}^2_{\sigma\tau}}$ |
|---|---|---|---|---|---|
| 16 | 0.9 | $9.43 \pm 0.15$ | $8.60 \pm 0.13$ | $9.23 \pm 0.15$ | $8.28 \pm 0.12$ |
| 32 | 0.9 | $16.00 \pm 0.33$ | $15.06 \pm 0.30$ | $15.39 \pm 0.31$ | $14.08 \pm 0.27$ |
| 64 | 0.9 | $26.40 \pm 0.70$ | $25.32 \pm 0.66$ | $24.98 \pm 0.65$ | $22.92 \pm 0.57$ |
| 128 | 0.9 | $44.97 \pm 1.56$ | $43.74 \pm 1.50$ | $41.25 \pm 1.37$ | $38.18 \pm 1.22$ |
| 256 | 0.9 | $76.35 \pm 3.45$ | $75.19 \pm 3.37$ | $70.28 \pm 3.05$ | $65.74 \pm 2.76$ |
| 512 | 1.9 | $119.02 \pm 4.62$ | $117.83 \pm 4.55$ | $110.03 \pm 4.11$ | $102.53 \pm 3.69$ |

Table 6: Autocorrelation times for the runs performed at the point ZF. For each lattice size $L$, we show the number of measurements (MCS) in units of $10^6$, the integrated autocorrelation times for the energies ($\tau_{\text{int},\mathcal{E}_\omega}$ and $\tau_{\text{int},\mathcal{E}_{\sigma\tau}}$) and the integrated autocorrelation times for the susceptibilities ($\tau_{\text{int},\mathcal{M}^2_\omega}$ and $\tau_{\text{int},\mathcal{M}^2_{\sigma\tau}}$).

| $L$ | $MCS$ | $\chi_\omega$ | $\chi_{\sigma\tau}$ | $\xi_\omega$ | $\xi_{\sigma\tau}$ | $C_{H,\max}$ | $C_{H,\min}$ |
|---|---|---|---|---|---|---|---|
| 16 | 0.9 | $145.93 \pm 0.18$ | $104.31 \pm 0.19$ | $15.815 \pm 0.036$ | $11.958 \pm 0.027$ | $8.793 \pm 0.030$ | $0.27326 \pm 0.00088$ |
| 32 | 0.9 | $493.30 \pm 0.74$ | $319.01 \pm 0.72$ | $31.482 \pm 0.081$ | $23.779 \pm 0.059$ | $12.602 \pm 0.052$ | $0.28820 \pm 0.00090$ |
| 64 | 0.9 | $1660.56 \pm 3.04$ | $970.81 \pm 2.63$ | $62.797 \pm 0.187$ | $47.488 \pm 0.134$ | $17.783 \pm 0.090$ | $0.29708 \pm 0.00140$ |
| 128 | 0.9 | $5595.77 \pm 12.41$ | $2958.76 \pm 9.63$ | $125.552 \pm 0.437$ | $94.974 \pm 0.312$ | $24.808 \pm 0.152$ | $0.30122 \pm 0.00250$ |
| 256 | 0.9 | $18772.61 \pm 48.29$ | $8951.11 \pm 33.94$ | $249.624 \pm 0.990$ | $188.436 \pm 0.696$ | $33.765 \pm 0.248$ | $0.30392 \pm 0.00316$ |
| 512 | 0.9 | $63352.87 \pm 189.64$ | $27341.35 \pm 120.27$ | $501.468 \pm 2.282$ | $378.482 \pm 1.609$ | $45.485 \pm 0.399$ | $0.30652 \pm 0.00462$ |

Table 7: The same as in Table 5 for the point X2.

| $L$ | MCS | $\tau_{\text{int},\mathcal{E}_\omega}$ | $\tau_{\text{int},\mathcal{E}_{\sigma\tau}}$ | $\tau_{\text{int},\mathcal{M}^2_\omega}$ | $\tau_{\text{int},\mathcal{M}^2_{\sigma\tau}}$ |
|---|---|---|---|---|---|
| 16 | 0.9 | $6.405 \pm 0.085$ | $5.965 \pm 0.076$ | $6.172 \pm 0.081$ | $5.606 \pm 0.069$ |
| 32 | 0.9 | $9.326 \pm 0.148$ | $8.815 \pm 0.136$ | $8.706 \pm 0.134$ | $7.978 \pm 0.117$ |
| 64 | 0.9 | $13.686 \pm 0.264$ | $13.169 \pm 0.249$ | $12.372 \pm 0.227$ | $11.370 \pm 0.200$ |
| 128 | 0.9 | $20.338 \pm 0.476$ | $19.719 \pm 0.454$ | $17.798 \pm 0.389$ | $16.232 \pm 0.340$ |
| 256 | 0.9 | $27.775 \pm 0.758$ | $27.196 \pm 0.735$ | $23.777 \pm 0.601$ | $21.839 \pm 0.530$ |
| 512 | 0.9 | $39.559 \pm 1.288$ | $38.928 \pm 1.257$ | $32.474 \pm 0.957$ | $29.712 \pm 0.839$ |

Table 8: The same as in Table 6 for the point X2.

| Ratio | 4-state Potts model | | ZF model | | X2 model | | Ising model | |
|---|---|---|---|---|---|---|---|---|
| | numerical | exact | numerical | exact | numerical | exact | numerical | exact |
| $\gamma/\nu$ | $1.744 \pm 0.001$ | 7/4 | $1.750 \pm 0.004$ | 7/4 | $1.751 \pm 0.001$ | 7/4 | $1.7501 \pm 0.0002$ | 7/4 |
| $\gamma'/\nu$ | $1.744 \pm 0.001$ | 7/4 | $1.668 \pm 0.005$ | 5/3 | $1.605 \pm 0.001$ | 1.6045 | | 1/2 |
| $\alpha/\nu$ | $0.768 \pm 0.009$ | $1 \times \log^{-3/2}$ | $0.663 \pm 0.006$ | 2/3 | $0.438 \pm 0.008$ | 0.4183 | | log |
| $z_{\text{int},\mathcal{E}}$ | $0.876 \pm 0.012$ | $\geq 1 \times \log^{-3/2}$ | $0.740 \pm 0.010$ | $\geq 2/3$ | $0.477 \pm 0.028$ | $\geq 0.4183$ | $0.240 \pm 0.004$ | $\geq$ log |

Table 9: Ratios of static critical exponents and dynamic critical exponent ($z_{\text{int},\mathcal{E}}$) coming from the power-law fits of the results contained in Tables 4–8. For the Ising model we include the fits to the dynamical data reported in Ref. [7]. For each model we present two columns, one with the MC results (the left one) and the other with the exact known results (the right one). The errors represent one standard deviation (i.e., confidence level of 68%). The symbol "$1 \times \log^{-3/2}$" means that the leading term of the specific heat for the 4-state Potts model behaves like $L \log^{-3/2} L$. Likewise, the symbol "log" means that the leading term of the specific heat for the Ising model is $\log L$.

| $L_{min}$ | $C_H \sim L^{\alpha/\nu}$ $\alpha/\nu$ | $\chi^2$ | $\tau_{\text{int},\mathcal{E}} \sim L^{z_{\text{int},\mathcal{E}}}$ $z_{\text{int},\mathcal{E}}$ | $\chi^2$ |
|---|---|---|---|---|
| 16 | $0.749 \pm 0.003$ | 13.82 (DF= 5, level= 2%) | $0.867 \pm 0.009$ | 4.32 (DF= 5, level= 50%) |
| 32 | $0.756 \pm 0.004$ | 5.66 (DF= 4, level= 23%) | $\mathbf{0.876 \pm 0.012}$ | **3.16 (DF= 4, level= 53%)** |
| 64 | $0.762 \pm 0.005$ | 1.84 (DF= 3, level= 61%) | $0.887 \pm 0.017$ | 2.26 (DF= 3, level= 52%) |
| 128 | $\mathbf{0.768 \pm 0.009}$ | **1.40 (DF= 2, level= 50%)** | $0.861 \pm 0.033$ | 1.31 (DF= 2, level= 52%) |
| 256 | $0.781 \pm 0.019$ | 0.71 (DF= 1, level= 40%) | $0.880 \pm 0.067$ | 1.20 (DF= 1, level= 27%) |
| 512 | $0.824 \pm 0.054$ | 0.00 (DF= 0, level= 100%) | $1.080 \pm 0.194$ | 0.00 (DF= 0, level= 100%) |

Table 10: Estimates of $\alpha/\nu$ and $z_{\text{int},\mathcal{E}}$ for the 4-state Potts model at criticality as a function of the points involved in the fit ($L \geq L_{min}$). Errors represent one standard deviation, DF stands for the number of degrees of freedom and "level" is the confidence level of the fit (i.e., the probability that $\chi^2$ would equal or exceed the observed value, assuming that the underlying statistical model is correct). The preferred fits are marked with boldface.

| $L_{min}$ | $C_H \sim L^{\alpha/\nu} \log^{-3/2} L$ $\alpha/\nu$ | $\chi^2$ | $C_H \sim L \log^{-p} L$ $p$ | $\chi^2$ |
|---|---|---|---|---|
| 16 | $1.118 \pm 0.003$ | 236.01 (DF= 5, level= $5 \times 10^{-49}$) | $1.008 \pm 0.011$ | 86.06 (DF= 5, level= $5 \times 10^{-17}$) |
| 32 | $1.086 \pm 0.004$ | 54.23 (DF= 4, level= $5 \times 10^{-11}$) | $1.102 \pm 0.016$ | 24.02 (DF= 4, level= 0.008%) |
| 64 | $1.062 \pm 0.005$ | 9.47 (DF= 3, level= 2%) | $1.185 \pm 0.025$ | 5.33 (DF= 3, level= 15%) |
| 128 | $\mathbf{1.039 \pm 0.009}$ | **0.45 (DF= 2, level= 80%)** | $\mathbf{1.286 \pm 0.052}$ | **0.34 (DF= 2, level= 84%)** |
| 256 | $1.030 \pm 0.019$ | 0.19 (DF= 1, level= 66%) | $1.320 \pm 0.115$ | 0.23 (DF= 1, level= 63%) |
| 512 | $1.052 \pm 0.054$ | 0.00 (DF= 0, level= 100%) | $1.158 \pm 0.355$ | 0.00 (DF= 0, level= 100%) |

Table 11: Results of the weighted least-squares fits for the specific heat of the 4-state Potts model at criticality to a function $C_H = AL^{\alpha/\nu} \log^{-3/2} L$ (first column) and to a function $C_H = A'L \log^{-p} L$ (second column). The exact results are $\alpha/\nu = 1$ and $p = \frac{3}{2}$ respectively [18,16,17]. For each fit we show the $\chi^2$, the number of degrees of freedom (DF) and the confidence level ("level").

| $L$ | ZF | X2 |
|---|---|---|
| 16 | $-0.86034 \pm 0.00065$ | $-0.77163 \pm 0.00052$ |
| 32 | $-0.88553 \pm 0.00059$ | $-0.79294 \pm 0.00042$ |
| 64 | $-0.89995 \pm 0.00052$ | $-0.80885 \pm 0.00037$ |
| 128 | $-0.90944 \pm 0.00039$ | $-0.81835 \pm 0.00034$ |
| 256 | $-0.91529 \pm 0.00035$ | $-0.82414 \pm 0.00031$ |
| 512 | $-0.91858 \pm 0.00021$ | $-0.82828 \pm 0.00028$ |
| $\infty$ | $-0.92388$ (exact) | $-0.83771$ (exact) |

Table 12: Eigenvectors of the specific-heat matrix $\widehat{C}_H$. They are parametrized such that $\vec{w}_{\text{min}} = (1, a)$ corresponds to the smaller eigenvalue $C_{H,\text{min}}$ and $\vec{w}_{\text{max}} = (a, -1)$ to the largest eigenvalue $C_{H,\text{max}}$. For the models ZF and X2 we give the measured values of the parameter $a$ as a function of the lattice size $L$. The bottom row ($L = \infty$) shows the theoretically predicted infinite-volume value of $a$ taken from the (4.26) [see text].

| $L_{min}$ | $C_{H,max} \sim L^{\alpha/\nu}$ | | $\tau_{\mathrm{int},\mathcal{E}_\omega} \sim L^{z_{\mathrm{int},\mathcal{E}_\omega}}$ | |
|---|---|---|---|---|
| | $\alpha/\nu$ | $\chi^2$ | $z_{\mathrm{int},\mathcal{E}_\omega}$ | $\chi^2$ |
| 16 | $0.484 \pm 0.002$ | 76.78 (DF= 4, level= $8 \times 10^{-16}$) | $0.534 \pm 0.007$ | 5.16 (DF= 4, level= 27%) |
| 32 | $0.469 \pm 0.003$ | 26.22 (DF= 3, level= $9 \times 10^{-6}$%) | $0.527 \pm 0.011$ | 4.29 (DF= 3, level= 23%) |
| 64 | $0.455 \pm 0.004$ | 7.50 (DF= 2, level= 2%) | $0.509 \pm 0.016$ | 2.38 (DF= 2, level= 30%) |
| 128 | $\mathbf{0.438 \pm 0.008}$ | **0.32 (DF= 1, level= 57%)** | $\mathbf{0.477 \pm 0.028}$ | **0.39 (DF= 1, level= 53%)** |
| 256 | $0.430 \pm 0.017$ | 0.00 (DF= 0, level= 100%) | $0.510 \pm 0.061$ | 0.00 (DF= 0, level= 100%) |

Table 13: Estimates of $\alpha/\nu$ and $z_{\mathrm{int},\mathcal{E}_\omega}$ for the point X2 as a function of the points involved in the fit ($L \geq L_{min}$). We also show the $\chi^2$, the number of degrees of freedom (DF) and the confidence level ("level") of each fit.

| Point | Type | $L_{min}$ | $x^\star$ $(= x^\star_\omega = x^\star_{\sigma\tau})$ | $\chi^2$ |
|---|---|---|---|---|
| Potts | C | 128 | $1.002 \pm 0.003$ | 2.56 (DF= 3, level= 46%) |
| Potts | L | 16 | $1.023 \pm 0.007$ | 1.94 (DF= 5, level= 86%) |

| Point | Type | $L_{min}$ | $x^\star_\omega$ | $\chi^2$ |
|---|---|---|---|---|
| ZF | C | 16 | $1.015 \pm 0.002$ | 4.12 (DF= 5, level= 53%) |
| X2 | C | 64 | $0.980 \pm 0.002$ | 1.80 (DF= 3, level= 61%) |
| X2 | L | 16 | $0.965 \pm 0.006$ | 1.15 (DF= 4, level= 89%) |
| X2 | $\Delta = \frac{1}{2}$ | 16 | $0.975 \pm 0.003$ | 1.08 (DF= 4, level= 80%) |

| Point | Type | $L_{min}$ | $x^\star_{\sigma\tau}$ | $\chi^2$ |
|---|---|---|---|---|
| ZF | C | 16 | $0.852 \pm 0.002$ | 2.26 (DF= 4, level= 69%) |
| X2 | C | 256 | $0.737 \pm 0.002$ | 0.57 (DF= 1, level= 45%) |
| X2 | L | 16 | $0.729 \pm 0.004$ | 2.08 (DF= 4, level= 72%) |
| X2 | $\Delta = \frac{1}{2}$ | 16 | $0.736 \pm 0.002$ | 2.09 (DF= 4, level= 72%) |

Table 14: Estimates of $x^\star$ for the three points of the AT self-dual curve considered in this paper. For each point we present the result of the least-square fit to a constant (C) or to a function of the type (4.44) (L). For the X2 model we also include the fit to constant plus corrections of order $\Delta = \frac{1}{2}$. We also include the $\chi^2$, the number of degrees of freedom (DF) and the confidence level ("level").

| $L_{min}$ | $x(L) = x^\star$ | | $x(L) = x^\star + A/\log 2L$ | |
|---|---|---|---|---|
| | $x^\star$ | $\chi^2$ | $x^\star$ | $\chi^2$ |
| 16 | $0.9943 \pm 0.0015$ | 19.00 (DF= 6, level= 0.4%) | $\mathbf{1.0228 \pm 0.0071}$ | **1.95 (DF= 5, level= 86%)** |
| 32 | $0.9965 \pm 0.0017$ | 10.19 (DF= 5, level= 7%) | $1.0274 \pm 0.0107$ | 1.62 (DF= 4, level= 80%) |
| 64 | $0.9993 \pm 0.0021$ | 3.98 (DF= 4, level= 41%) | $1.0247 \pm 0.0165$ | 1.58 (DF= 3, level= 66%) |
| 128 | $\mathbf{1.0023 \pm 0.0032}$ | **2.56 (DF= 3, level= 46%)** | $1.0380 \pm 0.0326$ | 1.35 (DF= 2, level= 51%) |
| 256 | $1.0050 \pm 0.0046$ | 1.92 (DF= 2, level= 38%) | $1.0683 \pm 0.0713$ | 1.12 (DF= 1, level= 29%) |
| 512 | $1.0118 \pm 0.0075$ | 0.57 (DF= 1, level= 45%) | $0.8436 \pm 0.2235$ | 0.00 (DF= 0, level= 100%) |
| 1024 | $0.9968 \pm 0.0211$ | 0.00 (DF= 0, level= 100%) | | |

| $L_{min}$ | $x(L) = x^\star + A/\sqrt{L}$ | | $x(L) = x^\star + A/L$ | |
|---|---|---|---|---|
| | $x^\star$ | $\chi^2$ | $x^\star$ | $\chi^2$ |
| 16 | $\mathbf{1.0098 \pm 0.0041}$ | **1.94 (DF= 5, level= 85%)** | $1.0023 \pm 0.0026$ | 3.38 (DF= 5, level= 64%) |
| 32 | $1.0118 \pm 0.0055$ | 1.65 (DF= 4, level= 80%) | $\mathbf{1.0051 \pm 0.0035}$ | **1.92 (DF= 4, level= 75%)** |
| 64 | $1.0109 \pm 0.0078$ | 1.62 (DF= 3, level= 65%) | $1.0060 \pm 0.0050$ | 1.86 (DF= 3, level= 60%) |
| 128 | $1.0163 \pm 0.0132$ | 1.36 (DF= 2, level= 51%) | $1.0100 \pm 0.0080$ | 1.44 (DF= 2, level= 49%) |
| 256 | $1.0283 \pm 0.0257$ | 1.07 (DF= 1, level= 30%) | $1.0189 \pm 0.0147$ | 0.93 (DF= 1, level= 33%) |
| 512 | $0.9557 \pm 0.0745$ | 0.00 (DF= 0, level= 100%) | $0.9798 \pm 0.0431$ | 0.00 (DF= 0, level= 100%) |

Table 15: Estimates of $x(L)$ as a function of the number of points included in the fit ($L \geq L_{min}$) for the 4-state Potts model at criticality. We show the fit of $x(L)$ to a pure constant, as well as to a constant plus some decreasing functions. We also include the $\chi^2$, the number of degrees of freedom (DF) and the confidence level ("level") for each fit.

| Ansatz | Ising | X2 | ZF | 4-state Potts |
|---|---|---|---|---|
| $AL^p$ | $p = 0.060 \pm 0.004$<br>$\chi^2 = 1.02$ (2 DF, 60%)<br>$L_{min} = 100$ | $p = 0.051 \pm 0.009$<br>$\chi^2 = 1.28$ (4 DF, 86%)<br>$L_{min} = 16$ | $p = 0.077 \pm 0.012$<br>$\chi^2 = 1.23$ (4 DF, 87%)<br>$L_{min} = 16$ | $p = 0.118 \pm 0.012$<br>$\chi^2 = 1.39$ (5 DF, 93%)<br>$L_{min} = 16$ |
| $A \log^p L$ | $p = 0.315 \pm 0.020$<br>$\chi^2 = 2.59$ (2 DF, 27%)<br>$L_{min} = 100$ | $p = 0.263 \pm 0.061$<br>$\chi^2 = 0.63$ (3 DF, 89%)<br>$L_{min} = 32$ | $p = 0.311 \pm 0.049$<br>$\chi^2 = 1.15$ (4 DF, 89%)<br>$L_{min} = 16$ | $p = 0.543 \pm 0.073$<br>$\chi^2 = 1.92$ (4 DF, 75%)<br>$L_{min} = 32$ |
| $A + B \log L$ | $\chi^2 = 1.43$ (2 DF, 49%)<br>$L_{min} = 100$ | $\chi^2 = 1.43$ (4 DF, 84%)<br>$L_{min} = 16$ | $\chi^2 = 1.03$ (4 DF, 91%)<br>$L_{min} = 16$ | $\chi^2 = 1.98$ (5 DF, 85%)<br>$L_{min} = 16$ |
| $A$ | $\chi^2 = 36.74$ (1 DF, $10^{-9}$)<br>$L_{min} = 256$ | $\chi^2 = 1.46$ (2 DF, 48%)<br>$L_{min} = 128$ | $\chi^2 = 0.77$ (3 DF, 68%)<br>$L_{min} = 128$ | $\chi^2 = 1.56$ (2 DF, 46%)<br>$L_{min} = 256$ |
| $A + \frac{B}{\log L}$ | $\chi^2 = 4.00$ (1 DF, 5%)<br>$L_{min} = 128$ | $\chi^2 = 0.90$ (3 DF, 76%)<br>$L_{min} = 32$ | $\chi^2 = 0.82$ (3 DF, 85%)<br>$L_{min} = 32$ | $\chi^2 = 0.60$ (3 DF, 75%)<br>$L_{min} = 64$ |
| $A + \frac{B}{L^{1/8}}$ | $\chi^2 = 2.57$ (2 DF, 28%)<br>$L_{min} = 100$ | $\chi^2 = 0.63$ (3 DF, 89%)<br>$L_{min} = 32$ | $\chi^2 = 1.01$ (4 DF, 91%)<br>$L_{min} = 16$ | $\chi^2 = 0.77$ (3 DF, 86%)<br>$L_{min} = 64$ |
| $A + \frac{B}{L^{1/4}}$ | $\chi^2 = 4.04$ (2 DF, 13%)<br>$L_{min} = 100$ | $\chi^2 = 0.67$ (3 DF, 88%)<br>$L_{min} = 32$ | $\chi^2 = 1.33$ (4 DF, 86%)<br>$L_{min} = 16$ | $\chi^2 = 0.66$ (3 DF, 88%)<br>$L_{min} = 64$ |
| $A + \frac{B}{\sqrt{L}}$ | $\chi^2 = 5.31$ (1 DF, 2%)<br>$L_{min} = 128$ | $\chi^2 = 0.99$ (3 DF, 80%)<br>$L_{min} = 32$ | $\chi^2 = 0.83$ (3 DF, 84%)<br>$L_{min} = 32$ | $\chi^2 = 0.64$ (3 DF, 89%)<br>$L_{min} = 64$ |
| $A + \frac{B}{L}$ | $\chi^2 = 10.67$ (1 DF, 0.1%)<br>$L_{min} = 128$ | $\chi^2 = 0.64$ (2 DF, 42%)<br>$L_{min} = 64$ | $\chi^2 = 0.18$ (2 DF, 91%)<br>$L_{min} = 64$ | $\chi^2 = 1.11$ (3 DF, 77%)<br>$L_{min} = 64$ |

Table 16: Results of fitting the ratio $\tau_{\text{int},\mathcal{E}}/C_H$ for different Ansätze (power-law, logarithmic, and bounded) for all the models. We only show the "best" fits for each case. For each of them we give the value of the $\chi^2$, the number of degrees of freedom (DF), the confidence level and the $L_{min}$ used. For the first two power-law fits we also give the estimate of that power.

| | $\tau_{\text{int},\mathcal{E}} \sim L \log^{-p} L$ | | $\tau_{\text{int},\mathcal{E}}/C_H \sim \log^p L$ | |
|---|---|---|---|---|
| $L_{min}$ | $p$ | $\chi^2$ | $p$ | $\chi^2$ |
| 16 | $0.538 \pm 0.038$ | 3.35 (DF= 5, level= 65%) | $0.472 \pm 0.049$ | 3.64 (DF= 5, level= 60%) |
| 32 | $0.563 \pm 0.056$ | 3.00 (DF= 4, level= 56%) | $\mathbf{0.543 \pm 0.073}$ | **1.92 (DF= 4, level= 75%)** |
| 64 | $0.558 \pm 0.087$ | 3.00 (DF= 3, level= 39%) | $0.630 \pm 0.112$ | 0.85 (DF= 3, level= 84%) |
| 128 | $\mathbf{0.776 \pm 0.180}$ | **1.09 (DF= 2, level= 58%)** | $0.515 \pm 0.232$ | 0.52 (DF= 2, level= 77%) |
| 256 | $0.736 \pm 0.402$ | 1.08 (DF= 1, level= 30%) | $0.593 \pm 0.518$ | 0.50 (DF= 1, level= 48%) |
| 512 | $-0.525 \pm 1.280$ | 0.00 (DF= 0, level= 100%) | $1.683 \pm 1.632$ | 0.00 (DF= 0, level= 100%) |

Table 17: Results for the 4-state Potts model at criticality of fitting the autocorrelation time $\tau_{\text{int},\mathcal{E}}$ to a function $AL \log^{-p}$ (first column) and of fitting the ratio $C_H/\tau_{\text{int},\mathcal{E}}$ to a function $B \log^{-p} L$ (second column). In both cases, we show the $\chi^2$, the number of degrees of freedom (DF) and the confidence level ("level").

| $t_{min}$ | $L = 16$ $t_{max} = 52$ | $L = 32$ $t_{max} = 92$ | $L = 64$ $t_{max} = 164$ | $L = 128$ $t_{max} = 320$ | $L = 256$ $t_{max} = 560$ | $L = 512$ $t_{max} = 1012$ |
|---|---|---|---|---|---|---|
| 1 | $\tau_{\exp,\mathcal{E}} = 12.32 \pm 0.07$ $\chi^2 = 312.84$, DF=50 level=$10^{-39}$ | $\tau_{\exp,\mathcal{E}} = 20.56 \pm 0.10$ $\chi^2 = 1315.1$, DF=90 level=$10^{-216}$ | $\tau_{\exp,\mathcal{E}} = 35.21 \pm 0.14$ $\chi^2 = 5411.0$, DF=162 level=$6 \times 10^{-1020}$ | $\tau_{\exp,\mathcal{E}} = 65.37 \pm 0.41$ $\chi^2 = 7266.9$, DF=318 level=$2 \times 10^{-1296}$ | $\tau_{\exp,\mathcal{E}} = 125.35 \pm 0.95$ $\chi^2 = 11584$, DF=558 level=$2 \times 10^{-2030}$ | |
| $0.5\tau_{\text{int},\mathcal{E}}$ | $\tau_{\exp,\mathcal{E}} = 13.43 \pm 0.12$ $\chi^2 = 48.84$, DF=45 level=32% | $\tau_{\exp,\mathcal{E}} = 23.62 \pm 0.21$ $\chi^2 = 111.17$, DF=79 level=1% | $\tau_{\exp,\mathcal{E}} = 42.99 \pm 0.32$ $\chi^2 = 148.32$, DF=142 level=34% | NO CONVERGENCE | NO CONVERGENCE | NO CONVERGENCE |
| $\tau_{\text{int},\mathcal{E}}$ | $\tau_{\exp,\mathcal{E}} = 13.69 \pm 0.21$ $\chi^2 = 31.24$, DF=38 level=77% | $\tau_{\exp,\mathcal{E}} = 24.33 \pm 0.34$ $\chi^2 = 82.06$, DF=68 level=12% | $\tau_{\exp,\mathcal{E}} = 45.01 \pm 0.54$ $\chi^2 = 130.94$, DF=122 level=27% | $\tau_{\exp,\mathcal{E}} = 100.1 \pm 2.0$ $\chi^2 = 193.13$, DF=239 level=99% | NO CONVERGENCE | $\tau_{\exp,\mathcal{E}} = 334 \pm 10$ $\chi^2 = 96.08$, DF=758 level=100% |
| $1.5\tau_{\text{int},\mathcal{E}}$ | $\tau_{\exp,\mathcal{E}} = 13.62 \pm 0.32$ $\chi^2 = 25.66$, DF=32 level=77% | $\tau_{\exp,\mathcal{E}} = 25.72 \pm 0.61$ $\chi^2 = 57.99$, DF=56 level=40% | $\tau_{\exp,\mathcal{E}} = 45.99 \pm 0.91$ $\chi^2 = 100.26$, DF=101 level=50% | $\tau_{\exp,\mathcal{E}} = 105.6 \pm 3.6$ $\chi^2 = 113.74$, DF=199 level=100% | $\tau_{\exp,\mathcal{E}} = 185.3 \pm 8.3$ $\chi^2 = 86.47$, DF=349 level=100% | $\tau_{\exp,\mathcal{E}} = 324 \pm 14$ $\chi^2 = 52.06$, DF=631 level=100% |
| $2\tau_{\text{int},\mathcal{E}}$ | $\tau_{\exp,\mathcal{E}} = 13.57 \pm 0.55$ $\chi^2 = 17.98$, DF=25 level=84% | $\tau_{\exp,\mathcal{E}} = 26.7 \pm 1.0$ $\chi^2 = 40.33$, DF=45 level=67% | $\tau_{\exp,\mathcal{E}} = 51.0 \pm 1.5$ $\chi^2 = 91.85$, DF=122 level=98% | $\tau_{\exp,\mathcal{E}} = 112.8 \pm 6.4$ $\chi^2 = 65.00$, DF=159 level=100% | $\tau_{\exp,\mathcal{E}} = 232 \pm 19$ $\chi^2 = 43.11$, DF=279 level=100% | $\tau_{\exp,\mathcal{E}} = 329 \pm 24$ $\chi^2 = 31.42$, DF=505 level=100% |

Table 18: Values of $\tau_{\exp,\mathcal{E}}$ for the 4-state Potts model, using $t_{max} = 4\tau_{\text{int},\mathcal{E}}$ (actually, the nearest integer to $4\tau_{\text{int},\mathcal{E}}$), and various values of $t_{min}$. In each case we present the estimate of $\tau_{\exp,\mathcal{E}}$, the $\chi^2$ value of the fit, the number of degrees of freedom (DF) and the confidence level. For each lattice size $L$ we have used as $t_{min}$ the integer nearest to the value shown in the first column. Those fits where our self-consistent process did not converge are marked with "NO CONVERGENCE". The fit for $L = 512$ and $t_{min} = 1$ has not been performed, as it is very memory-consuming and it is expected to be rather poor.

| $t_{min}$ | $L = 16$ $t_{max} = 39$ | $L = 32$ $t_{max} = 69$ | $L = 64$ $t_{max} = 123$ | $L = 128$ $t_{max} = 240$ | $L = 256$ $t_{max} = 420$ | $L = 512$ $t_{max} = 759$ |
|---|---|---|---|---|---|---|
| 1 | $\tau_{\exp,\mathcal{E}} = 12.31 \pm 0.07$ $\chi^2 = 301.70$, DF=37 level=$3 \times 10^{-43}$ | $\tau_{\exp,\mathcal{E}} = 20.48 \pm 0.10$ $\chi^2 = 1239.0$, DF=67 level=$3 \times 10^{-215}$ | $\tau_{\exp,\mathcal{E}} = 35.08 \pm 0.14$ $\chi^2 = 5269.0$, DF=121 level=$2 \times 10^{-1020}$ | $\tau_{\exp,\mathcal{E}} = 63.24 \pm 0.40$ $\chi^2 = 6390.1$, DF=238 level=$2 \times 10^{-1169}$ | $\tau_{\exp,\mathcal{E}} = 114.48 \pm 0.93$ $\chi^2 = 8987.9$, DF=418 level=$5 \times 10^{-1586}$ | |
| $0.5\tau_{\text{int},\mathcal{E}}$ | $\tau_{\exp,\mathcal{E}} = 13.40 \pm 0.12$ $\chi^2 = 42.82$, DF=32 level=10% | $\tau_{\exp,\mathcal{E}} = 23.44 \pm 0.20$ $\chi^2 = 75.97$, DF=56 level=4% | $\tau_{\exp,\mathcal{E}} = 42.70 \pm 0.32$ $\chi^2 = 107.2$, DF=101 level=32% | $\tau_{\exp,\mathcal{E}} = 89.5 \pm 1.1$ $\chi^2 = 404.50$, DF=199 level=$4 \times 10^{-16}$ | NO CONVERGENCE | NO CONVERGENCE |
| $\tau_{\text{int},\mathcal{E}}$ | $\tau_{\exp,\mathcal{E}} = 13.63 \pm 0.21$ $\chi^2 = 25.55$, DF=25 level=43% | $\tau_{\exp,\mathcal{E}} = 23.95 \pm 0.34$ $\chi^2 = 50.84$, DF=45 level=25% | $\tau_{\exp,\mathcal{E}} = 44.28 \pm 0.54$ $\chi^2 = 85.97$, DF=81 level=33% | $\tau_{\exp,\mathcal{E}} = 94.0 \pm 2.0$ $\chi^2 = 156.5$, DF=159 level=54% | $\tau_{\exp,\mathcal{E}} = 163.9 \pm 4.6$ $\chi^2 = 125.8$, DF=279 level=100% | $\tau_{\exp,\mathcal{E}} = 329 \pm 12$ $\chi^2 = 77.82$, DF=503 level=100% |
| $1.5\tau_{\text{int},\mathcal{E}}$ | $\tau_{\exp,\mathcal{E}} = 13.50 \pm 0.32$ $\chi^2 = 20.41$, DF=19 level=37% | $\tau_{\exp,\mathcal{E}} = 24.88 \pm 0.59$ $\chi^2 = 31.04$, DF=33 level=56% | $\tau_{\exp,\mathcal{E}} = 44.46 \pm 0.88$ $\chi^2 = 58.50$, DF=60 level=53% | $\tau_{\exp,\mathcal{E}} = 97.0 \pm 3.5$ $\chi^2 = 88.59$, DF=119 level=98% | $\tau_{\exp,\mathcal{E}} = 164.1 \pm 7.6$ $\chi^2 = 63.83$, DF=209 level=100% | $\tau_{\exp,\mathcal{E}} = 314 \pm 18$ $\chi^2 = 43.11$, DF=376 level=100% |
| $2\tau_{\text{int},\mathcal{E}}$ | $\tau_{\exp,\mathcal{E}} = 13.30 \pm 0.54$ $\chi^2 = 12.18$, DF=12 level=43% | $\tau_{\exp,\mathcal{E}} = 25.10 \pm 0.94$ $\chi^2 = 16.30$, DF=22 level=80% | $\tau_{\exp,\mathcal{E}} = 45.4 \pm 1.4$ $\chi^2 = 43.52$, DF=40 level=32% | $\tau_{\exp,\mathcal{E}} = 101.8 \pm 6.0$ $\chi^2 = 57.22$, DF=79 level=97% | $\tau_{\exp,\mathcal{E}} = 194 \pm 17$ $\chi^2 = 36.46$, DF=139 level=100% | $\tau_{\exp,\mathcal{E}} = 313 \pm 31$ $\chi^2 = 27.20$, DF=250 level=100% |

Table 19: Values of $\tau_{\exp,\mathcal{E}}$ for the 4-state Potts model, using $t_{max} = 3\tau_{\text{int},\mathcal{E}}$ (more precisely, the nearest integer to that value). Notation as in Table 18.

| $t_{min}$ | $L=16$<br>$t_{max}=26$ | $L=32$<br>$t_{max}=37$ | $L=64$<br>$t_{max}=55$ | $L=128$<br>$t_{max}=81$ | $L=256$<br>$t_{max}=111$ | $L=512$<br>$t_{max}=158$ |
|---|---|---|---|---|---|---|
| 1 | $\tau_{\exp,\mathcal{E}_\omega}=6.34\pm0.03$<br>$\chi^2=47.38$, DF=24<br>level= 0.03% | $\tau_{\exp,\mathcal{E}_\omega}=9.11\pm0.05$<br>$\chi^2=150.26$, DF=35<br>level=$3\times10^{-16}$ | $\tau_{\exp,\mathcal{E}_\omega}=13.29\pm0.08$<br>$\chi^2=265.08$, DF=53<br>level=$6\times10^{-30}$ | $\tau_{\exp,\mathcal{E}_\omega}=18.94\pm0.13$<br>$\chi^2=504.72$, DF=79<br>level=$3\times10^{-63}$ | $\tau_{\exp,\mathcal{E}_\omega}=26.45\pm0.21$<br>$\chi^2=727.28$, DF=109<br>level=$4\times10^{-92}$ | $\tau_{\exp,\mathcal{E}_\omega}=36.86\pm0.33$<br>$\chi^2=958.10$, DF=156<br>level=$2\times10^{-115}$ |
| $0.5\tau_{\mathrm{int},\mathcal{E}_\omega}$ | $\tau_{\exp,\mathcal{E}_\omega}=6.45\pm0.04$<br>$\chi^2=24.46$, DF=22<br>level=32% | $\tau_{\exp,\mathcal{E}_\omega}=9.62\pm0.08$<br>$\chi^2=24.85$, DF=31<br>level=77% | $\tau_{\exp,\mathcal{E}_\omega}=14.06\pm0.13$<br>$\chi^2=80.84$, DF=47<br>level=0.2% | $\tau_{\exp,\mathcal{E}_\omega}=21.03\pm0.24$<br>$\chi^2=95.18$, DF=70<br>level=2% | $\tau_{\exp,\mathcal{E}_\omega}=30.61\pm0.40$<br>$\chi^2=152.19$, DF=96<br>level=$2\times10^{-4}$ | NO<br>CONVERGENCE |
| $\tau_{\mathrm{int},\mathcal{E}_\omega}$ | $\tau_{\exp,\mathcal{E}_\omega}=6.58\pm0.07$<br>$\chi^2=17.79$, DF=19<br>level=54% | $\tau_{\exp,\mathcal{E}_\omega}=9.78\pm0.12$<br>$\chi^2=20.36$, DF=27<br>level=82% | $\tau_{\exp,\mathcal{E}_\omega}=14.45\pm0.23$<br>$\chi^2=69.09$, DF=40<br>level=0.3% | $\tau_{\exp,\mathcal{E}_\omega}=21.74\pm0.40$<br>$\chi^2=78.17$, DF=60<br>level=6% | $\tau_{\exp,\mathcal{E}_\omega}=33.45\pm0.70$<br>$\chi^2=194.72$, DF=72<br>level=$3\times10^{-13}$ | $\tau_{\exp,\mathcal{E}_\omega}=47.8\pm1.2$<br>$\chi^2=104.12$, DF=117<br>level=90% |
| $1.5\tau_{\mathrm{int},\mathcal{E}_\omega}$ | $\tau_{\exp,\mathcal{E}_\omega}=6.68\pm0.13$<br>$\chi^2=15.67$, DF=15<br>level=40% | $\tau_{\exp,\mathcal{E}_\omega}=9.96\pm0.21$<br>$\chi^2=14.73$, DF=22<br>level=87% | $\tau_{\exp,\mathcal{E}_\omega}=15.23\pm0.41$<br>$\chi^2=59.35$, DF=33<br>level=0.3% | $\tau_{\exp,\mathcal{E}_\omega}=23.14\pm0.73$<br>$\chi^2=58.74$, DF=49<br>level=16% | $\tau_{\exp,\mathcal{E}_\omega}=34.2\pm1.2$<br>$\chi^2=78.72$, DF=68<br>level=18% | $\tau_{\exp,\mathcal{E}_\omega}=48.0\pm1.9$<br>$\chi^2=74.83$, DF=98<br>level=96% |
| $2\tau_{\mathrm{int},\mathcal{E}_\omega}$ | $\tau_{\exp,\mathcal{E}_\omega}=6.64\pm0.20$<br>$\chi^2=9.41$, DF=12<br>level=67% | $\tau_{\exp,\mathcal{E}_\omega}=9.77\pm0.34$<br>$\chi^2=16.30$, DF=22<br>level=80% | $\tau_{\exp,\mathcal{E}_\omega}=16.16\pm0.67$<br>$\chi^2=44.94$, DF=27<br>level=2% | $\tau_{\exp,\mathcal{E}_\omega}=26.8\pm1.4$<br>$\chi^2=37.44$, DF=39<br>level=54% | $\tau_{\exp,\mathcal{E}_\omega}=38.3\pm2.3$<br>$\chi^2=44.37$, DF=54<br>level=82% | $\tau_{\exp,\mathcal{E}_\omega}=48.1\pm3.3$<br>$\chi^2=43.41$, DF=78<br>level=100% |

Table 20: Values of $\tau_{\exp,\mathcal{E}_\omega}$ for the X2 model, using $t_{max}=4\tau_{\mathrm{int},\mathcal{E}_\omega}$. Notation as in Table 18.

| $t_{min}$ | $L=16$<br>$t_{max}=19$ | $L=32$<br>$t_{max}=28$ | $L=64$<br>$t_{max}=41$ | $L=128$<br>$t_{max}=61$ | $L=256$<br>$t_{max}=83$ | $L=512$<br>$t_{max}=119$ |
|---|---|---|---|---|---|---|
| 1 | $\tau_{\exp,\mathcal{E}_\omega}=6.34\pm0.03$<br>$\chi^2=38.71$, DF=17<br>level= 0.2% | $\tau_{\exp,\mathcal{E}_\omega}=9.10\pm0.05$<br>$\chi^2=141.89$, DF=26<br>level=$6\times10^{-18}$ | $\tau_{\exp,\mathcal{E}_\omega}=13.23\pm0.08$<br>$\chi^2=215.92$, DF=39<br>level=$2\times10^{-26}$ | $\tau_{\exp,\mathcal{E}_\omega}=18.72\pm0.13$<br>$\chi^2=402.09$, DF=59<br>level=$2\times10^{-52}$ | $\tau_{\exp,\mathcal{E}_\omega}=26.09\pm0.21$<br>$\chi^2=624.09$, DF=81<br>level=$9\times10^{-85}$ | $\tau_{\exp,\mathcal{E}_\omega}=36.09\pm0.33$<br>$\chi^2=826.07$, DF=117<br>level=$4\times10^{-107}$ |
| $0.5\tau_{\mathrm{int},\mathcal{E}_\omega}$ | $\tau_{\exp,\mathcal{E}_\omega}=6.45\pm0.04$<br>$\chi^2=16.62$, DF=15<br>level=34% | $\tau_{\exp,\mathcal{E}_\omega}=9.61\pm0.08$<br>$\chi^2=18.96$, DF=22<br>level=65% | $\tau_{\exp,\mathcal{E}_\omega}=13.94\pm0.13$<br>$\chi^2=44.77$, DF=33<br>level=8% | $\tau_{\exp,\mathcal{E}_\omega}=20.60\pm0.24$<br>$\chi^2=40.85$, DF=50<br>level=82% | $\tau_{\exp,\mathcal{E}_\omega}=29.68\pm0.40$<br>$\chi^2=98.71$, DF=68<br>level=0.9% | $\tau_{\exp,\mathcal{E}_\omega}=41.98\pm0.67$<br>$\chi^2=86.30$, DF=98<br>level=79% |
| $\tau_{\mathrm{int},\mathcal{E}_\omega}$ | $\tau_{\exp,\mathcal{E}_\omega}=6.57\pm0.07$<br>$\chi^2=10.77$, DF=12<br>level=55% | $\tau_{\exp,\mathcal{E}_\omega}=9.75\pm0.12$<br>$\chi^2=14.96$, DF=18<br>level=66% | $\tau_{\exp,\mathcal{E}_\omega}=14.17\pm0.23$<br>$\chi^2=37.19$, DF=26<br>level=7% | $\tau_{\exp,\mathcal{E}_\omega}=20.80\pm0.39$<br>$\chi^2=28.85$, DF=40<br>level=90% | $\tau_{\exp,\mathcal{E}_\omega}=29.82\pm0.65$<br>$\chi^2=79.36$, DF=54<br>level=1% | $\tau_{\exp,\mathcal{E}_\omega}=43.7\pm1.2$<br>$\chi^2=60.62$, DF=78<br>level=93% |
| $1.5\tau_{\mathrm{int},\mathcal{E}_\omega}$ | $\tau_{\exp,\mathcal{E}_\omega}=6.64\pm0.13$<br>$\chi^2=8.90$, DF=8<br>level=35% | $\tau_{\exp,\mathcal{E}_\omega}=9.89\pm0.21$<br>$\chi^2=9.82$, DF=13<br>level=71% | $\tau_{\exp,\mathcal{E}_\omega}=14.59\pm0.39$<br>$\chi^2=28.89$, DF=19<br>level=7% | $\tau_{\exp,\mathcal{E}_\omega}=21.09\pm0.68$<br>$\chi^2=19.59$, DF=29<br>level=91% | $\tau_{\exp,\mathcal{E}_\omega}=31.1\pm1.1$<br>$\chi^2=50.02$, DF=40<br>level=13% | $\tau_{\exp,\mathcal{E}_\omega}=42.3\pm1.8$<br>$\chi^2=46.00$, DF=59<br>level=89% |
| $2\tau_{\mathrm{int},\mathcal{E}_\omega}$ | $\tau_{\exp,\mathcal{E}_\omega}=6.56\pm0.20$<br>$\chi^2=2.58$, DF=5<br>level=76% | $\tau_{\exp,\mathcal{E}_\omega}=9.63\pm0.34$<br>$\chi^2=8.18$, DF=8<br>level=42% | $\tau_{\exp,\mathcal{E}_\omega}=15.08\pm0.62$<br>$\chi^2=20.69$, DF=13<br>level=8% | $\tau_{\exp,\mathcal{E}_\omega}=22.9\pm1.2$<br>$\chi^2=10.32$, DF=19<br>level=94% | $\tau_{\exp,\mathcal{E}_\omega}=33.5\pm2.1$<br>$\chi^2=27.71$, DF=26<br>level=37% | $\tau_{\exp,\mathcal{E}_\omega}=39.9\pm2.8$<br>$\chi^2=30.36$, DF=39<br>level=84% |

Table 21: Values of $\tau_{\exp,\mathcal{E}_\omega}$ for the X2 model, using $t_{max}=3\tau_{\mathrm{int},\mathcal{E}_\omega}$. Notation as in Table 18.

| $t_{min}$ | $L=16$<br>$t_{max}=38$ | $L=32$<br>$t_{max}=64$ | $L=64$<br>$t_{max}=106$ | $L=128$<br>$t_{max}=180$ | $L=256$<br>$t_{max}=306$ | $L=512$<br>$t_{max}=476$ |
|---|---|---|---|---|---|---|
| 1 | $\tau_{\exp,\mathcal{E}_\omega}=9.30\pm0.05$<br>$\chi^2=116.93$, DF=36<br>level=$2\times10^{-6}$ | $\tau_{\exp,\mathcal{E}_\omega}=15.15\pm0.10$<br>$\chi^2=358.79$, DF=62<br>level=$2\times10^{-23}$ | $\tau_{\exp,\mathcal{E}_\omega}=24.39\pm0.19$<br>$\chi^2=667.10$, DF=104<br>level=$5\times10^{-42}$ | $\tau_{\exp,\mathcal{E}_\omega}=40.28\pm0.37$<br>$\chi^2=1257.9$, DF=178<br>level=$8\times10^{-82}$ | $\tau_{\exp,\mathcal{E}_\omega}=72.64\pm0.78$<br>$\chi^2=2192.6$, DF=304<br>level=$10^{-282}$ | $\tau_{\exp,\mathcal{E}_\omega}=113.5\pm1.1$<br>$\chi^2=4274.1$, DF=474<br>level=$5\times10^{-602}$ |
| $0.5\tau_{\mathrm{int},\mathcal{E}_\omega}$ | $\tau_{\exp,\mathcal{E}_\omega}=9.75\pm0.08$<br>$\chi^2=21.77$, DF=32<br>level=91% | $\tau_{\exp,\mathcal{E}_\omega}=16.64\pm0.17$<br>$\chi^2=58.02$, DF=55<br>level=36% | $\tau_{\exp,\mathcal{E}_\omega}=28.08\pm0.36$<br>$\chi^2=124.13$, DF=92<br>level=1% | $\tau_{\exp,\mathcal{E}_\omega}=52.25\pm0.83$<br>$\chi^2=368.20$, DF=156<br>level=$3\times10^{-19}$ | NO<br>CONVERGENCE | NO<br>CONVERGENCE |
| $\tau_{\mathrm{int},\mathcal{E}_\omega}$ | $\tau_{\exp,\mathcal{E}_\omega}=9.83\pm0.12$<br>$\chi^2=21.06$, DF=28<br>level=82% | $\tau_{\exp,\mathcal{E}_\omega}=17.22\pm0.29$<br>$\chi^2=48.94$, DF=47<br>level=40% | $\tau_{\exp,\mathcal{E}_\omega}=29.81\pm0.63$<br>$\chi^2=124.72$, DF=79<br>level=0.08% | $\tau_{\exp,\mathcal{E}_\omega}=57.9\pm1.6$<br>$\chi^2=137.91$, DF=134<br>level=39% | $\tau_{\exp,\mathcal{E}_\omega}=99.71\pm3.2$<br>$\chi^2=99.73$, DF=229<br>level=100% | $\tau_{\exp,\mathcal{E}_\omega}=147.1\pm4.2$<br>$\chi^2=121.00$, DF=356<br>level=100% |
| $1.5\tau_{\mathrm{int},\mathcal{E}_\omega}$ | $\tau_{\exp,\mathcal{E}_\omega}=9.85\pm0.21$<br>$\chi^2=19.65$, DF=23<br>level=66% | $\tau_{\exp,\mathcal{E}_\omega}=17.80\pm0.49$<br>$\chi^2=44.00$, DF=39<br>level=27% | $\tau_{\exp,\mathcal{E}_\omega}=31.2\pm1.1$<br>$\chi^2=85.57$, DF=65<br>level=4% | $\tau_{\exp,\mathcal{E}_\omega}=60.1\pm2.6$<br>$\chi^2=71.20$, DF=111<br>level=100% | $\tau_{\exp,\mathcal{E}_\omega}=104.7\pm5.6$<br>$\chi^2=53.61$, DF=190<br>level=100% | $\tau_{\exp,\mathcal{E}_\omega}=160.5\pm7.1$<br>$\chi^2=58.86$, DF=296<br>level=100% |
| $2\tau_{\mathrm{int},\mathcal{E}_\omega}$ | $\tau_{\exp,\mathcal{E}_\omega}=10.12\pm0.36$<br>$\chi^2=17.26$, DF=18<br>level=51% | $\tau_{\exp,\mathcal{E}_\omega}=18.18\pm0.83$<br>$\chi^2=39.46$, DF=31<br>level=14% | $\tau_{\exp,\mathcal{E}_\omega}=35.9\pm2.3$<br>$\chi^2=46.99$, DF=52<br>level=67% | $\tau_{\exp,\mathcal{E}_\omega}=63.5\pm4.7$<br>$\chi^2=39.02$, DF=89<br>level=100% | $\tau_{\exp,\mathcal{E}_\omega}=114\pm10$<br>$\chi^2$=25.90, DF=152<br>level=100% | $\tau_{\exp,\mathcal{E}_\omega}=163\pm13$<br>$\chi^2=33.66$, DF=237<br>level=100% |

Table 22: Values of $\tau_{\exp,\mathcal{E}_\omega}$ for the ZF model, using $t_{max}=4\tau_{\mathrm{int},\mathcal{E}_\omega}$. Notation as in Table 18.

| $t_{min}$ | $L=16$ <br> $t_{max}=28$ | $L=32$ <br> $t_{max}=48$ | $L=64$ <br> $t_{max}=79$ | $L=128$ <br> $t_{max}=135$ | $L=256$ <br> $t_{max}=229$ | $L=512$ <br> $t_{max}=357$ |
|---|---|---|---|---|---|---|
| 1 | $\tau_{\exp,\mathcal{E}_\omega}=9.29\pm0.05$ <br> $\chi^2=99.77$, DF=26 <br> level=$10^{-10}$ | $\tau_{\exp,\mathcal{E}_\omega}=15.10\pm0.10$ <br> $\chi^2=327.13$, DF=46 <br> level=$5\times10^{-44}$ | $\tau_{\exp,\mathcal{E}_\omega}=24.14\pm0.19$ <br> $\chi^2=579.33$, DF=77 <br> level=$5\times10^{-78}$ | $\tau_{\exp,\mathcal{E}_\omega}=38.93\pm0.37$ <br> $\chi^2=1020.9$, DF=133 <br> level=$8\times10^{-137}$ | $\tau_{\exp,\mathcal{E}_\omega}=67.49\pm0.79$ <br> $\chi^2=1692.7$, DF=227 <br> level=$3\times10^{-222}$ | $\tau_{\exp,\mathcal{E}_\omega}=107.5\pm1.1$ <br> $\chi^2=3642.5$, DF=355 <br> level=$10^{-537}$ |
| $0.5\tau_{\mathrm{int},\mathcal{E}_\omega}$ | $\tau_{\exp,\mathcal{E}_\omega}=9.72\pm0.08$ <br> $\chi^2=9.15$, DF=22 <br> level=99% | $\tau_{\exp,\mathcal{E}_\omega}=16.55\pm0.17$ <br> $\chi^2=41.27$, DF=39 <br> level=37% | $\tau_{\exp,\mathcal{E}_\omega}=27.54\pm0.36$ <br> $\chi^2=81.69$, DF=65 <br> level=8% | $\tau_{\exp,\mathcal{E}_\omega}=50.57\pm0.86$ <br> $\chi^2=314.16$, DF=111 <br> level=$3\times10^{-21}$ | NO <br> CONVERGENCE | $\tau_{\exp,\mathcal{E}_\omega}=139.5\pm2.6$ <br> $\chi^2=213.48$, DF=296 <br> level=100% |
| $\tau_{\mathrm{int},\mathcal{E}_\omega}$ | $\tau_{\exp,\mathcal{E}_\omega}=9.78\pm0.12$ <br> $\chi^2=8.74$, DF=18 <br> level=97% | $\tau_{\exp,\mathcal{E}_\omega}=17.01\pm0.29$ <br> $\chi^2=34.29$, DF=31 <br> level=31% | $\tau_{\exp,\mathcal{E}_\omega}=28.17\pm0.60$ <br> $\chi^2=69.80$, DF=52 <br> level=5% | $\tau_{\exp,\mathcal{E}_\omega}=51.7\pm1.5$ <br> $\chi^2=81.87$, DF=89 <br> level=69% | $\tau_{\exp,\mathcal{E}_\omega}=93.9\pm3.5$ <br> $\chi^2=91.88$, DF=152 <br> level=100% | $\tau_{\exp,\mathcal{E}_\omega}=137.6\pm4.2$ <br> $\chi^2=101.06$, DF=237 <br> level=100% |
| $1.5\tau_{\mathrm{int},\mathcal{E}_\omega}$ | $\tau_{\exp,\mathcal{E}_\omega}=9.72\pm0.21$ <br> $\chi^2=7.37$, DF=13 <br> level=88% | $\tau_{\exp,\mathcal{E}_\omega}=17.36\pm0.48$ <br> $\chi^2=29.87$, DF=23 <br> level=15% | $\tau_{\exp,\mathcal{E}_\omega}=28.8\pm1.1$ <br> $\chi^2=53.00$, DF=38 <br> level=5% | $\tau_{\exp,\mathcal{E}_\omega}=51.8\pm2.5$ <br> $\chi^2=47.82$, DF=66 <br> level=96% | $\tau_{\exp,\mathcal{E}_\omega}=93.9\pm5.8$ <br> $\chi^2=43.53$, DF=113 <br> level=100% | $\tau_{\exp,\mathcal{E}_\omega}=148.0\pm8.0$ <br> $\chi^2=50.63$, DF=177 <br> level=100% |
| $2\tau_{\mathrm{int},\mathcal{E}_\omega}$ | $\tau_{\exp,\mathcal{E}_\omega}=9.84\pm0.35$ <br> $\chi^2=5.54$, DF=8 <br> level=70% | $\tau_{\exp,\mathcal{E}_\omega}=17.24\pm0.79$ <br> $\chi^2=26.92$, DF=15 <br> level=3% | $\tau_{\exp,\mathcal{E}_\omega}=31.6\pm2.0$ <br> $\chi^2=29.63$, DF=25 <br> level=24% | $\tau_{\exp,\mathcal{E}_\omega}=52.3\pm4.3$ <br> $\chi^2=31.20$, DF=44 <br> level=93% | $\tau_{\exp,\mathcal{E}_\omega}=100\pm10$ <br> $\chi^2=24.34$, DF=75 <br> level=100% | $\tau_{\exp,\mathcal{E}_\omega}=145\pm13$ <br> $\chi^2=32.14$, DF=118 <br> level=100% |

Table 23: Values of $\tau_{\exp,\mathcal{E}_\omega}$ for the ZF model, using $t_{max}=3\tau_{\mathrm{int},\mathcal{E}_\omega}$. Notation as in Table 18.

| | 4-state Potts model | | | ZF model | | | X2 model | | |
|---|---|---|---|---|---|---|---|---|---|
| $L$ | $t_{min}$ | $t_{max}$ | $\tau_{\mathrm{int},\mathcal{E}}/\tau_{\exp,\mathcal{E}}$ | $t_{min}$ | $t_{max}$ | $\tau_{\mathrm{int},\mathcal{E}_\omega}/\tau_{\exp,\mathcal{E}_\omega}$ | $t_{min}$ | $t_{max}$ | $\tau_{\mathrm{int},\mathcal{E}_\omega}/\tau_{\exp,\mathcal{E}_\omega}$ |
| 16 | $\frac{1}{2}\tau_{\mathrm{int},\mathcal{E}}$ | $4\tau_{\mathrm{int},\mathcal{E}}$ | $0.96\pm0.03$ | $\frac{1}{2}\tau_{\mathrm{int},\mathcal{E}_\omega}$ | $4\tau_{\mathrm{int},\mathcal{E}_\omega}$ | $0.97\pm0.02$ | $\frac{1}{2}\tau_{\mathrm{int},\mathcal{E}_\omega}$ | $4\tau_{\mathrm{int},\mathcal{E}_\omega}$ | $0.99\pm0.02$ |
| 32 | $\tau_{\mathrm{int},\mathcal{E}}$ | $4\tau_{\mathrm{int},\mathcal{E}}$ | $0.95\pm0.04$ | $\frac{1}{2}\tau_{\mathrm{int},\mathcal{E}_\omega}$ | $4\tau_{\mathrm{int},\mathcal{E}_\omega}$ | $0.96\pm0.03$ | $\frac{1}{2}\tau_{\mathrm{int},\mathcal{E}_\omega}$ | $4\tau_{\mathrm{int},\mathcal{E}_\omega}$ | $0.97\pm0.02$ |
| 64 | $\tau_{\mathrm{int},\mathcal{E}}$ | $4\tau_{\mathrm{int},\mathcal{E}}$ | $0.92\pm0.02$ | $\frac{1}{2}\tau_{\mathrm{int},\mathcal{E}_\omega}$ | $3\tau_{\mathrm{int},\mathcal{E}_\omega}$ | $0.96\pm0.04$ | $\frac{1}{2}\tau_{\mathrm{int},\mathcal{E}_\omega}$ | $3\tau_{\mathrm{int},\mathcal{E}_\omega}$ | $0.98\pm0.03$ |
| 128 | $\tau_{\mathrm{int},\mathcal{E}}$ | $3\tau_{\mathrm{int},\mathcal{E}}$ | $0.84\pm0.04$ | $\tau_{\mathrm{int},\mathcal{E}_\omega}$ | $3\tau_{\mathrm{int},\mathcal{E}_\omega}$ | $0.87\pm0.06$ | $\frac{1}{2}\tau_{\mathrm{int},\mathcal{E}_\omega}$ | $3\tau_{\mathrm{int},\mathcal{E}_\omega}$ | $0.99\pm0.03$ |
| 256 | $\tau_{\mathrm{int},\mathcal{E}}$ | $3\tau_{\mathrm{int},\mathcal{E}}$ | $0.87\pm0.05$ | $\tau_{\mathrm{int},\mathcal{E}_\omega}$ | $3\tau_{\mathrm{int},\mathcal{E}_\omega}$ | $0.81\pm0.07$ | $\frac{1}{2}\tau_{\mathrm{int},\mathcal{E}_\omega}$ | $3\tau_{\mathrm{int},\mathcal{E}_\omega}$ | $0.94\pm0.06$ |
| 512 | $\tau_{\mathrm{int},\mathcal{E}}$ | $3\tau_{\mathrm{int},\mathcal{E}}$ | $0.77\pm0.06$ | $\frac{1}{2}\tau_{\mathrm{int},\mathcal{E}_\omega}$ | $3\tau_{\mathrm{int},\mathcal{E}_\omega}$ | $0.85\pm0.05$ | $\frac{1}{2}\tau_{\mathrm{int},\mathcal{E}_\omega}$ | $3\tau_{\mathrm{int},\mathcal{E}_\omega}$ | $0.94\pm0.05$ |

Table 24: Ratios $\tau_{\mathrm{int},\mathcal{E}}/\tau_{\exp,\mathcal{E}}$ for the three models considered in this paper. For each lattice size $L$ we give the ratio and the interval $[t_{min}, t_{max}]$ used for its computation.

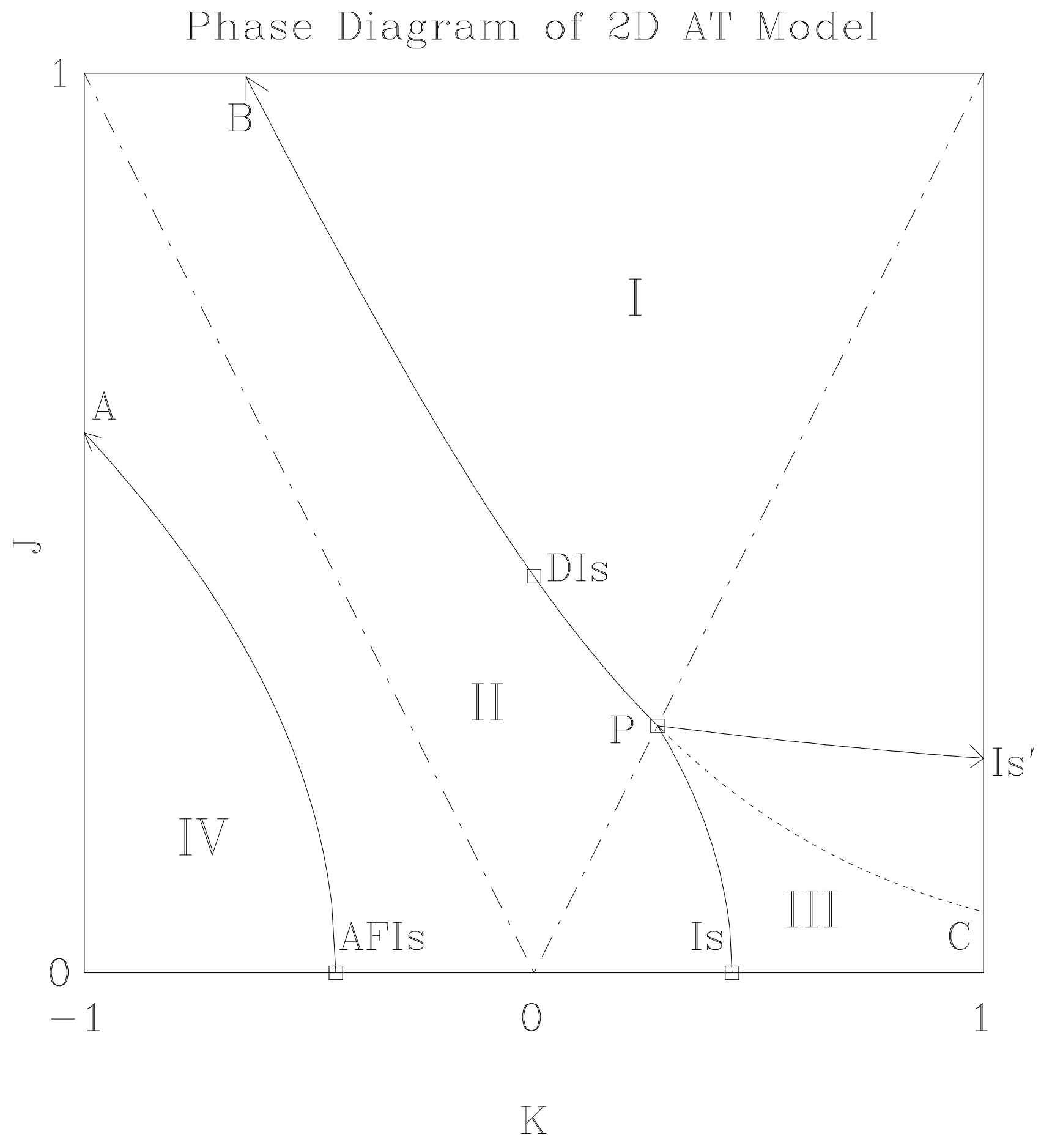

Figure 1: Phase diagram of the symmetric Ashkin–Teller model on the square lattice. The self-dual curve is B–DIs–P–C. The solid curves represent second-order phase transitions, the dash-dotted ones the 4-state Potts-model subspace, and the dotted one the non-critical part of the self-dual curve. The Roman numerals designate the different phases of the model (see text).

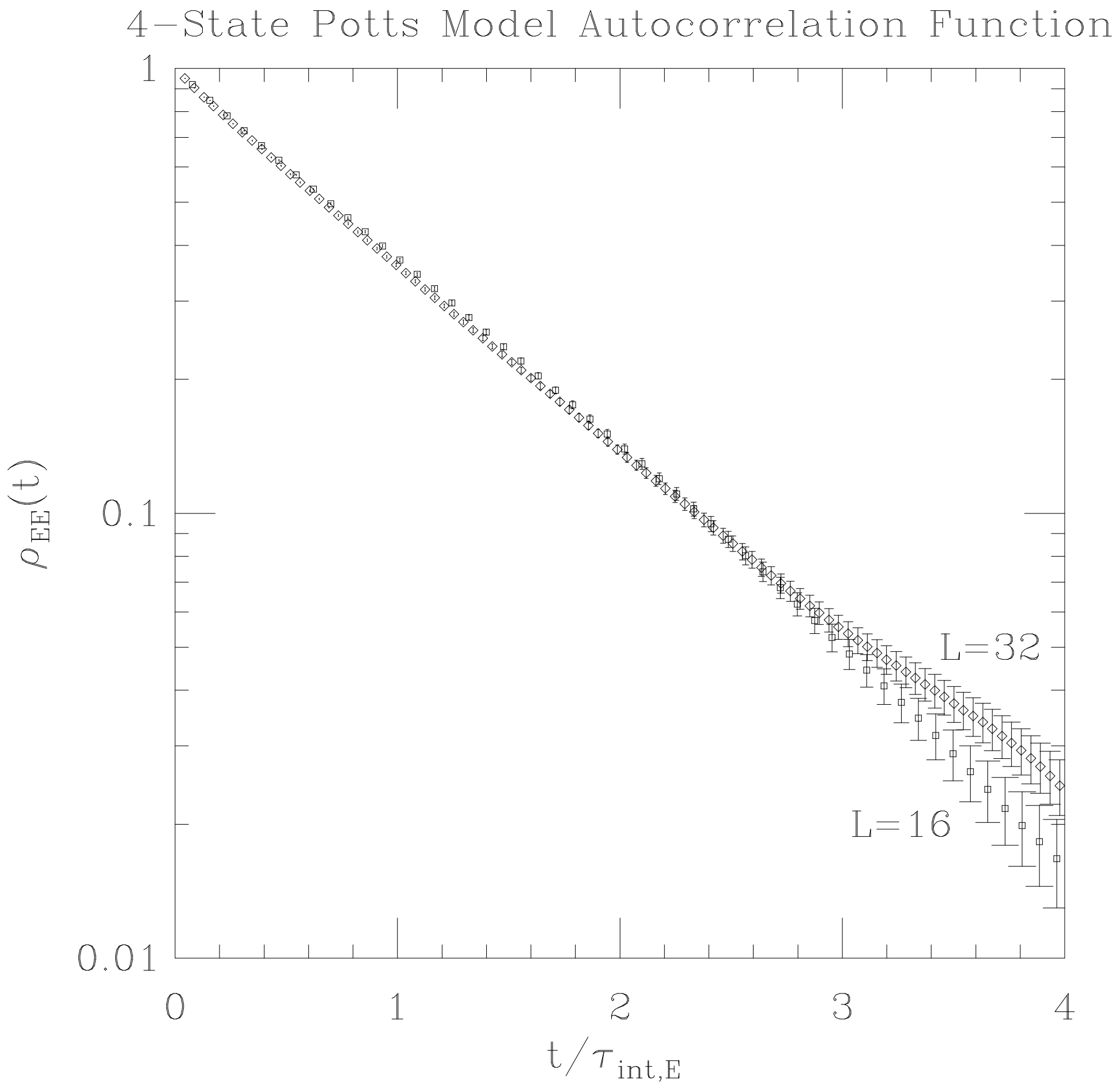

Figure 2: Autocorrelation function $\rho_{\mathcal{E}\mathcal{E}}(t)$ for the 4-state Potts model and $L = 16$ ($\square$) and $L = 32$ ($\diamond$), with the abscissa scaled by $\tau_{\mathrm{int},\mathcal{E}}$. The error bars are the square root of the diagonal terms of the covariance matrix (see Appendix B).

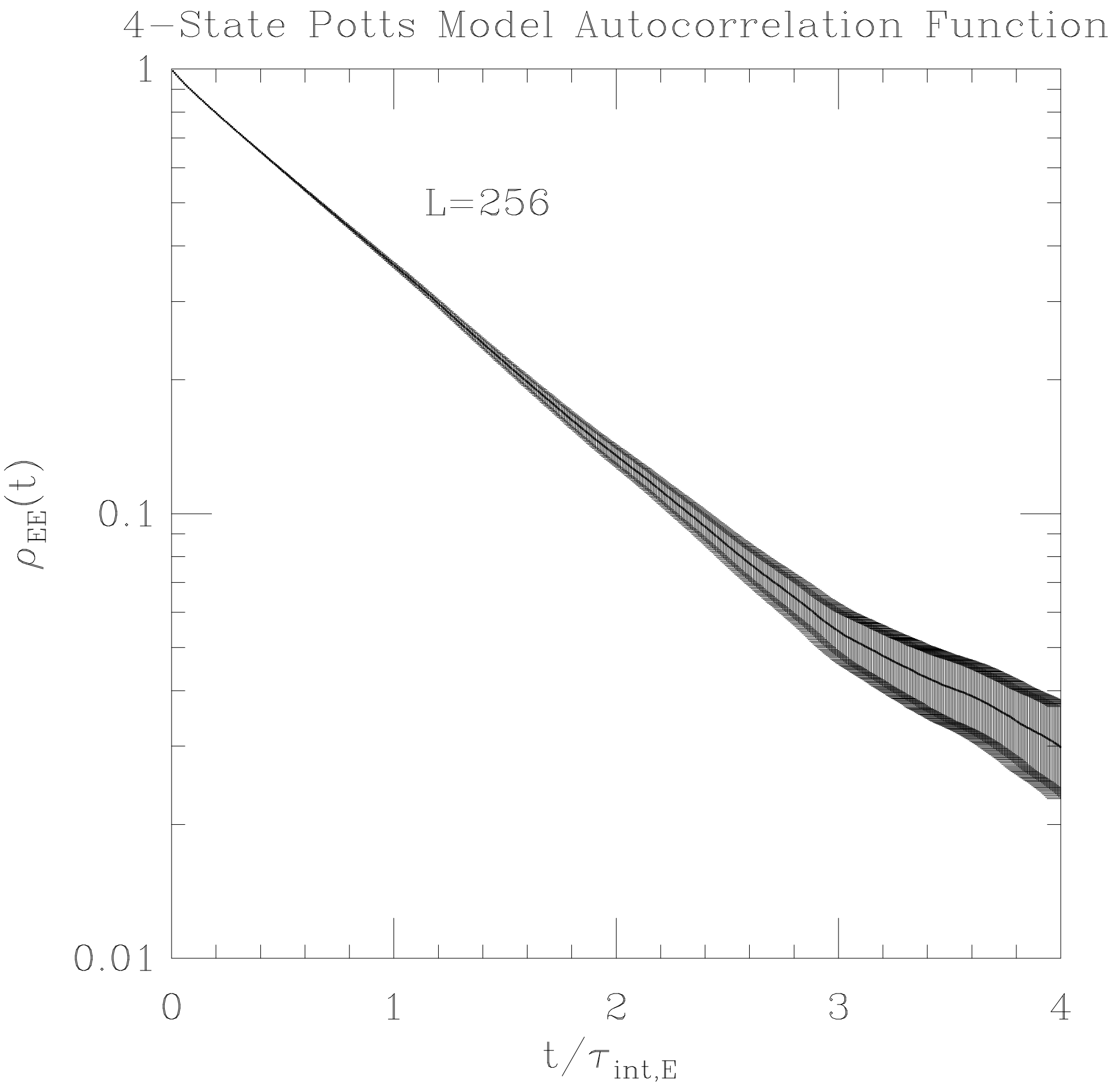

Figure 3: The same as in Figure 2, for $L = 256$.